\documentclass[twocolumn]{aastex63}

\usepackage{amsmath}

\usepackage{graphicx}
\usepackage{enumitem}
\usepackage{mathrsfs}

\shorttitle{Quenching Fast, Slow, Early, Late}
\shortauthors{Tacchella et al.}

\begin{document}

\title{\vspace{-0.2cm}Fast, Slow, Early, Late: Quenching Massive Galaxies at $z\sim0.8$}

\correspondingauthor{Sandro Tacchella}
\email{tacchella@unist.ac.kr}

\author[0000-0002-8224-4505]{Sandro Tacchella}
\affil{Center for Astrophysics $\vert$ Harvard \& Smithsonian, 60 Garden St., Cambridge, MA 02138, USA}
\affil{Department of Physics, Ulsan National Institute of Science and Technology (UNIST), Ulsan 44919, Republic of Korea}
\author[0000-0002-1590-8551]{Charlie Conroy}
\affil{Center for Astrophysics $\vert$ Harvard \& Smithsonian, 60 Garden St, Cambridge, MA 02138, USA}
\author[0000-0002-1590-8551]{S. M. Faber}
\affil{Department of Astronomy and Astrophysics, University of California, Santa Cruz, CA 95064, USA}
\author[0000-0002-9280-7594]{Benjamin D. Johnson}
\affil{Center for Astrophysics $\vert$ Harvard \& Smithsonian, 60 Garden St, Cambridge, MA 02138, USA}
\author[0000-0001-6755-1315]{Joel Leja}
\affil{Department of Astronomy \& Astrophysics, The Pennsylvania State University, University Park, PA 16802, USA}
\affil{Institute for Computational \& Data Sciences, The Pennsylvania State University, University Park, PA, USA}
\affil{Institute for Gravitation and the Cosmos, The Pennsylvania State University, University Park, PA 16802, USA}
\author[0000-0001-6813-875X]{Guillermo Barro}
\affil{Department of Physics, University of the Pacific, Stockton, CA 95211, USA}
\author[0000-0002-6993-0826]{Emily C. Cunningham}
\affil{Center for Computational Astrophysics, Flatiron Institute, 162 5th Ave., New York, NY 10010, USA}
\author[0000-0001-6146-2645]{Alis J. Deason}
\affil{Institute for Computational Cosmology, Department of Physics, University of Durham, South Road, Durham DH1 3LE, UK}
\affil{Centre for Extragalactic Astronomy, Department of Physics, University of Durham, South Road, Durham DH1 3LE, UK}
\author[0000-0001-8867-4234]{Puragra Guhathakurta}
\affil{Department of Astronomy and Astrophysics, University of California, Santa Cruz, CA 95064, USA}
\author[0000-0003-2775-2002]{Yicheng Guo}
\affil{Department of Physics and Astronomy, University of Missouri,
Columbia, MO 65211, USA}
\author[0000-0001-6950-1629]{Lars Hernquist}
\affil{Center for Astrophysics $\vert$ Harvard \& Smithsonian, 60 Garden St, Cambridge, MA 02138, USA}
\author[0000-0003-3385-6799]{David C. Koo}
\affil{Department of Astronomy and Astrophysics, University of California, Santa Cruz, CA 95064, USA}
\author[0000-0001-7494-5910]{Kevin McKinnon}
\affil{Department of Astronomy and Astrophysics, University of California, Santa Cruz, CA 95064, USA}
\author[0000-0001-7494-5910]{Constance M. Rockosi}
\affil{Department of Astronomy and Astrophysics, University of California, Santa Cruz, CA 95064, USA}
\author[0000-0002-5065-9896]{Joshua S. Speagle}
\affil{Department of Statistical Sciences, University of Toronto, Toronto, ON M5S 3G3, Canada}
\affil{David A. Dunlap Department of Astronomy \& Astrophysics, University of Toronto, Toronto, ON M5S 3H4, Canada}
\affil{Dunlap Institute for Astronomy \& Astrophysics, University of Toronto, Toronto, ON M5S 3H4, Canada}
\author[0000-0002-8282-9888]{Pieter van Dokkum}
\affil{Astronomy Department, Yale University, New Haven, CT 06511, USA}
\author[0000-0002-4176-9145]{Hassen M. Yesuf}
\affil{Kavli Institute for Astronomy and Astrophysics, Peking University, Beijing 100871, China}
\affil{Kavli Institute for the Physics and Mathematics of the Universe, The University of Tokyo, Kashiwa, Japan 277-8583}

\begin{abstract}
We investigate the stellar populations for a sample of 161 massive, mainly quiescent galaxies at $\langle z_{\rm obs} \rangle=0.8$ with deep Keck/DEIMOS rest-frame optical spectroscopy (HALO7D survey). With the fully Bayesian framework \texttt{Prospector}, we simultaneously fit the spectroscopic and photometric data with an advanced physical model (including non-parametric star-formation histories, emission lines, variable dust attenuation law, and dust and AGN emission) together with an uncertainty and outlier model. We show that both spectroscopy and photometry are needed to break the dust-age-metallicity degeneracy. We find a large diversity of star-formation histories: although the most massive ($M_{\star}>2\times10^{11}~M_{\odot}$) galaxies formed the earliest (formation redshift of $z_{\rm f}\approx5-10$ with a short star-formation timescale of $\tau_{\rm SF}\lesssim1~\mathrm{Gyr}$), lower-mass galaxies have a wide range of formation redshifts, leading to only a weak trend of $z_{\rm f}$ with $M_{\star}$. Interestingly, several low-mass galaxies with have formation redshifts of $z_{\rm f}\approx5-8$. Star-forming galaxies evolve about the star-forming main sequence, crossing the ridgeline several times in their past. Quiescent galaxies show a wide range and continuous distribution of quenching timescales ($\tau_{\rm quench}\approx0-5~\mathrm{Gyr}$) with a median of $\langle\tau_{\rm quench}\rangle=1.0_{-0.9}^{+0.8}~\mathrm{Gyr}$ and of quenching epochs of $z_{\rm quench}\approx0.8-5.0$ ($\langle z_{\rm quench}\rangle=1.3_{-0.4}^{+0.7}$). This large diversity of quenching timescales and epochs points toward a combination of internal and external quenching mechanisms. In our sample, rejuvenation and ``late bloomers'' are uncommon. In summary, our analysis supports the ``grow \& quench'' framework and is consistent with a wide and continuously populated diversity of quenching timescales.
\end{abstract}


\section{Introduction}
\label{sec:introduction}

How and why galaxies grow in stellar mass and cease their star formation are key open questions of galaxy formation and evolution. Although scaling relations between the star formation activity and other galaxy properties such as stellar mass, morphology, and environment exist, it is challenging observationally to constrain how \textit{individual} galaxies evolve about these scaling relations. The goal of this paper is to measure detailed star-formation histories (SFHs) of individual galaxies (including that of prior merged galaxies) at early cosmic times to assess on which timescales galaxies form their stars and then cease their star formation.

Over the past two decades, both observational and theoretical studies have motivated a paradigm shift in galaxy evolution, in which smooth gas accretion plays a major role compared to galaxy-galaxy mergers in driving high star-formation rates (SFRs) at early cosmic times (redshifts of $z>1$; see \citealt{forster-schreiber20} for a review). Observations show that a majority of these early star-forming galaxies are rotating disks without any sign of ongoing merging \citep{genzel06, genzel08, wisnioski15, simons17, forster-schreiber18_SINS} and that their SFRs are tightly correlated with their stellar mass ($M_{\star}$) over several orders of magnitude, a correlation often called the star-forming main sequence (SFMS; \citealt{noeske07, daddi07, whitaker12, renzini15, speagle16}). 

This empirical evidence of galaxies sustaining their SFRs over prolonged periods of time through continuous gas accretion is supported by cosmological models. Numerical simulations show that massive galaxies can acquire a large fraction of their gas via steady cold inflows that penetrate effectively through the shock-heated media of massive dark matter halos \citep{keres05, dekel06, dekel08, faucher-giguere11a}. Furthermore, simulations and (semi)analytical models naturally reproduce the observed star-forming main sequence, indicating that galaxies -- even at early cosmic times -- self-regulate and grow along the evolving SFMS \citep{bouche10, lilly13_bathtub, dekel14_bathtub, mitchell14, sparre15_MS, rodriguez-puebla16, tacchella16_MS, tacchella18, donnari19}.

While the majority of massive galaxies are star-forming at $z\sim2$, a population of quiescent galaxies is building up with cosmic time and dominates the massive end of the galaxy stellar mass function at $z\sim0$ \citep{ilbert13, muzzin13, davidzon17}. Therefore, with passing cosmic time, galaxies transition from being star-forming to being quiescent \citep{bell04, faber07}, a process often called ``quenching''. The evolving SFMS with a simple prescription for quenching (for example, at fixed stellar mass) and merging is indeed able to explain the evolution of the mass functions of star-forming and quiescent galaxies with cosmic time \citep{peng10_Cont}. Besides these observations, this ``grow \& quench'' framework together with the buildup of lower mass quiescent galaxies in high-density environments (sometimes referred to as satellite quenching; e.g., \citealt{peng12}) can explain that the sites of active star formation shift from high-mass galaxies at early times to lower mass systems at later epochs (``downsizing''; \citealt{cowie96, gallazzi05, gallazzi21, bundy06}), why more massive galaxies are older while their halos have assembled more recently \citep{thomas99, graves09a}, and the morphological landscape of galaxies \citep{carollo13a, bluck14, damjanov14, damjanov19, lilly16, barro17, mosleh17, tacchella17, tacchella19, osborne20, chen20}.

While the grow \& quench framework is able to successfully explain a wide variety of observations, it has recently been called into question. The fundamental problem is that we cannot observe individual galaxies growing and then quenching: as observers, we are bound to observe different galaxies at different cosmic epochs, which allows us to do cross-sectional studies, but not longitudinal ones \citep{abramson16}. Therefore, the SFMS as a fundamental pillar of the grow \& quench framework may not indicate a scaling law about which individual galaxies grow but could arise instead from a diverse family of log-normal SFHs that look significantly different from simply following SFMS \citep{kelson16, abramson16}. Generally, this raises the question of whether and how galaxies evolve about the SFMS \citep{kelson14, abramson15, munoz15, caplar19, tacchella20}. On the other hand, the archaeological record of the galaxies' stellar populations ought in principle to encode how \textit{individual} galaxies evolve with cosmic time \citep[e.g.,][]{thomas99, renzini06, graves09a, trager09, pacifici16, morishita19, webb20} -- an avenue we follow in this paper. As we also note later in the paper, this archaeological approach gives the integrated evolution of all the stellar components in a galaxy, which may have assembled via different evolutionary tracks.

Another open question related to the grow \& quench framework is about quenching: which physical mechanism(s) are responsible for shutting down the star formation? Are galaxies quenching ``fast'' or ``slow''? From semianalytical and cosmological models it clear that some process is needed to inhibit the growth of too massive galaxies, possibly pointing to black hole feedback \citep{di-matteo05, croton06, bower06, hopkins06a}, supernova feedback \citep{springel05b, cox06_fb, dalla-vecchia12, lagos13} or virial shock heating of gaseous halos \citep{birnboim03, keres09}. Furthermore, different mechanisms could interact with each other leading to a complex interplay. For example, a hot halo might be required for quenching but only quenches a galaxy in cooperation with stellar or black hole feedback \citep[e.g.,][]{voit15, tacchella16_MS, bower17, chen20}. Since different processes could act on distinct timescales and spatial scales, observationally constraining the epoch of quenching and the quenching timescale could help with pinning down the quenching mechanism \citep[e.g.,][]{rodriguez-montero19, wright19, park21}. 

Focusing first on the epoch of quenching, observations show that quenching is happening continuously over cosmic time, starting back at $z\gtrsim3$ \citep[e.g.,][]{gobat12, kriek16, valentino20} and continuing to today \citep[e.g.,][]{bell04, faber07, peng10_Cont, barro13, muzzin13, ilbert13}. Importantly, quiescent galaxies retain information about the time and manner of their quenching, as manifested in ($i$) structural scaling laws obeyed by quenching galaxies back in time \citep[e.g.,][]{van-der-wel14a, barro17, chen20} and ($ii$) relationships between structure and stellar population properties (i.e., ages, metallicities) in the Fundamental Plane space of quiescent galaxies today \citep[e.g.,][]{graves10b, cappellari16}. In particular, \citet{graves10b} predicted the duration of the star-forming phase and the onset of quenching in different parts of the Fundamental Plane from stacks of SDSS spectra of $z\sim0$ quiescent galaxies. They find that the local Fundamental Plane reveals a wide range of quenching histories at a given $M_{\star}$ back in time in the form of a wide range of stellar ages, while these diverse histories seem to tighten up when considering velocity dispersion instead of $M_{\star}$. 

There has also been a large effort to constrain the quenching timescale at low redshifts \citep[e.g.,][]{wetzel13, schawinski14, yesuf14, peng15, hahn17, smethurst18, trussler20} as well as higher redshifts \citep[e.g.,][]{barro13, belli15, belli19, belli21, tacchella15_sci, fossati17, wu18, herrera-camus19, estrada-carpenter20, wild20}. These studies employ a wide range of different methods and quenching definitions, making it difficult to compare them to each other, and also to theoretical predictions. Broadly speaking, at lower redshifts ($z<0.5$), massive galaxies quench on timescales of $\tau_{\rm quench}\approx1-4$ Gyr, while galaxies at higher redshifts ($z>1$) quench on shorter timescales $\tau_{\rm quench}<1$ Gyr. Furthermore, at all epochs, a population of quenched post-starburst galaxies, also known as K+A or E+A galaxies, exists, which recently quenched on short timescales \citep{dressler83, quintero04, wild09, wild20, yesuf14}. These studies highlight that there is a wide range of quenching timescales, usually referred to as a ``slow'' and a ``fast'' quenching channel. However, it is not clear whether quenching timescales are really following a bi-modal distribution.

In this paper, we focus on constraining the SFHs of galaxies at an epoch when the universe was half of its current age ($z\sim0.8$). Accurate measurements of SFHs rely on high-quality data, both photometric and spectral. Even with high-quality data, predictions of SFHs are increasingly less accurate the farther back in time one extrapolates from the epoch of observation. High-quality spectral data have typically been available for \textit{local} galaxies because they are bright, but observations taken at today's epoch mean that early epochs remain shrouded in mystery. The present data set moves the epoch of observation back in time, closer to key evolutionary events, allowing us to focus on the following two questions: (1) do galaxies grow along the SFMS during their star-forming phase, and (2) when and how rapidly does star-formation cease?

We present deep Keck/DEIMOS rest-frame optical spectroscopy of 161 massive galaxies at $z\sim0.8$. We combine this high spectral resolution broadband spectroscopy with accurate photometry in the key wavelength range from $\sim2000$ \AA~ to 12 $\mu$m (rest-frame). With the fully Bayesian framework \texttt{Prospector} \citep{johnson21}, we simultaneously fit spectroscopic and photometric data in order to break the dust-SFH-metallicity degeneracy. We measure a large diversity of SFHs, giving rise to a wide range in star-formation and quenching timescales. Nevertheless, our results are consistent with the grow \& quench framework, where galaxies evolve about the SFMS ridgeline while star forming, followed by quenching. We find that rejuvenation plays only a minor role. These results have important implications for structuring future galaxy modeling programs, both nearby and distant. In the future, we will build on this analysis, relating SFHs from this work to the galaxies' morphology, structural parameters, and metal abundance.

Throughout this work, we will use a rather broad definition of quenching. Specifically, quenching is defined as the process in which galaxies cease their star formation and transition from star-forming to quiescent. In the literature, a wide range of different criteria have been used to distinguish star-forming and quiescent galaxies, ranging from cuts in color to specific SFR ($\mathrm{sSFR}=\mathrm{SFR}/M_{\star}$; see, e.g., \citealt{leja19_uvj} for a comparison of sSFR and color cuts). Here we consider a cut in sSFR because it quantifies best whether a galaxy is still increasing its stellar mass owing to star formation or not. In particular, one can write:

\begin{equation}
    M_{\star}(t) = M_{\star}(t_0)\times e^{\int_{t_0}^{t} \mathrm{sSFR}(t') \mathrm{d}t'},
\label{eq:stellar_mass}
\end{equation}

\noindent
where $M_{\star}(t_0)>0$ is the stellar mass of the galaxy at some earlier time $t_0$ (i.e. $t_0<t$). From this, one can derive the mass-doubling number $\mathscr{D}$

\begin{equation}
    \mathscr{D}(z) = \mathrm{sSFR}(z) \times t_{\rm H}(z),
\label{eq:mass_double}
\end{equation}

\noindent
which is the number of times the stellar mass doubles within the age of the universe at redshift $z$, $t_{\rm H}(z)$, assuming a constant sSFR. Throughout this work, we classify galaxies as star-forming, transitioning, and quiescent if $\mathscr{D}(z)>1/3$, $1/20<\mathscr{D}(z)<1/3$, and $\mathscr{D}(z)<1/20$, respectively. The motivations for these cuts are given in Section~\ref{subsec:sample}. It is important to note that a large fraction of the ``green valley'' at $z=0$ has $\mathscr{D}<1/20$, indicating that these galaxies can be considered quiescent. This is not the case at earlier cosmic times since the population sSFRs are overall higher relative to the age of the universe. 

The outline of this paper is as follows. Section~\ref{sec:observational_data} describes the galaxy sample and observational data. Sections~\ref{sec:physical_model} and \ref{sec:fitting} describe the physical model adopted to describe the observational data and the fitting procedure itself, respectively. Section~\ref{sec:results} presents the results. We discuss the results in Section~\ref{sec:discussion} and conclude in Section~\ref{sec:conclusion}. Throughout this work, we assume the cosmological parameters of WMAP-7 \citep{komatsu11}.

\section{Sample and data}
\label{sec:observational_data}

In this section we describe the spectroscopic and photometric data used in our analysis (Sections~\ref{subsec:spectroscopy} and \ref{subsec:photometry}). In Section~\ref{subsec:sample}, we discuss how our sample of galaxies relates to the underlying galaxy population at $z\sim0.8$.

\subsection{Spectroscopy}
\label{subsec:spectroscopy}

The spectroscopic data have been taken as part of the HALO7D program, a survey conducted in CANDELS fields \citep{grogin11, koekemoer11} with the Keck II/DEIMOS instrument \citep{faber03}. HALO7D is a multi-semester program with the main goal of surveying faint halo stars with Hubble Space Telescope (HST) measured proper motions in order to measure their line-of-sight velocities and chemical abundances, giving 6D phase-space information and chemical abundances for hundreds of remote Milky Way halo stars \citep{cunningham19a, cunningham19b}. The targeted fields, together with the deep exposures necessary to reach the faintest stars in the Milky Way halo, are an opportunity for a novel synergy of extragalactic and Galactic science. In addition to the primary halo star targets -- which only occupy about a quarter of slitlets on a given DEIMOS mask -- spectra for extragalactic targets have been taken. These data have been used to study galactic winds in $z\sim1$ \citep[][Wang et al. in prep.]{yesuf17}, internal galaxy kinematics (Barro et al. in prep.), and dwarf galaxies (Guo et al. in prep.). Here we focus on the highest-priority filler sample of galaxies, i.e., massive star-forming and quiescent galaxies at $z\sim0.8$.

\subsubsection{Sample selection}
\label{subsec:selection}

This filler sample of massive star-forming and quiescent $z\sim0.8$ galaxies has been selected from the CANDELS survey and an extended region around EGS with IRAC imaging. This latter, EGS IRAC-selected galaxy sample is drawn from the Rainbow database\footnote{\url{http://rainbowx.fis.ucm.es/Rainbow_navigator_public/}} \citep{barro11_phot}, which covers an area of 1728 arcmin$^2$ centered on the EGS and provides spectral energy distributions (SEDs) ranging from the UV to the mid-IR (MIR) regime \citep{barro11}. The highest-priority targets are galaxies with stellar masses of $M_{\star}>10^{11}~\mathrm{M_{\odot}}$, including both star-forming and quiescent galaxies. The second-highest priority includes galaxies with $M_{\star}=10^{10}-10^{11}~\mathrm{M_{\odot}}$ and UVJ-quiescent colors. We emphasize that this sample is not volume or mass complete but traces the massive galaxy population around $z\sim0.8$. It is an unbiased sample of galaxies above $M_{\star}>10^{11}~\mathrm{M_{\odot}}$ but biased toward quiescent galaxies below this mass limit. 

\subsubsection{Observations and data reduction}
\label{subsec:observations}

\begin{figure}
    \centering
    \includegraphics[width=\linewidth]{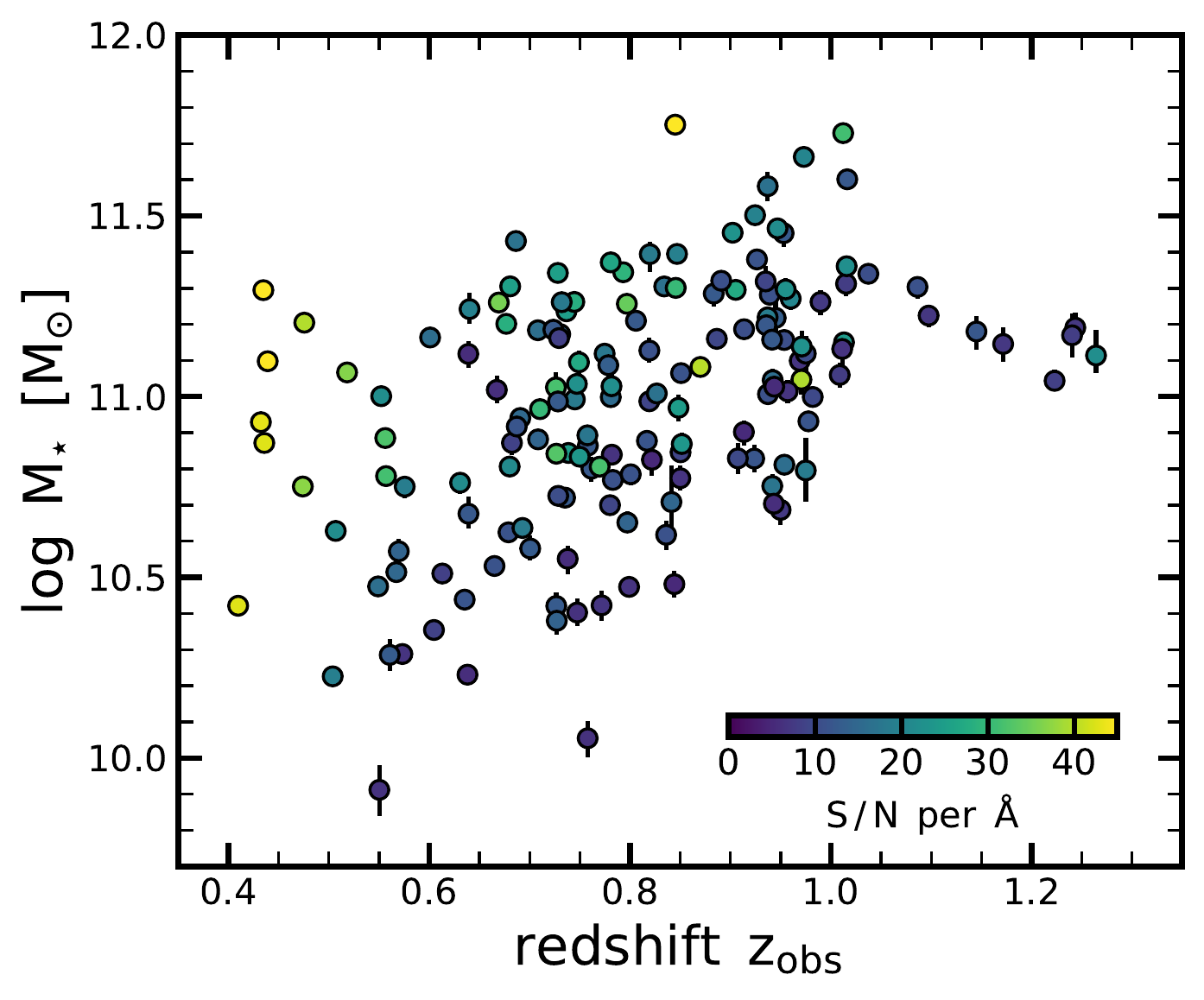}
    \caption{Signal-to-noise (S/N) ratio of our galaxies in the $M_{\star}-z_{\rm obs}$ plane. Our sample of 161 galaxies covers a wide redshift range ($z_{\rm obs}\approx0.4-1.2$) and spans about two orders of magnitude in stellar mass ($M_{\star}\approx10^{10}-10^{12}~M_{\odot}$). The average (and median) redshift of our sample is $\langle z_{\rm obs} \rangle=0.8$. The color-coding of the points corresponds to S/N of the spectrum, measured in wavelength window of $\lambda_{\rm obs}=7000-9200~\mathrm{\AA}$. We only consider galaxies with $\mathrm{S/N}>5$ per $\mathrm{\AA}$. Galaxies in our sample have spectra with S/N between $5.2-62.0$ per $\mathrm{\AA}$, with an average of 17.4 per $\mathrm{\AA}$.}
    \label{fig:snr}
\end{figure}

The observations and data reductions are described in detail in \citet{cunningham19a}. In summary, the HALO7D observations use the 600 line/millimeter grating on DEIMOS centered around $7200~\mathrm{\AA}$ with the GG455 order-blocking filter. This setup gives a nominal wavelength coverage of $4600-9500~\mathrm{\AA}$ at a resolution (FWHM) of $\sim3.5~\mathrm{\AA}$ for a $1\arcsec$ slit width and $0.65~\mathrm{\AA}/\mathrm{pixel}$ dispersion. The slit position angles are set to within $\pm30\%$ of the parallactic angle to minimize light loss in the blue due to atmospheric dispersion. In total, 232 galaxies were observed. Useful data (continuum signal-to-noise [S/N] ratio of at least 5 per $\mathrm{\AA}$, no major artifacts in the data, and no active galactic nucleus (AGN) with point sources or with broad lines) have been collected for 161 galaxies. For the remainder of this paper, we focus on those galaxies. The exposure times of these galaxies range between 4.0 and 48.6 hr, with an average of 11.1 hr. The HALO7D observations were reduced using the automated DEEP2/DEIMOS spec2d pipeline developed by the DEEP2 team \citep{cooper12, newman13}, which, among other things, also performs the sky subtraction. Wavelength regions that are heavily affected by skylines are masked, making up about 15\% of all the pixels. Calibrations were done using a quartz lamp for flat-fielding and red NeKrArXe lamps for wavelength calibration. We do not perform any flux calibration of the spectra since we directly model the spectroscopic flux calibration during fitting (Section~\ref{subsec:spec_cali}). The instrumental line-spread function (LSF) has been measured from these arc lamps as well as the night skylines, as described in Appendix~\ref{app_sec:lsf}. Importantly, our analyses of both photometry and spectroscopy in this work make the simplifying assumption that all galaxies are spatially uniform. In future work, we will account for and exploit spatial variations in colors and stellar populations that do exist \citep[e.g.,][]{szomoru13, tacchella15, mosleh17, suess19}.

Fig.~\ref{fig:snr} shows the stellar mass ($M_{\star}$) as a function of observed redshift ($z_{\rm obs}$) for our sample of 161 galaxies. The stellar masses are obtained from our SED modeling, as described in Section~\ref{sec:fitting}. The color scaling of the points indicates the S/N ratio of the spectra, measured in an observed-frame wavelength window of $\lambda_{\rm obs}=7000-9200~\mathrm{\AA}$. Our sample spans a wide range in redshift with $z_{\rm obs}\approx0.4-1.2$ and about two orders in stellar mass ($M_{\star}\approx10^{10}-10^{12}~M_{\odot}$). The core of the sample lies in the redshift interval $0.6-1.0$, with an average redshift of $\langle z_{\rm obs} \rangle=0.8$. The median stellar mass is $\log M_{\star}/M_{\odot} = 11.0$. The galaxies at $z_{\rm obs}<0.6$ have the highest S/N. In the core redshift range of our sample ($z_{\rm obs}=0.6-1.0$), no clear trend of redshift with S/N exists. The quality of our spectra is comparable to that of the LEGA-C survey \citep{van-der-wel16}, but our sample is smaller while probing a larger redshift range. Importantly, although the redshift range probed by our galaxies is rather large, thanks to the broad wavelength coverage ($\lambda_{\rm obs}=4600-9500~\mathrm{\AA}$), key absorption features are covered by all galaxies. Specifically, the core sample of our galaxies probes the hydrogen absorption lines from H10 (found at $3799~\mathrm{\AA}$) to H$\beta$ (at $4863~\mathrm{\AA}$), the Calcium H and K lines (at $3934~\mathrm{\AA}$ and $3969~\mathrm{\AA}$), the CN line (at $4160~\mathrm{\AA}$), the MgIb triplet (at $5176~\mathrm{\AA}$), and several other Mg (at $5530~\mathrm{\AA}$), Ca (including at $4227~\mathrm{\AA}$ and $4455~\mathrm{\AA}$) and Fe lines (including at $4383~\mathrm{\AA}$, $4531~\mathrm{\AA}$, $4668~\mathrm{\AA}$, and $5270~\mathrm{\AA}$). Only galaxies probing the highest redshifts ($z_{\rm obs}>1$) do not have coverage of the MgIb triplet and the H$\beta$ line.

\subsection{Photometry}
\label{subsec:photometry}

\begin{figure*}
    \centering
    \includegraphics[width=\textwidth]{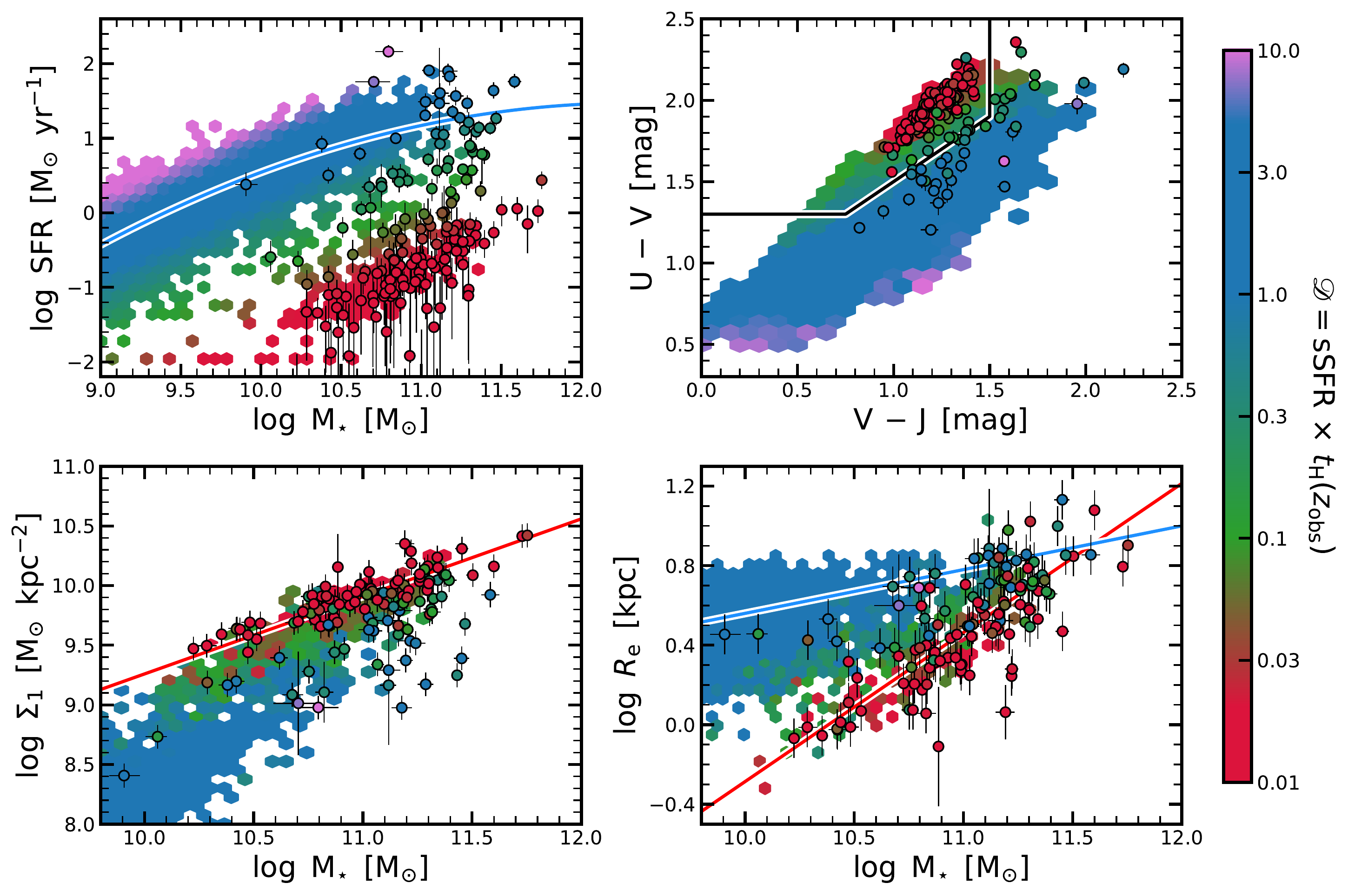}
    \caption{Our sample of 161 galaxies in comparison with the whole galaxy population at $z=0.6-1.0$. The panels from top left to bottom right show the planes of $\mathrm{SFR}-M_{\star}$, rest-frame UVJ colors, $\Sigma_1-M_{\star}$ and $R_{\rm e}-M_{\star}$. Our galaxies are shown as circles, while the whole galaxy population from 3D-HST is shown with the underlying hexbins (a total of $\sim10,000$ objects at $z=0.6-1.0$). The color-coding in all panels corresponds to the doubling number $\mathscr{D}$, as defined by Eq.~\ref{eq:mass_double}. The solid blue line in the top left panel shows the SFMS from \citet{leja21}, the solid red line in the bottom left panel indicates the $\Sigma_1-M_{\star}$ relation for quiescent galaxies from \citet{barro17}, and the solid blue and red lines show the $R_{\rm e}-M_{\star}$ relations for star-forming and quiescent galaxies from \citet{van-der-wel14a}, respectively. Our galaxy sample follows the typical trends of the galaxy population (i.e., more quiescent galaxies lie in the UVJ-quiescent box, have higher central stellar mass densities $\Sigma_1$, and have smaller sizes $R_{\rm e}$). By selection (Section~\ref{subsec:selection}), our sample focuses on massive, quiescent galaxies and only contains a few star-forming galaxies.}
    \label{fig:sample}
\end{figure*}

We match the 161 HALO7D galaxies to photometric catalogs. As described in the previous section, most of our galaxies ($82.8\%$ of the sample) lie in the CANDELS survey footprint. Specifically, $19.0\%$, $20.9\%$, and $42.9\%$ of the sample lies in COSMOS, EGS, and GOODS-N, respectively. We match those galaxies with 3D-HST photometric catalogs \citep{skelton14}. Specifically, our galaxies are covered by between 17 (the EGS field) and 44 (the COSMOS field) photometric bands spanning a range of $0.3-8\mu\mathrm{m}$ in the observed frame. The photometry is supplemented by Spitzer/MIPS $24\mu\mathrm{m}$ fluxes from \citet{whitaker14}. The MIPS $24\mu\mathrm{m}$ coverage is important because the rest-frame MIR wavelengths are dominated by warm dust emission, a key empirical proxy for obscured star formation \citep{kennicutt98}. The other 28 galaxies ($17.2\%$ of the sample) are not lying in the 3D-HST footprint but in the extended EGS region. We use the UV-IR photometry in the Rainbow database\footnote{\url{http://rainbowx.fis.ucm.es/Rainbow_navigator_public/}} published by \citet{barro11_phot} for those objects. 

The 3D-HST team self-consistently rederives zero-points for each instrument and filter in order to bring data from different instruments onto a common flux scale. Details are described in \citet{skelton14}. Since this process is imperfect, we adopt the procedure by \citet{leja19} and add the zero-point correction for each band of photometry to the flux errors in quadrature. This effect varies from $0\%$ to $28\%$ of the total flux, depending on the photometric band. Additionally, a $5\%$ minimum error is enforced for each band of photometry to allow for systematic errors in the physical models for stellar, gas, and dust emission.

\subsection{Galaxy Sample}
\label{subsec:sample}

In this section, we compare our galaxy sample to the underlying galaxy population at $z=0.6-1.0$, the core redshift range of our sample. Fig.~\ref{fig:sample} shows from top left to bottom right the planes of $\mathrm{SFR}-M_{\star}$, rest-frame UVJ colors, $\Sigma_1-M_{\star}$ and $R_{\rm e}-M_{\star}$. The circles indicate our sample, while the background hexbins show the whole galaxy population of CANDELS/3D-HST. Specifically, the stellar population parameters of the CANDELS/3D-HST comparison sample ($M_{\star}$, SFR, and UVJ rest-frame colors) have been taken from \citet{leja19}, while $H$-band half-light size $R_{\rm e}$ and the central stellar mass surface density within 1 kpc $\Sigma_1$ are obtained from \citet{van-der-wel14a}. In particular, we estimate $\Sigma_1$ \citep{cheung12, saracco12, fang13, van-dokkum14_dense_cores, tacchella15_sci, tacchella17, woo17, barro17} by computing the fraction of the total luminosity in the $H$-band within 1 kpc from the single S\'{e}rsic fits by \citet{van-der-wel14a}, assuming a constant mass-to-light ratio throughout the galaxy. The blue line in the upper left panel shows the SFMS \citep{leja21}, the black line in the upper right panel shows the UVJ quiescent box as defined in \citet{whitaker12b}, the red line in the bottom left panel marks the $\Sigma_1-M{\star}$ relation for quiescent galaxies as measured in \citet{barro17}, and the blue and red lines show the $R_{\rm e}-M_{\star}$ relations for star-forming and quiescent galaxies from \citet{van-der-wel14a}, respectively. The color-coding in all panels corresponds to the doubling number $\mathscr{D}$, a measure of the number of times the stellar mass would double over the age of the universe at the current sSFR (see Eq.~\ref{eq:mass_double}). As mentioned in the Introduction, we classify galaxies as star-forming, transitioning, and quiescent if $\mathscr{D}(z)>1/3$, $1/20<\mathscr{D}(z)<1/3$, and $\mathscr{D}(z)<1/20$, respectively. These cuts correspond to blue, green, and red coloring in Fig.~\ref{fig:sample}, verifying that these cuts are meaningful.

As shown in Fig.~\ref{fig:sample}, the galaxies of our sample follow the overall trends of the massive galaxy population. At a given $M_{\star}$, star-forming galaxies have lower $\Sigma_1$ and larger $R_{\rm e}$ than quiescent galaxies. By selection (Section~\ref{subsec:selection}), our sample is unbiased above $10^{11}~M_{\odot}$, while below this limit it is skewed toward quiescent galaxies (low SFRs and UVJ quiescent colors). Importantly, our quiescent galaxies span the whole parameter space of $\Sigma_1-M_{\star}$ and $R_{\rm e}-M_{\star}$ of quiescent galaxies. The star-forming galaxies in our sample are typically massive with $M_{\star}>10^{11}~M_{\odot}$. Several star-forming galaxies (based on the doubling number $\mathscr{D}$) in our sample lie in the UVJ-quiescent box, consistent with the expected contamination rate of roughly 20\% \citep{leja19_uvj}. We take a more detailed look at those objects and the UVJ color-color diagram in Appendix~\ref{app_sec:uvj}. 

\subsection{Comparison to IllustrisTNG}
\label{subsec:tng}

Throughout this paper, we compare our observational measurements with similar measurements from the cosmological, hydrodynamical simulation IllustrisTNG \citep{marinacci18, naiman18, nelson18_color, pillepich18, springel18, nelson19_dr}. A detailed comparison between IllustrisTNG and our observations will allow us to assess the validity of adopted subgrid models in IllustrisTNG. In particular, we will focus on processes that are related to shaping SFHs, such as quenching, which in IllustrisTNG is mainly driven by black hole feedback:  galaxies in IllustrisTNG quench once the energy from black hole kinetic winds at low accretion rates becomes larger than the gravitational binding energy of gas within the galaxy stellar radius \citep[e.g.,][]{terrazas20}. This occurs at a particular black hole mass threshold. 

Throughout this paper, we focus on the intermediate-sized box TNG100-1 (TNG100 for the remainder of the paper), which combines a moderate resolution with a large volume to allow us to track the evolution of massive quiescent galaxies at $z\approx0.5-2$. The baryonic mass resolution of this box is $1.4 \times 10^6~M_{\odot}$, and the box size is $110.7^3~\mathrm{Mpc}^3$. We randomly select a sample of 1340 out of the 5703 galaxies with $M_{\star}>10^{10}~M_{\odot}$ from the snapshot at $z=0.7$. For each galaxy in TNG100, the stellar properties are extracted by considering all the bound particles. About a third (475 galaxies) of those galaxies are quiescent, consistent with observational estimates \citep{donnari19}. In order to perform a fair comparison between our observational measurements and the measurements from TNG100, we ``project'' the theoretical quantities into the observational space (spectra and photometry). Specifically, we predict the DEIMOS spectra and photometry by using the stellar particles' ages and metallicities and by drawing the other parameters of the galaxy SED, such as dust attenuation, dust emission, and AGN emission, from the prior (see Section~\ref{sec:physical_model}). For this, we use the same stellar population synthesis models as in the fitting (Section~\ref{subsec:sps}). We add noise to the spectra and photometry by drawing from the noise distribution of our observational data (see Fig.~\ref{fig:snr}). After predicting realistic spectroscopic and photometric data from TNG100 galaxies, we run the same analysis on them as on our observations (Sections~\ref{sec:physical_model} and \ref{sec:fitting}) and apply the same UVJ/sSFR selection as in the observational sample when comparing to the observations in the individual figures.

Although we perform this ``apples-to-apples'' comparison between observations and IllustrisTNG, the comparison has still its limitations. Firstly, even though the observations and simulations probe the same comoving volume within a factor of 2, cosmic variance and sample selection (both observational and simulated samples are not mass complete) can affect the interpretation of the rarest objects. Secondly, because of the finite mass resolution, the intermediate-sized box TNG100 is not able to resolve the first-forming galaxies \citep[e.g.,][]{vogelsberger20}. The high-resolution box TNG50 \citep{nelson19, pillepich19} has an order of magnitude higher baryonic mass resolution, but its volume is a factor $\sim8$ smaller. Hence, massive quiescent galaxies are not well probed, in particular at higher redshifts. Thirdly, aperture effects might play a role: even though our spectra include most of the light (at least 80\% for most of our galaxies as estimated from the slit geometry and the HST $H$-band morphology), this may still be different from including all the bound stellar particles in IllustrisTNG. This should be kept in mind when we interpret the comparison of the observations with IllustrisTNG.

Additionally, these synthetic mock observations of the IllustrisTNG galaxies allow us to assess systematic uncertainties in the estimated parameters compared to the true values. A preliminary examination reveals that our inferred mass-weighted ages and star-formation timescales are on average overestimated by $\sim20\%$ and $\sim10\%$, respectively. We do not find any bias for the quenching epoch or quenching timescale over the whole sample, though there is a fraction of galaxies (about $20\%$) that quench rapidly in the simulation ($\tau_{\rm quench}<200~\mathrm{Myr}$) for which we overestimate the quenching timescale ($\tau_{\rm quench}\approx0.5-1.0~\mathrm{Gyr}$). These galaxies typically quench early (more than 1 Gyr before the epoch of observation), making it difficult for us to pick up the signal of ``fast-quenching'' in the data (in addition to having SFH bins on scales of Gyr on long look-back times; see the next section). Although these limitations do not alter our conclusions, they are of course important to consider in more detail. We postpone a more thorough analysis of this to an upcoming publication.

\section{Physical model for the galaxy SED}
\label{sec:physical_model}

\begin{deluxetable*}{p{0.1\textwidth} p{0.4\textwidth} p{0.4\textwidth}}
\tablecaption{Free parameters and their associated priors for the fiducial physical model within \texttt{Prospector}. \label{tab:parameters}}
\tablehead{
\colhead{Parameter} & \colhead{Description} & \colhead{Prior}
}
\startdata
\hline
$z_{\rm obs}$ & redshift & uniform: $\mathrm{min}=z_{\rm spec}-0.005$, $\mathrm{max}=z_{\rm spec}+0.005$, where $z_{\rm spec}$ obtained from spectrum \\
$\sigma_{\star}/(\mathrm{km}\mathrm{s}^{-1})$ & velocity dispersion of stars & uniform: $\mathrm{min}=40.0$, $\mathrm{max}=400.0$ \\
$\log(\mathrm{M}_{\star}/\mathrm{M}_{\odot})$ & total stellar mass formed & uniform: $\mathrm{min}=9.5$, $\mathrm{max}=12$\\
$\log(\mathrm{Z}_{\star}/\mathrm{Z}_{\odot})$ & stellar metallicity & uniform: $\mathrm{min}=-1.0$, $\mathrm{max}=0.19$ \\
SFR ratios & ratio of the SFRs in adjacent bins of the $N_{\rm SFH}$-bin non-parametric SFH ($N_{\rm SFH}-1$ parameters total); default choice $N_{\rm SFH}=10$ & Student's-t distribution with $\sigma=0.3$ and $\nu=2$. \\
$n$ & power-law modifier to shape of the \citet{calzetti00} attenuation curve of the diffuse dust  (Eq.~\ref{eq:diffuse_dust}) & uniform: $\mathrm{min}=-1.0$, $\mathrm{max}=0.4$ \\
$\hat{\tau}_{\rm dust, 2}$ & diffuse dust optical depth (Eq.~\ref{eq:diffuse_dust}) & clipped normal: $\mathrm{min}=0$, $\mathrm{max}=4$, $\mu=0.3$, $\sigma$=1\\
$\hat{\tau}_{\rm dust, 1}$ & birth-cloud dust optical depth (Eq.~\ref{eq:dust_birth}) & clipped normal in ($\tau_{\rm dust, 1}/\tau_{\rm dust, 2}$): $\mathrm{min}=0$, $\mathrm{max}=2$, $\mu=1$, $\sigma=0.3$ \\
$\gamma_{\rm e}$ & mass fraction of dust in high radiation intensity & log-uniform: $\mathrm{min}=10^{-4}$, $\mathrm{max}=0.1$ \\
$\mathrm{U}_{\rm min}$ & minimum starlight intensity to which the dust mass is exposed & clipped normal: $\mathrm{min}=0.1$, $\mathrm{max}=15$, $\mu=2.0$, $\sigma=1.0$ \\
$\mathrm{q}_{\rm PAH}$ & percent mass fraction of PAHs in dust & uniform: $\mathrm{min}=0.5$, $\mathrm{max}=7.0$ \\
$f_{\rm AGN}$ & AGN luminosity as a fraction of the galaxy bolometric luminosity & log-uniform: $\mathrm{min}=10^{-5}$, $\mathrm{max}=3$ \\
$\tau_{\rm AGN}$ & optical depth of AGN torus dust & log-uniform: $\mathrm{min}=5$, $\mathrm{max}=150$ \\
$\sigma_{\rm gas}/(\mathrm{km}\mathrm{s}^{-1})$ & velocity dispersion of gas & uniform: $\mathrm{min}=30$, $\mathrm{max}=300$ \\
$\log(\mathrm{Z}_{\mathrm{gas}}/\mathrm{Z}_{\odot})$ & gas-phase metallicity & uniform: $\mathrm{min}=-2.0$, $\mathrm{max}=0.5$ \\
$\log(\mathrm{U})$ & ionization parameter for the nebular emission & uniform: $\mathrm{min}=-4.0$, $\mathrm{max}=-1$ \\
$f_{\rm out}$ & fraction of spectral pixels that are considered outliers by the mixture model & uniform: $\mathrm{min}=10^{-5}$, $\mathrm{max}=0.5$ \\
$j_{\rm spec}$ & multiplicative noise inflation term for spectrum & uniform: $\mathrm{min}=1.0$, $\mathrm{max}=5.0$ \\
\hline
\enddata
\end{deluxetable*}

In this section, we introduce the adopted physical model to describe the aforementioned observational data. We generate this physical model within \texttt{Prospector} \citep{leja17, johnson17_prospector, johnson21}. \texttt{Prospector} is a code to conduct fully Bayesian inference of stellar population properties from photometric and/or spectroscopic data. A strength of \texttt{Prospector} is the flexible spectroscopic calibration model, which allows us to combine photometric and spectroscopic data from the UV to IR while accounting for spectrophotometric calibration errors. Furthermore, \texttt{Prospector} includes flexible SFH parameterizations, which is important for understanding the diversity of evolutionary pathways of galaxies. A summary of the parameters and priors of our physical model can be found in Table~\ref{tab:parameters}. The fitting procedure is described in detail in Section~\ref{sec:fitting}. In addition to the material presented here, key features and further details on SED modeling with \texttt{Prospector} can also be found in \citet{leja18_AGN}, \citet{leja19_nonparm} and \citet{leja19}.

\subsection{Stellar population model}
\label{subsec:sps}

For stellar population synthesis, the Flexible Stellar Population Synthesis (\texttt{FSPS}) package\footnote{https://github.com/cconroy20/fsps} is used \citep{conroy09a}. In this work we use the MIST stellar evolutionary tracks and isochrones \citep{choi16, dotter16} with the MILES stellar spectral library \citep{falcon-barroso11}. The MIST models are based on MESA, an open-source stellar evolution package \citep{paxton11, paxton13, paxton15, paxton18}. 

We model the chemical enrichment histories of our galaxies with a delta function, assuming that all stars within the galaxy have the same metal content with scaled-solar abundances. This single metallicity is varied with a prior that is uniform in $\log(Z_{\star}/Z_{\odot})$ between $-1.0$ and 0.19, where $Z_{\odot}=0.0142$. The upper limit of the stellar metallicity prior is given by the MILES stellar templates. This upper limit might lead to an underestimation of the stellar metallicity in the fitting, which itself would result in an overestimation of the stellar ages. Although we cannot rule this out completely, we find that our mass-weighted ages are on average overestimated by only $\sim20\%$ (0.05 dex) using the mock spectra of IllustrisTNG (Section~\ref{subsec:tng}). We intend to constrain the abundance pattern in more detail in an upcoming publication. Finally, a \citet{chabrier03} initial mass function is assumed throughout this work.

\subsection{Star-formation history}
\label{subsec:SFH_assumption}

\begin{figure*}
    \centering
    \includegraphics[width=\textwidth]{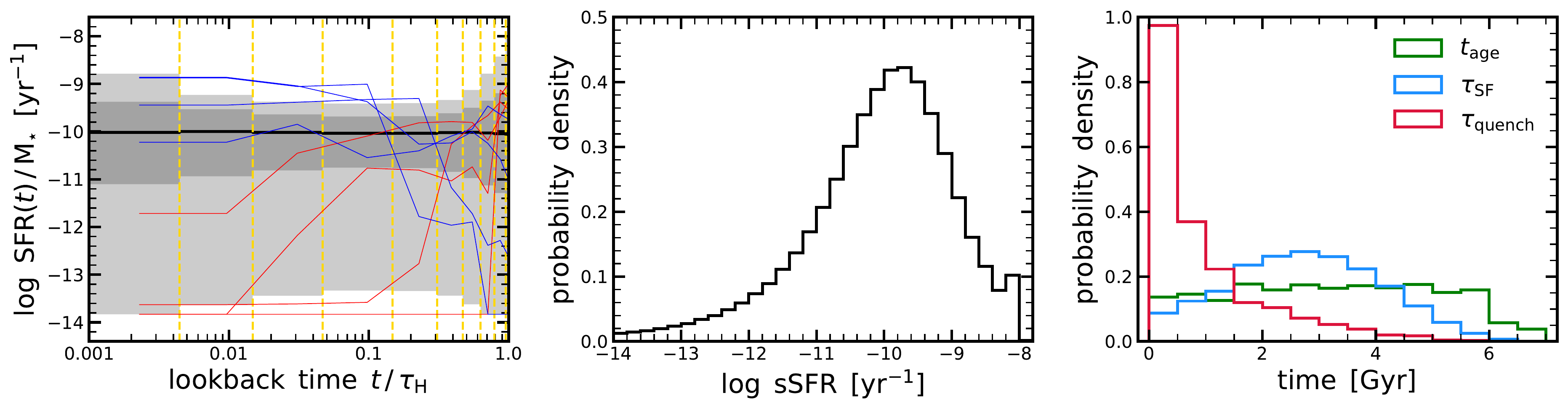}
    \caption{Star-formation history (SFH) prior adopted in our SED modeling. The left panel shows the prior density for $\mathrm{SFR}(t)/M_{\star}$ (where $M_{\star}$ is the total stellar mass formed), with four random draws from the prior that are quiescent (shown in red) and star-forming (shown in blue) at the epoch of observation. The solid black line shows the median, while the light (dark) shaded regions show the $2^{\rm nd}-98^{\rm th}$ ($16^{\rm th}-84^{\rm th}$) percentiles. The middle panel shows the prior density in sSFR (measured over the previous 100 Myr). The right panel shows the prior in mass-weighted age ($t_{\rm age}$; green line), star-formation timescale ($\tau_{\rm SF}$; blue line), and quenching timescale ($\tau_{\rm quench}$; red line). The middle and right panels assume the average redshift of our sample ($z_{\rm obs}=0.8$).}
    \label{fig:prior_sfh}
\end{figure*}

In our fiducial setup, we assume a ``non-parametric'', piece-wise constant SFH \citep{cid-fernandes05, ocvirk06, tojeiro07}. ``Non-parametric'' here means that no particular shape for the SFH is assumed and that an arbitrary function in SFH space can be reasonably approximated. \citet{lower20} showed that this flexible non-parametric approach outperforms traditional parametric forms (such as exponentially declining or log-normal SFHs) in capturing variations in galaxy SFHs, leading to significantly improved stellar masses in SED fitting. We assume that the SFH can be described by $N_{\rm SFH}$ time bins, where the SFR within each bin is constant. We fix $N_{\rm SFH}=10$ for the science analysis; we explore varying the number of time bins in Appendix~\ref{app_sec:variations} and Appendix~\ref{app_sec:prior_impact}. Increasing the $N_{\rm SFH}$ to 14 does not affect our results (see also \citealt{leja19_nonparm}), but the fits of 30 galaxies ($19\%$ of the sample) do not converge within a reasonable amount of time (i.e., 14 days on a single CPU). There are approaches that determine the appropriate number of time bins on the fly, such as adaptively binning in time \citep{tojeiro07} or using evidence comparison to determine the optimal number of bins \citep{dye08, iyer19}. As discussed in \citet{leja19_nonparm}, we use a piecewise model with a fixed number of bins because it is more scalable in a sampling framework.

The $N_{\rm SFH}$ time bins are specified in look-back time. Throughout the paper, we define the look-back time to be the time prior to the epoch of observation. Four bins are fixed at $0-30$ Myr, $30-100$ Myr, $100-300$ Myr, and $300-1000$ Myr to capture variation in the recent SFH of galaxies. We model a maximally old population with a fifth bin at $(0.95t_{\rm H}-t_{\rm H})$, where $t_{\rm H}$ is the age of the universe at the observed redshift. The remaining $N_{\rm SFH}-5$ bins are spaced equally in logarithmic time between 1 Gyr and $0.95t_{\rm H}$.

Since we include  ``more bins than the data warrant'' and let the sampler fully map the interbin covariances allowed by the prior and the data, a potential failure mode for this is overfitting, which is caused by an excess of model flexibility and results in overestimated uncertainties. This is in contrast to the classic dangers of ``underfitting'', whereby model parameters are overly constrained when too little parameter space is permitted. The danger posed by overly flexible models can be alleviated by choosing a prior that weights for physically plausible SFH forms. This is a complex problem that has been explored in detail in \citet[][see also \citealt{carnall19_sfh}]{leja19_nonparm}. We follow this analysis by adopting a continuity prior, which enforces smoothness by weighting against sharp transitions in $\mathrm{SFR}(t)$, similar to the regularization schemes from \citet{ocvirk06} and \citet{tojeiro07}. The prior is tuned to allow similar transitions in SFR to those of galaxies in the Illustris hydrodynamical simulations \citep{vogelsberger14, diemer17}, though it is deliberately designed to include broader behavior than seen in these simulations since we do not want to assume these models to be the truth. The resulting prior probability densities for $\mathrm{SFR}(t)$, $\mathrm{sSFR}$, mass-weighted age, star-formation timescale, and quenching timescale are shown in Fig.~\ref{fig:prior_sfh}.

In addition to this non-parametric approach, we also consider parametric SFHs in order to be able to compare to previous literature and to explore possible biases. In the parametric approach, we assume that the SFH follows a delayed $\tau$-model:

\begin{equation}
    \mathrm{SFR}(t) = (t-t_{\rm a})e^{-(t-t_{\rm a})/\tau}
    \label{eq:delayed_tau}
\end{equation}

\noindent
The parameter $\tau$ is varied with a logarithmic prior between $-1.0<\log(\tau)<10.0$, and the parameter $t_{\rm a}$ is varied with a uniform prior between 0 and the age of the universe at galaxies $z_{\rm obs}$ ($t_{\rm H}(z_{\rm obs})$). The results for the parametric SFHs are shown and discussed in Appendices~\ref{app_sec:variations} and \ref{app_sec:prior_impact}. Briefly, parametric SFHs typically lead to younger ages than non-parametric SFHs, which biases the stellar mass low and the sSFR high.

\subsection{Dust attenuation model}
\label{subsec:dust_attenuation_model}

We model dust attenuation using a two-component dust attenuation model with a flexible attenuation curve. Specifically, we use the two-component \citet{charlot00} dust attenuation model, which postulates separate birth-cloud and diffuse dust screens. The birth-cloud component ($\tau_{\rm dust, 1}(\lambda)$) in our model attenuates nebular emission and stellar emission only from stars formed in the past 10 Myr:

\begin{equation}
    \tau_{\rm dust, 1}(\lambda) = \hat{\tau}_{\rm dust, 1}\left(\frac{\lambda}{5500~\mathrm{\AA}}\right)^{-1}.
    \label{eq:dust_birth}
\end{equation}

\noindent
The diffuse component ($\tau_{\rm dust, 2}(\lambda)$) has a variable attenuation curve and attenuates all stellar and nebular emission from the galaxy. We use the prescription from \citet{noll09}:

\begin{equation}
    \tau_{\rm dust, 2}(\lambda) = \frac{\hat{\tau}_{\rm dust, 2}}{4.05}(k'(\lambda)+D(\lambda)) \left(\frac{\lambda}{5500~\mathrm{\AA}}\right)^n.
    \label{eq:diffuse_dust}
\end{equation}

\noindent
$\hat{\tau}_{\rm dust, 2}$ controls the normalization of the diffuse dust, $n$ is the diffuse dust attenuation index, $k'(\lambda)$ is the \citet{calzetti00} attenuation curve, and $D(\lambda)$ is a Lorentzian-like Drude profile describing the UV dust bump. We tie the strength of the UV dust absorption bump to the best-fit diffuse dust attenuation index, following the results of \citet{kriek13}. The free parameters in Eq.~\ref{eq:diffuse_dust} are therefore $\hat{\tau}_{\rm dust, 2}$ and $n$. We adopt a flat prior for $\hat{\tau}_{\rm dust, 2}$ ($0<\hat{\tau}_{\rm dust, 2}<4.0$) and $n$ ($-1.0<n<0.4$). The upper limit on $n$ is chosen to disallow a flat attenuation curve, which would cause $\hat{\tau}_{\rm dust, 2}$ to be nearly fully degenerate with the normalization of the SED.

Although $\hat{\tau}_{\rm dust, 1}$ and $\hat{\tau}_{\rm dust, 2}$ have a similar effect on the SED and are often degenerate, it is important to distinguish between these parameters to properly predict emission lines, in particular the line equivalent widths. The total optical depth toward nebular emission lines is roughly twice that of the stellar continuum \citep{calzetti94, kashino13, price14}. In our dust attenuation model, this means $\hat{\tau}_{\rm dust, 1}\sim\hat{\tau}_{\rm dust, 2}$, since $\hat{\tau}_{\rm dust, 2}$ is applied to the entire emission from the galaxy. We adopt a joint prior on the ratio of the two in order to allow for some reasonable variation around the fiducial results in the literature: a clipped normal centered on 1 with width of 0.3 in the range of $0 < \hat{\tau}_{\rm dust, 1}/\hat{\tau}_{\rm dust, 2} < 2.0$.

\subsection{Dust emission model}

We assume energy balance, i.e., all the energy attenuated by dust is reemitted in the IR \citep{da-cunha08}. Thanks to this assumption, the MIR photometry delivers additional constraints on the total amount of dust attenuation and on the dust-free stellar SED. However, in order to apply energy balance and to compute $L_{\rm IR}$ from the UV-MIR SED, we need to make some assumptions about the shape of the IR SED.

We use the \citet{draine07} dust emission templates to describe the shape of the IR SED, which are based on the silicate-graphite-PAH model of interstellar dust \citep{mathis77, draine84}. These templates have three free parameters controlling the shape of the IR SED: $U_{\rm min}$, $\gamma_{\rm e}$, and $\mathrm{q}_{\rm PAH}$. $U_{\rm min}$ and $\gamma_{\rm e}$ together control the shape and location of the thermal dust emission bump in the IR SED, while $\mathrm{q}_{\rm PAH}$ describes the fraction of total dust mass that is in PAHs. This last parameter is particularly important because a substantial fraction, or even a majority, of the MIR emission comes from strong PAH emission features. Since our photometry only includes bands up to MIPS 24$\mu$m, we use informative priors for $U_{\rm min}$ and $\gamma_{\rm e}$, while assuming a flat prior for $\mathrm{q}_{\rm PAH}$ (see Table~\ref{tab:parameters}). The adopted priors are consistent with both the SINGS sample \citep{draine07_sings} and the \citet{brown14} galaxies with Herschel photometry and lead to a minimal amount of bias in $L_{\rm IR}$, SFR, and dust attenuation in galaxies without far-IR photometry. This is discussed in detail in Appendix C of \citet{leja17}.

\subsection{AGN model}

Building on \citet{leja18_AGN}, we adopt the AGN templates from \citet{nenkova08a} and \citet{nenkova08b}. The CLUMPY AGN templates are incorporated in \texttt{FSPS} \citep{conroy09a}, and a detailed description is given in the \texttt{FSPS} documentation. Only the dust emission from the central torus is included in this model; it is assumed that the UV and optical emission from the central engine is fully obscured by the AGN dust torus. This is a viable assumption since we discarded AGNs with point sources or with broad lines. Our AGN model has two free parameters: $f_{\rm AGN}$, the ratio of the bolometric luminosity between the galaxy and the AGN, and $\tau_{\rm AGN}$, the optical depth of an individual dust clump at $5500~\mathrm{\AA}$. A log-uniform prior is adopted for $f_{\rm AGN}$, with an allowed range of $10^{-5}<f_{\rm AGN}<3$. A log-uniform prior describes the observed power-law distribution of black hole accretion rates \citep{aird17, georgakakis17, caplar18}. A log-uniform prior on $\tau_{\rm AGN}$ is adopted between $5<\tau_{\rm AGN}<150$, as the SED response to logarithmic changes in $\tau_{\rm AGN}$ is approximately linear (see Figure 1 in \citealt{leja18_AGN}).

\subsection{Nebular emission model}
\label{subsec:EL_model}

We adopt the standard approach to generating nebular emission in \texttt{FSPS}, whereby the ionizing continuum from the model stellar populations is assumed to be fully absorbed by the gas and emitted as both line and continuum emission. The nebular line and continuum emission is generated using a \texttt{CLOUDY} \citep{ferland98, ferland13} grid within \texttt{FSPS}, as described in \citet{byler17}. We assume for the gas-phase metallicity a uniform prior between $-2.0<\log(Z_{\rm gas}/Z_{\odot})<0.5$ and for the ionization parameter $U$ a uniform prior between $-4.0<\log(U)<-1.0$. Furthermore, we assume a flat prior for the gas-phase velocity dispersion ($30<\sigma_{\rm gas}/(\mathrm{km}\mathrm{s}^{-1})<300$). In addition, motivated by the complexity of the physics that produce nebular emission lines (\citealt{kewley19_review} and references therein), we take a flexible approach to model the nebular line amplitudes (Section~\ref{subsec:EL_marginalization}).

\section{Measuring galaxy properties from spectroscopy and photometry}
\label{sec:fitting}

We have described the physical model for the galaxy SEDs, including all parameters and priors, in the previous section. In this section, we describe the details of the fitting procedure. We use \texttt{Prospector} to perform the fitting, since it allows a rigorous combination of the photometric and spectroscopic data by including a spectroscopic calibration model (Section~\ref{subsec:spec_cali}), a noise and outlier model (Section~\ref{subsec:outlier}), and an emission-line marginalization routine (Section~\ref{subsec:EL_marginalization}). After describing the joint fitting of photometric and spectroscopic data (Section~\ref{subsec:joint}), we present the fitting results (Section~\ref{subsec:fitting_results}) and discuss the gain in fitting both photometry and spectroscopy together (Appendix~\ref{app_sec:gain}).

\subsection{Spectroscopic calibration model}
\label{subsec:spec_cali}

As described in Section~\ref{subsec:observations}, the spectra are not flux-calibrated. At each likelihood call, we match the model spectrum to the normalization of the spectroscopic data by fitting a polynomial in wavelength to their ratio. Our approach implements the Chebyshev polynomial calibration model, computed at each likelihood call from a simple least-squares maximum likelihood fit to the ratio of the data to the calibrated model spectrum, excluding the regions where emission lines may be present.  The order of the polynomial $m$ is determined by $m=(\lambda_{\rm max}-\lambda_{\rm min})/100\mathrm{\AA}$ within each wavelength interval (see also \citealt{kelson00} and \citealt{conroy18}.), with typically $m\approx20$. This order is chosen to account for broad continuum mismatch issues but is not so flexible that it could over-fit broad absorption features. We have experimented with this approach by changing the order of the polynomial and find that the results are generally insensitive to this choice.

Using the maximum-likelihood fit for the calibration has the advantage of computational speed. However, ideally one would marginalize the likelihood of the data over all possible calibration polynomials for each model call.  Naively, this can be done at the cost of introducing $m$ additional model parameters describing the polynomial coefficients, but this is computationally prohibitive at present.  It is possible to analytically marginalize the likelihood over all possible coefficient values, but this has not yet been implemented \citep[but see][]{carnall19}.

The net effect of our approach is that the large-scale continuum shape and normalization of the model are set by the photometry. By fitting a moderate-order polynomial to the ratio between the observed and physical model spectrum at each likelihood call, the spectroscopic calibration model basically removes all information content from the continuum shape of the spectroscopic data. This means that the continuum shape of the observed spectrum does not inform any of the galaxy’s physical parameters. Instead, information about physical parameters that affect the continuum shape derives from the photometry, which does not include any multiplicative calibration model. Therefore, there is no degeneracy between the spectroscopic calibration model and the galaxy's parameters, such as the dust content or the SFH.

\subsection{Noise and outlier model}
\label{subsec:outlier}

We find that the standard fitting procedure is sensitive to outliers, that is, spectroscopic data points that are not well described by our model, because of inaccurate uncertainties or limitations of the model itself. We mitigate this problem by becoming insensitive to ``bad'' spectral data points. Specifically, we use a mixture model to describe outliers, following the approach described in \citet[][see also \citealt{sharma17_review} and \citealt{press97}]{hogg10}. These outlier pixels do not include the masked pixels (Section~\ref{subsec:observations}).

This model alters the likelihood by assuming some possibility that any given spectral pixel is an outlier, $f_{\rm out}$. The likelihood is calculated by marginalizing over $f_{\rm out}$ for each pixel; thus, no individual pixels are uniquely identified as outliers (see Eq.~\ref{eqn:spec_likelihood} below). It is assumed that outlier pixels have their uncertainties inflated by a factor of 50. $f_{\rm out}$ is a free parameter in the model, and typically $0.1\%$ of pixels in each fit are outliers. 

In addition, to account for possible under- or overestimates of the spectroscopic uncertainties, we introduce a parameter that multiplies the spectroscopic uncertainties by some constant factor ($j_{\rm spec}$) before calculating the likelihood. This is a free parameter in the model; in general, we find values very close to 1, indicating that the spectroscopic noise is not broadly under- or overestimated.

\subsection{Emission-line marginalization}
\label{subsec:EL_marginalization}

\begin{figure*}
    \centering
    \includegraphics[width=1.0\textwidth]{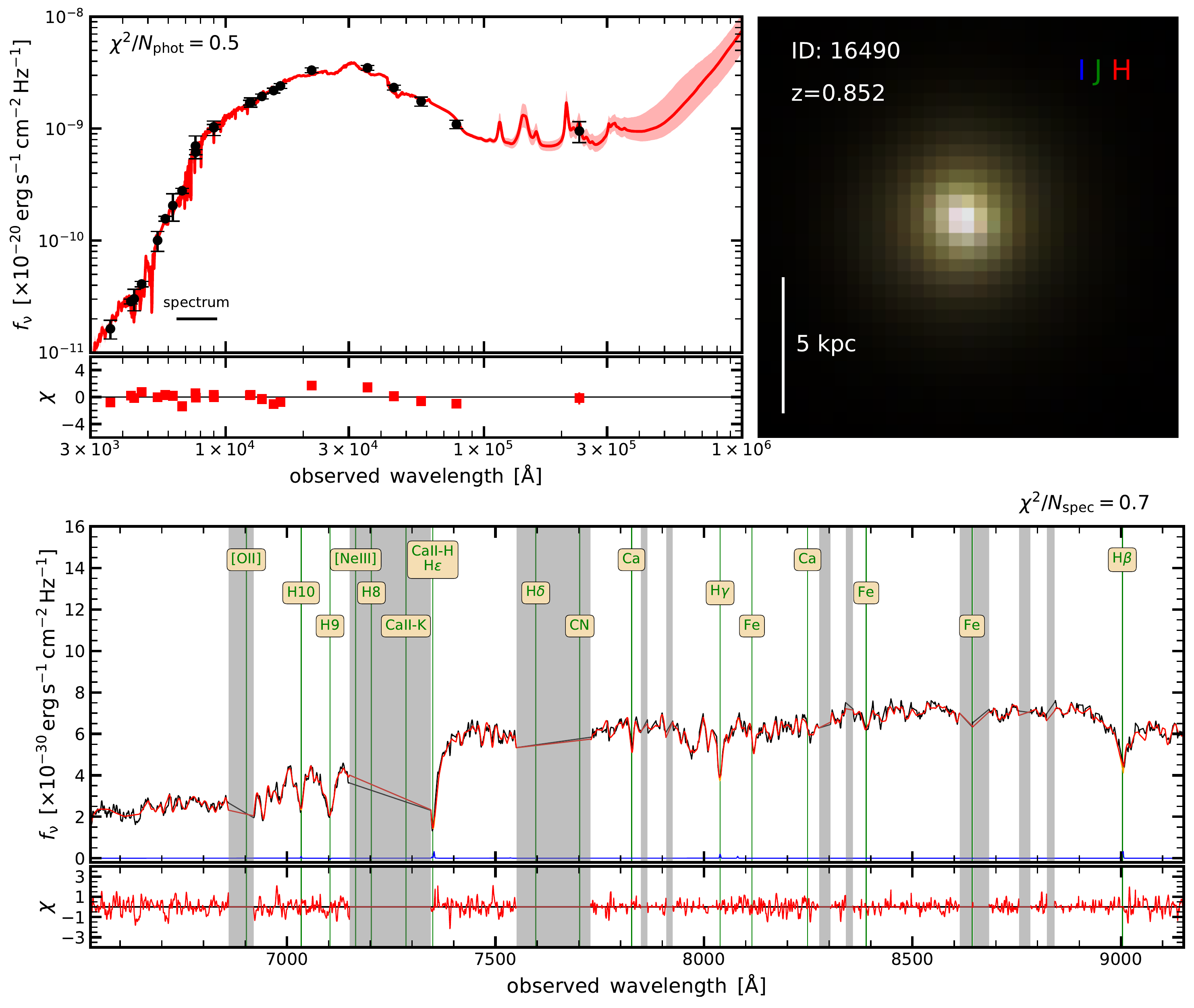}
    \caption{Observational data with the model for an example galaxy at $z_{\rm obs}=0.852$. The top left panel shows the photometry; the top right panel shows the HST $I$-, $J$- and $H$-band color-composite image; and the bottom panel shows the spectroscopic data ($\mathrm{S}/\mathrm{N}=23.8$ with a total exposure time of 9.3 hr). The observational data are indicated with black, while the model fit (drawn from the posterior) is shown in red. The blue line shows the emission line spectrum (emission lines for this galaxy are weak). The gray regions show the masked region in the spectra, where sky emission dominates. The smaller panels associated with the top left and bottom panels show $\chi$, defined by $(\mathrm{model}-\mathrm{data})/\sigma$. The overall reduced $\chi^2$ for the photometry and spectroscopy are 0.5 and 0.7, respectively.}
    \label{fig:example_spec}
\end{figure*}

\begin{figure*}
    \centering
    \includegraphics[width=0.7\textwidth]{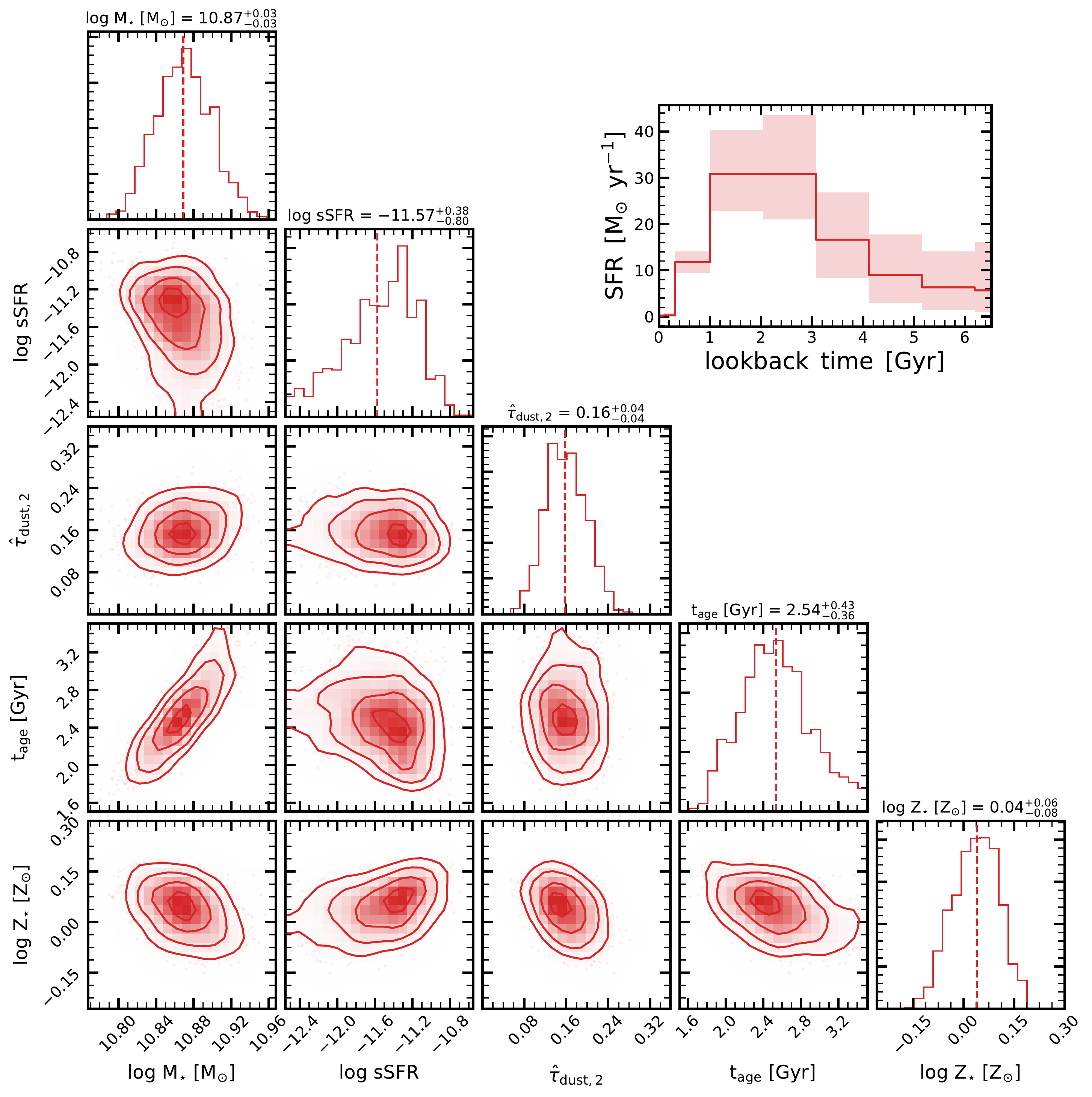}
    \caption{Joint posterior plot of the posteriors of key quantities from the galaxy fit shown in Fig.~\ref{fig:example_spec}. The plotted quantities include stellar mass ($M_{\star}$), sSFR, dust opacity of the diffuse dust component in the $V$-band ($\hat{\tau}_{\rm dust, 2}$), stellar age ($t_{\rm age}$), and stellar metallicity ($Z_{\star}$). The inset in the upper right corner shows the SFH posterior. The posteriors are well converged. We are overall able to break the dust-age-metallicity and get meaningful constraints on those quantities.}
    \label{fig:example_corner}
\end{figure*}

As mentioned in Section~\ref{subsec:EL_model}, nebular emission lines are included in the model spectrum. Each emission line is modeled as a Gaussian with a variable width and amplitude. We fit for the velocity dispersion of the gas ($\sigma_{\rm gas}$), while the emission-line amplitudes are marginalized over in each fitting step, as described in \citet{johnson21}. In our fiducial fitting, we use the maximum-likelihood amplitude for each emission line, which means that we totally decouple the emission lines from the SFH since the emission lines for most of our quiescent galaxies are believed to be emitted from LIERs, shocks, and/or AGNs. 

\subsection{Joint fitting of photometric and spectroscopic data}
\label{subsec:joint}

We describe here the fitting methodology. We fit the physical model described in Section~\ref{sec:physical_model} to the observational data (spectroscopy with photometry) presented in Section~\ref{sec:observational_data}, together with the models for systematic effects described in Sections~\ref{subsec:spec_cali} and \ref{subsec:outlier}. We denote the parameters of the model with $\Theta$.

We assume that the uncertainties on the photometric fluxes $\sigma_i$ are Gaussian and independent. In this case, the scatter of observed fluxes $f_i$ about their true values follows a $\chi^2$ distribution. The log-likelihood function for the photometry then follows:

\begin{equation}
\label{eqn:chisq_likelihood}
    \ln(\mathcal{L}_{\rm phot}) = -0.5\sum_{i}^{N_{\rm band}} \left(\frac{f_i-m_i(\Theta)}{\sigma_i}\right)^2
\end{equation}

\noindent
where $m_i(\Theta)$ is the model prediction for the observed flux $f_i$ and $N_{\rm band}$ is the number of photometric bands.

We make the same assumptions about the distribution of uncertainties when fitting the spectrum. However, the likelihood equation is modified by the outlier model such that

\begin{equation}
\label{eqn:spec_likelihood}
    \mathcal{L}_{\rm spec} = (1-f_{\rm out})\mathcal{L}(f,m,\sigma) + f_{\rm out}\mathcal{L}(f,m,\sigma_{\rm out})
\end{equation}
where $\mathcal{L}$ is calculated following Eq.~\ref{eqn:chisq_likelihood} (instead of summing over the photometric bands, we sum over spectral pixels in wavelength), and $\sigma_{\rm out}=50\sigma$ as described in Section~\ref{subsec:outlier}. 

The total likelihood is thus:
\begin{equation}
    \ln\mathcal{L}_{\mathrm{total}} = \ln\mathcal{L}_{\mathrm{phot}} + \ln\mathcal{L}_{\mathrm{spec}} + \ln\mathcal{L}_{\mathrm{eline}}.
\end{equation}

\noindent
where $\mathcal{L}_{\mathrm{eline}}$ represents a penalty term that takes into account the prior on emission-line amplitudes. We set this term to 0 in our fidicual analysis.

Our fiducial model assumes 10 SFHs bins, leading to a total of 27 free parameters (Table~\ref{tab:parameters}). Sampling our posterior distribution with the dynamic nested sampling algorithm \texttt{dynesty} \citep{speagle20} therefore requires several million evaluations of our log-likelihood function. Each likelihood call takes about 50 ms. Fitting each galaxy therefore requires roughly 100 CPU hours. 

\subsection{Fitting results}
\label{subsec:fitting_results}

We verified the fits to the photometry and spectroscopy for each individual galaxy. We checked that none of the posteriors pile up at the edges of the priors. This is particularly true for the stellar metallicity, which alleviates the concern that our stellar age estimates are biased high. The median reduced $\chi^2$ calculated with the model drawn from the posteriors for the photometry and for the spectroscopy are $1.08^{+1.02}_{-0.52}$ and $0.83^{+0.22}_{-0.17}$, respectively. The $\chi^2$ for the spectroscopy is slightly below 1, which is consistent with our obtained jitter terms $j_{\rm spec}$ to be clustered at 1 (we do not allow for $j_{\rm spec}<1.0$). We have also inspected the stacked residuals of the photometry, finding that the model does reproduce the data well. We found a weak, but significant trend related to the $K$-band and IRAC photometry. Specifically, the average $K$-band has a $\chi$ of about 1 (i.e., model underestimated the $K$-band flux), while the IRAC bands show a gradient so that the model $K$-to-IRAC colors are too red with respect to observations. A possible cause for this trend could be thermally pulsing AGB (TP-AGB) stars. In principle, \texttt{FSPS} allows to choose the normalization of the TP-AGB stars, though it is currently unfeasible to marginalize over this within \texttt{Prospector}. Nevertheless, this feature should be investigated in the future. We also investigated the stacked residual of the observed- and rest-frame spectra, finding no significant trend, which shows that the removing of skylines and modeling of emission lines overall worked well.

We show the observational data of an example galaxy along with the fitted model in Fig.~\ref{fig:example_spec}. The galaxy has a redshift of $z_{\rm obs}=0.852$. The spectroscopic data have $\mathrm{S}/\mathrm{N}=23.8$ per $\mathrm{\AA}$ with an exposure time of 9.3 hr. The model fits the photometric and spectroscopic data well. The residuals are distributed around 0. The reduced $\chi^2$ for the photometry and spectroscopy are 0.5 and 0.7, respectively.

The resulting posteriors of some key quantities of this fit are shown in Fig.~\ref{fig:example_corner}. This joint posterior plot shows the stellar mass ($M_{\star}$), sSFR, dust opacity in the $V$-band ($\hat{\tau}_{\rm dust, 2}$), stellar age ($t_{\rm age}$), and stellar metallicity ($Z_{\star}$). We find for this galaxy a stellar mass of $\log M_{\star}/M_{\odot}=10.87^{+0.03}_{-0.03}$, a low sSFR with $\log \mathrm{sSFR}/\mathrm{yr}^{-1}=-11.59^{+0.38}_{-0.73}$, a dust opacity of $\hat{\tau}_{\rm dust, 2}=0.16^{+0.04}_{-0.04}$, an age of $t_{\rm age}/\mathrm{Gyr}=2.51^{+0.45}_{-0.36}$, and roughly solar metallicity with $\log Z_{\star}/Z_{\odot}=0.04^{+0.06}_{-0.07}$. Here and throughout the paper the age $t_{\rm age}$ corresponds to the mass-weighted age.  From this fit, we find that this galaxy is quiescent with a doubling number of $\mathscr{D}=0.02$, i.e., it takes this galaxy about 50 Hubble times (age of the universe at $z_{\rm obs}$) to double its mass with its current sSFR. Although this galaxy is not actively forming stars, the galaxy is overall young, consistent with an SFH that rose through most of cosmic time and only declined in the past $\sim1$ Gyr.

\begin{figure*}
    \centering
    \includegraphics[width=\textwidth]{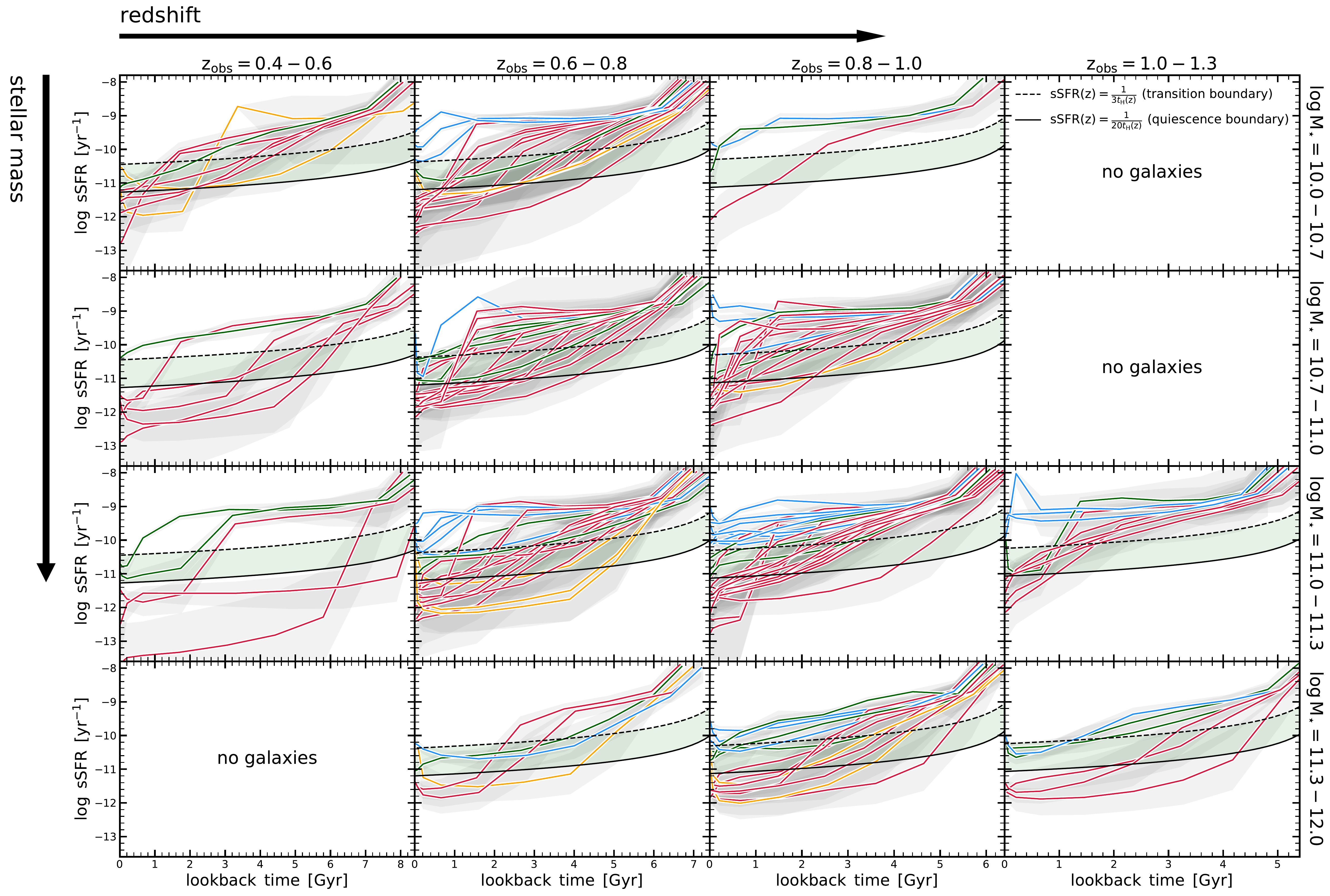}
    \caption{Star-formation histories (SFHs) of all galaxies in our sample. The panels show increasing observed redshifts from left to right, while stellar mass increases from top to bottom. For each galaxy, we plot $\mathrm{sSFR}(t)$ as a function of look-back time from the epoch of observations, i.e. the SFR at each epoch is normalized by the stellar mass formed by that epoch. The SFHs are color-coded by whether galaxies are star-forming (blue), transitioning (green), quiescent (red), or rejuvenating (orange) at the epoch of observation. In each panel, we show the boundary between the star-forming and transition regime as a dashed line and the boundary between the transition and quiescent regime as a solid line, which are defined by comparing the sSFR to the age of the universe (Eq.~\ref{eq:mass_double}). The transition region itself is highlighted in green. At a given stellar mass and observed redshift (i.e., in a given panel), we find a large diversity of SFHs. Even focusing on quiescent galaxies only, we find a large diversity of pathways: some galaxies cease their star formation early, some galaxies late, some galaxies cease their star formation quickly, some galaxies on long timescales (see also Fig.~\ref{fig:sfh_massive_QG}). Despite this diversity, the SFHs can be divided into three phases: a star-forming, a transition, and a quiescent phase (see also Fig.~\ref{fig:sfh_avg}).}
    \label{fig:sfh_all}
\end{figure*}

\begin{figure*}
    \centering
    \includegraphics[width=\textwidth]{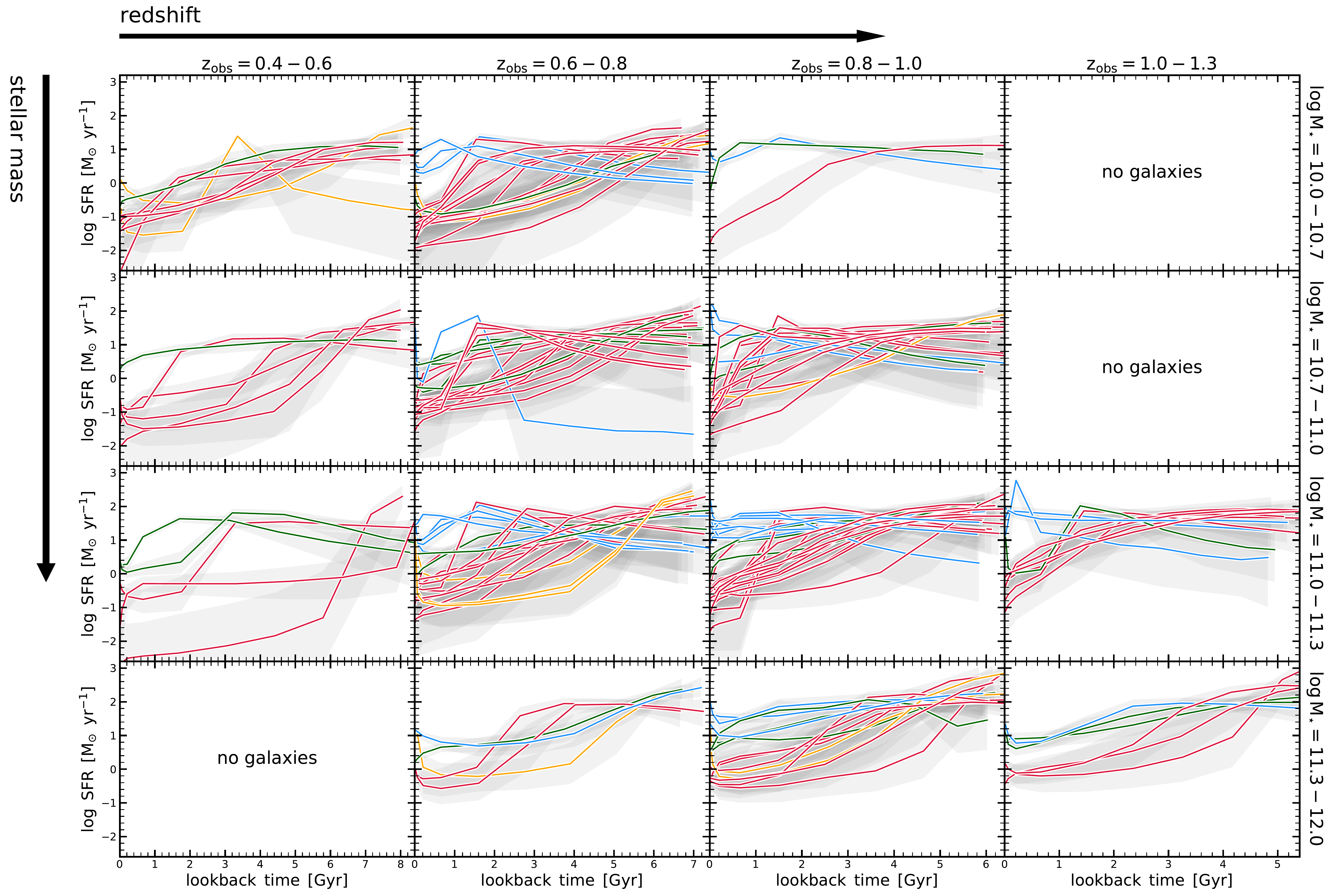}
    \caption{SFHs of all galaxies in our sample, plotted as SFR versus look-back time. The figure follows the same schema as Fig.~\ref{fig:sfh_all}. This figure highlights that star-forming galaxies' SFHs peak more recently than quiescent ones. Furthermore, despite the similar sSFR tracks at early times in Fig.~\ref{fig:sfh_all}, we show here that the SFRs actually span a wide range.}
    \label{fig:sfh_all_sfr}
\end{figure*}

In addition, we discuss in Appendix~\ref{app_sec:gain} the gain in fitting both photometry and spectroscopy. In summary, the spectroscopy constrains the metallicity, while the photometry constrains the dust attenuation. However, both the photometry and the spectroscopy are needed to break the dust-age-metallicity degeneracy and derive SFHs.

\section{Results}
\label{sec:results}

\begin{deluxetable}{p{0.1\linewidth} p{0.85\linewidth}}
\tablecaption{Definitions of the key timescales. \label{tab:timescales}}
\tablehead{\colhead{Symbol} & \colhead{Description}}
\startdata
\hline
$t_{\rm age}$ & mass-weighted age$^{a}$ \\
$z_{\rm f}$ & formation redshift: redshift of look-back time $t_{\rm age}$; i.e., $z(t_{\rm H}-t_{\rm age})=z_{\rm f}$ \\
$\tau_{\rm SF}$ & star-formation timescale: time between when 20\% and 80\% of the stellar mass formed \\
$\tau_{\rm quench}$ & quenching timescale: time to transition through the ``green valley'' ($1/[20t_{\rm H}(z)]<\mathrm{sSFR}<1/[3t_{\rm H}(z)]$) \\ 
$z_{\rm quench}$ & epoch of quenching: redshift when the galaxy transitions through the ``green valley'', i.e., redshift of the average cosmic time between entering ($\mathrm{sSFR}=1/[3t_{\rm H}(z)]$) and leaving ($\mathrm{sSFR}=1/[20t_{\rm H}(z)]$) the transition region \\ 
\hline
\enddata
\tablenotetext{a}{The mass-weighted age (a weighted average) is similar to $t_{50}$, which is the look-back time when 50\% of the stellar mass has been formed and therefore the median age. We adopt the former throughout this work, but note that the difference between $t_{\rm age}$ and $t_{50}$ is small (at most 16\%).}
\end{deluxetable}

We present the main results in this section. We start by showing the reconstructed SFHs in Section~\ref{subsec:sfh}. Sections~\ref{subsec:age}, \ref{subsec:tau_sf} and \ref{subsec:t_quench} focus on interesting aspects of the SFHs, specifically the mass-weighted age ($t_{\rm age}$), star-formation timescale ($\tau_{\rm SF}$), quenching timescale ($\tau_{\rm quench}$), and the epoch of quenching ($z_{\rm quench}$). The definitions of these key quantities are summarized in Table~\ref{tab:timescales}.

\subsection{Reconstructed star-formation histories}
\label{subsec:sfh}

\begin{figure}
    \centering
    \includegraphics[width=\linewidth]{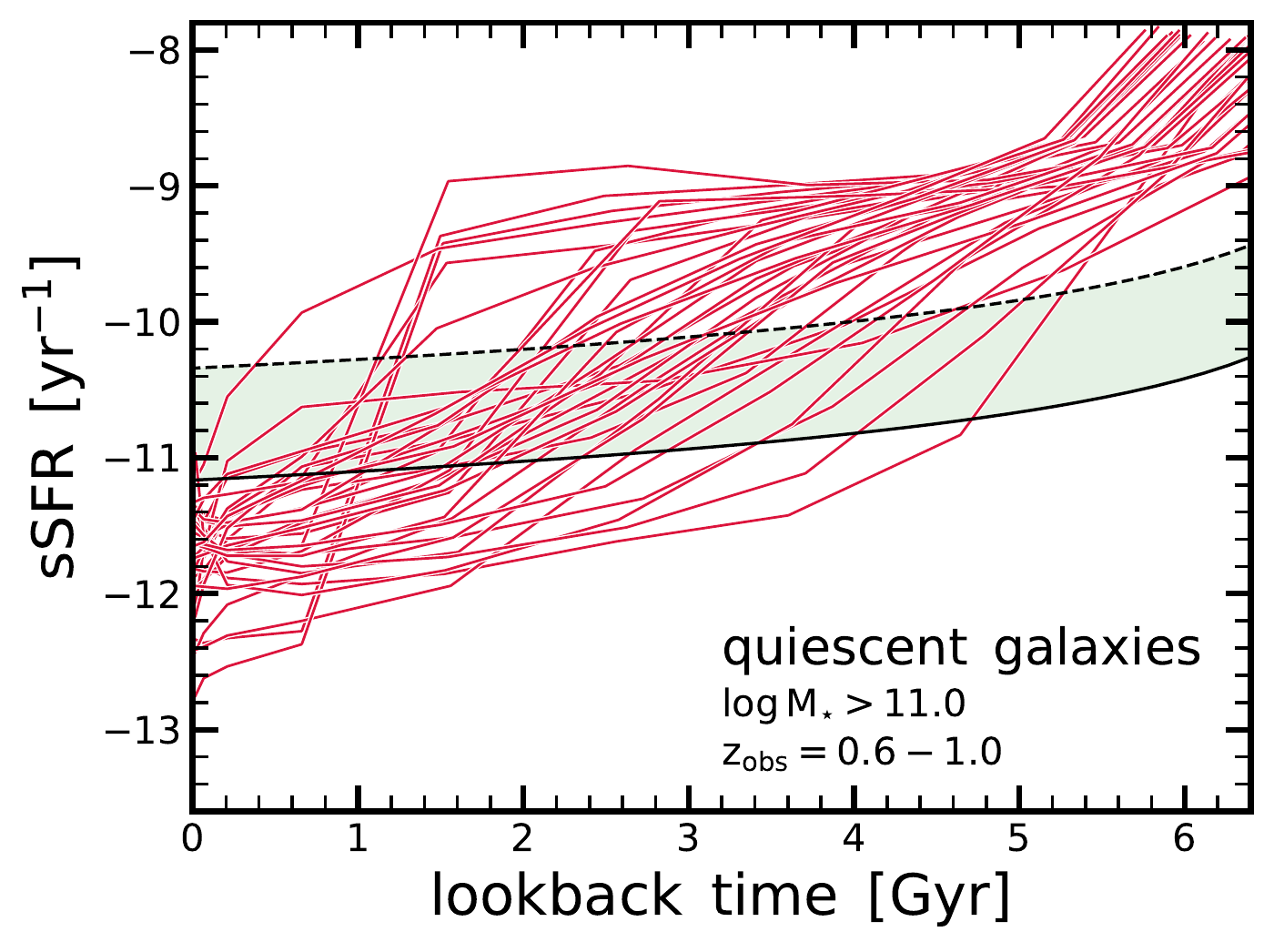}
    \caption{Large diversity of SFHs of massive quiescent galaxies at $z_{\rm obs}=0.6-1.0$. Following Fig.~\ref{fig:sfh_all}, we plot the tracks of sSFR versus look-back time for individual massive ($\log(M_{\star}/M_{\odot})>11.0$) and quiescent galaxies at $z_{\rm obs}=0.6-1.0$. This figure highlights the large diversity: galaxies quench fast, slow, early and late.}
    \label{fig:sfh_massive_QG}
\end{figure}

\begin{figure}
    \centering
    \includegraphics[width=\linewidth]{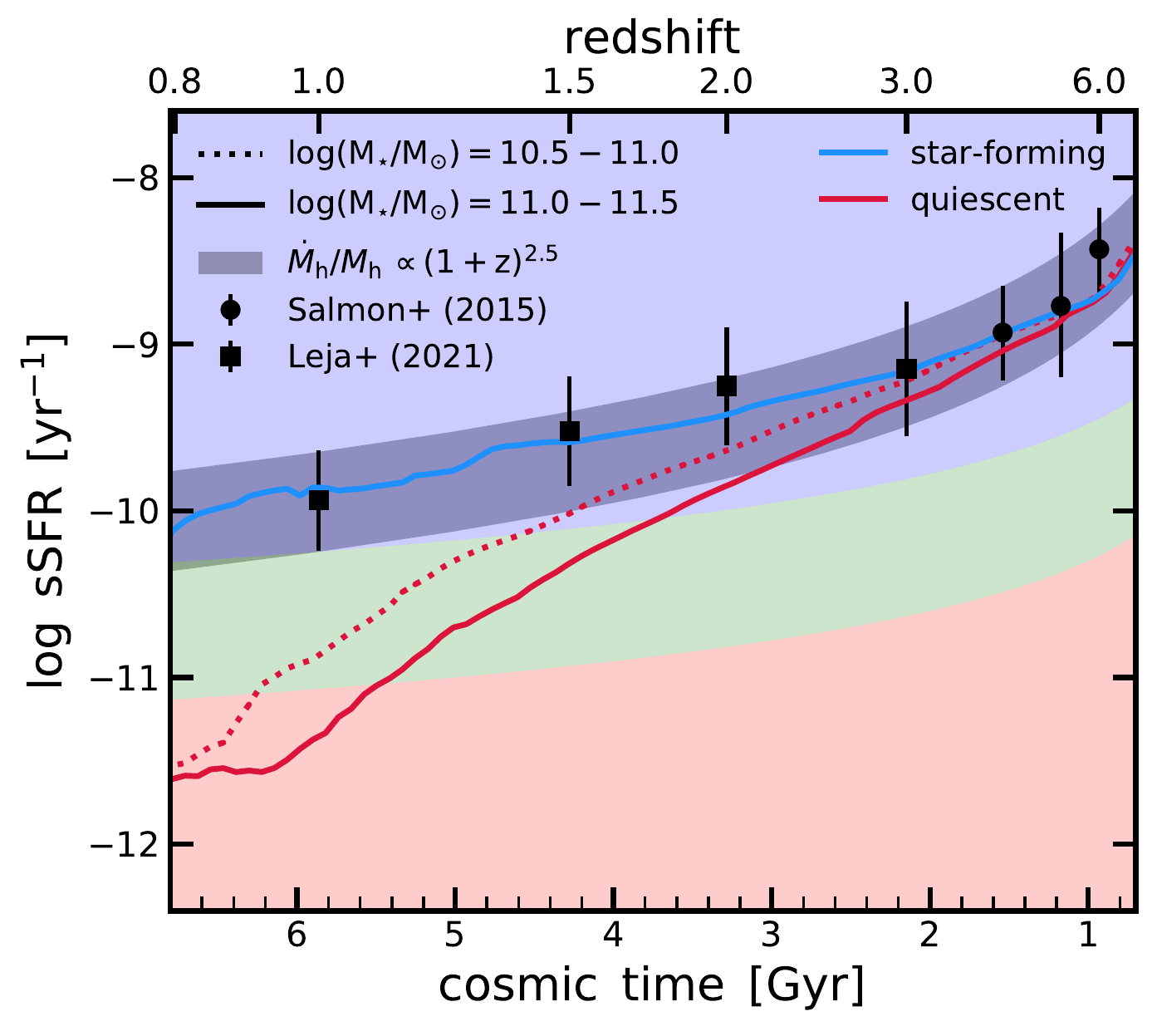}
    \caption{Median SFHs for star-forming (blue) and quiescent (red) galaxies. The dotted and solid lines indicate lower-mass ($\log~\mathrm{M}_{\star}/M_{\odot}=10.5-11.0$) and higher-mass ($\log~\mathrm{M}_{\star}/M_{\odot}=11.0-11.5$) galaxies. The blue, green, and red shaded regions mark the star-forming, transition, and quiescent regime as defined by comparing the sSFR to the age of the universe (Eq.~\ref{eq:mass_double}). At early times ($z=3-6$), star-forming and quiescent galaxies in both mass bins have similar sSFRs, consistent with direct high-$z$ observations by \citet{salmon16} and \citet{leja21}. The star-forming galaxies follow roughly the specific accretion rate of dark matter halos ($\dot{M}_{\rm h}/M_{\rm h}\propto(1+z)^{2.5}$). Massive quiescent galaxies cease their star formation on average earlier, but not faster than lower-mass quiescent galaxies.}
    \label{fig:sfh_avg}
\end{figure}

Fig.~\ref{fig:sfh_all} presents the SFH of all galaxies in our sample as a function of observed redshift ($z_{\rm obs}$; from left to right) and of stellar mass ($M_{\star}$; from top to bottom). We plot the SFHs as sSFR versus look-back time (SFR versus look-back time is shown in Fig.~\ref{fig:sfh_all_sfr}). The solid lines show individual galaxies, while the shaded region shows the $16\mathrm{th}-84\mathrm{th}$ percentiles. The coloring of the lines corresponds to whether galaxies at the time of observations ($z_{\rm obs}$) are star-forming (blue), transitioning (green) or quiescent (red). Rejuvenating galaxies, i.e., galaxies that were quiescent in the past and at the epoch of observation in the transition or star-forming region, are shown as orange lines. Plotting in $\mathrm{sSFR}(t)$ gives the advantage of being able to directly read off from the figure whether a galaxy is still actively growing owing to star formation. In order to help guide the eye, the dashed and solid black lines indicate the boundary between the star-forming and transition regions ($\mathrm{sSFR}=1/[3t_{\rm H}(z)]$) and between the transition and quiescent regions ($\mathrm{sSFR}=1/[20t_{\rm H}(z)]$), meaning that galaxies that at least double their mass within three times the age of the universe are considered star forming, while galaxies that do not double their mass within 20 times the age of the universe are considered quiescent (see Eq.~\ref{eq:mass_double}). The green shaded region marks the transition region.

The key result presented in Fig.~\ref{fig:sfh_all} is the diversity of SFHs in our sample. At a given observed redshift and stellar mass, we find a large diversity of pathways through the sSFR$-$time space. This is consistent with other studies that find that massive galaxies form a highly diverse population at $z\gtrsim1$ \citep[e.g.,][]{van-dokkum11}. By definition, galaxies start with high sSFRs at early times with $\mathrm{sSFR}\approx10^{-8}-10^{-9}~\mathrm{yr}^{-1}$, which means that these galaxies double their mass every few hundred million years. This phase of star formation lasts only a few hundred Myr in some galaxies, while it lasts for several Gyr in other galaxies. There is a large diversity in both when galaxies start transitioning to the quiescent region (i.e., when they start quenching) and how long this transition takes (quenching timescale). We quantify both the duration of the star-forming phase (i.e., star-formation timescale) and the quenching timescale in more detail in the upcoming sections.

Similarly, Fig.~\ref{fig:sfh_all_sfr} shows the SFR versus look-back time tracks for our sample, again highlighting the diversity of pathways. In particular, this figure highlights the range of different SFRs at early cosmic times. Although there is the overall trend that star-forming galaxies form later than quiescent galaxies (i.e., they peak at later cosmic times), there are several outliers that do not follow this trend. Furthermore, even at fixed $M_{\star}$ and $z_{\rm obs}$, quiescent galaxies themselves show a large diversity in early SFRs and the peak times, consistent with many pathways to quiescence. This is also highlighted separately in Fig.~\ref{fig:sfh_massive_QG}, which shows the sSFR tracks for massive ($\log(M_{\star}/M_{\odot})>11.0$) quiescent galaxies with $z_{\rm obs}=0.6-1.0$. Quiescent galaxies quench fast, slow, early and late -- and not all galaxies that quench early quench fast, nor do all galaxies that quench late quench slowly. 

Despite this diversity, there is some overall coherence, in particular regarding the sSFR tracks in the star-forming phase. Therefore, it is worth studying the median SFHs for different samples. Fig.~\ref{fig:sfh_avg} shows the median SFH (sSFR versus cosmic time) for our sample, focusing now on galaxies with $z_{\rm obs}=0.6-1.0$. We show the median SFH for quiescent galaxies in the mass bins $\log(M_{\star}/M_{\odot})=10.5-11.0$ and $\log(M_{\star}/M_{\odot})=11.0-11.5$ as dotted and solid lines, respectively. The median SFH of star-forming galaxies with $\log(M_{\star}/M_{\odot})=11.0-11.5$ is shown as a blue line. We do not show the low-mass star-forming galaxies because only a few galaxies are in this bin (see Fig.~\ref{fig:sample}). The gray band indicates the rescaled specific dark matter accretion rate with $\dot{M}_{\rm h}/M_{\rm h}\propto(1+z)^{2.5}$ \citep{wechsler02, neistein08, dekel13}.

Fig.~\ref{fig:sfh_avg} shows that during this early phase of star formation (within 2 Gyr of the Big Bang, i.e., $z>3$), star-forming and quiescent galaxies have similar sSFR and are consistent with direct measurements of SFR and $M_{\star}$ (i.e., SFMS) at high $z$ by \citet{salmon15}. This shows that one can in principle use this archaeological approach to estimate the SFR and $M_{\star}$ of the galaxy population in observationally inaccessible parameter space (i.e., high $z$ and low $M_{\star}$; e.g. \citealt{iyer18}), though caution must be exercised since these SFHs include all stellar mass ever accreted, i.e., it is difficult to correct for the effects of merging (see Section~\ref{subsec:mainsequence}). The median sSFR track of star-forming galaxies follows the independent estimates of the SFMS also at lower redshifts \citep{leja21}, which is an important consistency and validation check of our obtained SFHs. Furthermore, the median SFH of star-forming galaxies lies within the gray band at all times, indicating that the sSFR evolution is consistent with the evolution of the specific mass accretion rate of dark matter halos. 

\begin{figure*}
    \centering
    \includegraphics[width=\textwidth]{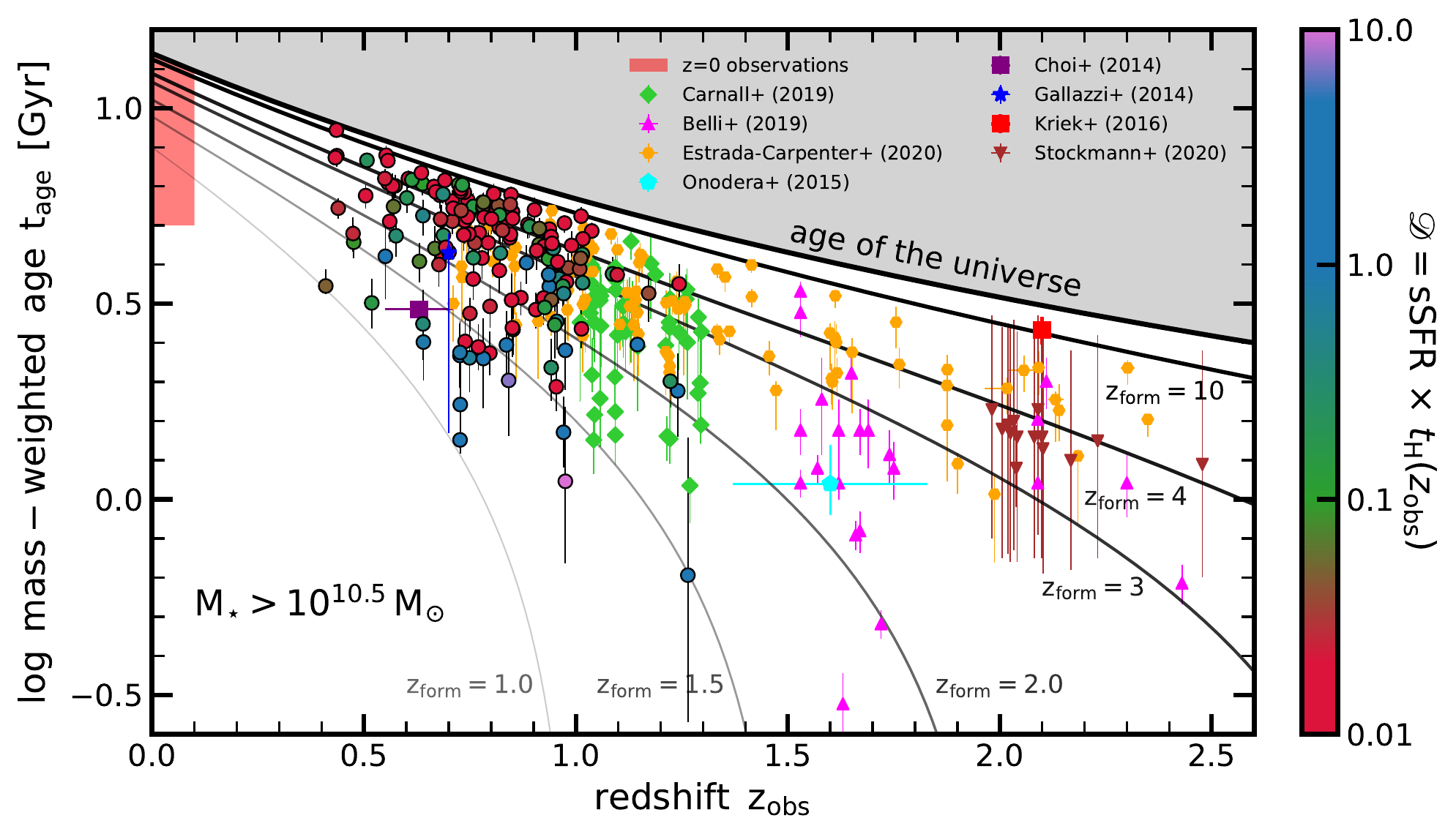}
    \caption{Mass-weighted age ($t_{\rm age}$) as a function of observed redshift ($z_{\rm obs}$). The gray to black lines indicate the age of simple stellar populations (SSPs) formed at different formation redshifts ($z_{\rm form}=1$, 1.5, 2.0, 3.0, 4.0, and 10.0). Our measurements are shown as large circles, color-coded by the doubling number $\mathscr{D}$, and are compared to a wide range of literature data of massive, quiescent galaxies \citep{belli19, carnall19, choi14, estrada-carpenter20, gallazzi05, gallazzi14, gallazzi21, kriek16, onodera15, spolaor10, stockmann20}. In agreement with previous studies, we find that massive, quiescent galaxies have a wide range of formation redshifts, ranging from $z_{\rm form}=10$ down to $z_{\rm form}=1.5$.}
    \label{fig:redshift_versus_age}
\end{figure*}

\begin{figure*}
    \centering
    \includegraphics[width=\textwidth]{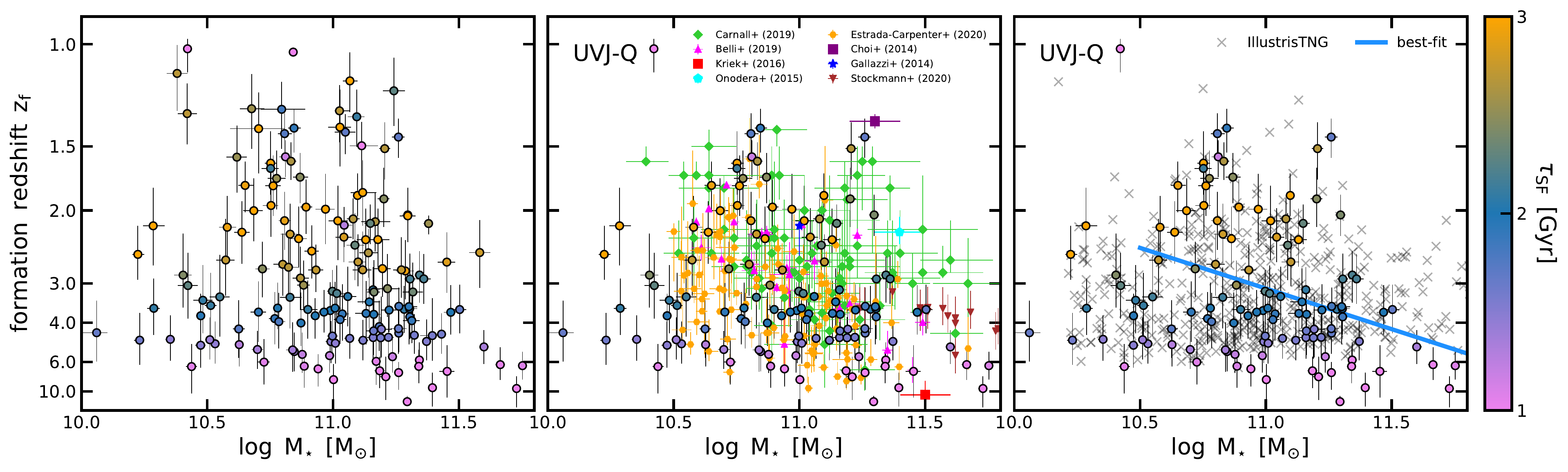}
    \caption{Formation redshift $z_{\rm f}$ (redshift at which half of the stellar mass is formed) as a function of mass. The left panel shows all galaxies in our sample, the middle panel shows only UVJ-quiescent galaxies in comparison with literature measurements \citep{belli19, carnall19, choi14, estrada-carpenter20, gallazzi14, kriek16, onodera15, stockmann20}, and the right panel compares our UVJ-quiescent galaxies with the ones from IllustrisTNG (TNG100). The color-coding corresponds to the star-formation timescales $\tau_{\rm SF}$, which is the time between when 20\% and 80\% of the mass is formed. The most massive galaxies in our sample ($M_{\star}>2\times10^{11}~M_{\odot}$) form early ($z_{\rm f}>4$) and on short timescales ($\tau_{\rm SF}\lesssim1~\mathrm{Gyr}$). At intermediate masses, we find a large diversity of formation redshifts. Galaxies that formed more recently typically have a long star-formation timescale. At low stellar masses, we find a few objects with surprisingly early formation redshifts. Overall, our formation times are consistent with \citet{estrada-carpenter20} but earlier than \citet{carnall19}. The IllustrisTNG (TNG100) galaxies, processed in the same way as our observations, are shown as gray crosses, pointing toward no trend with $M_{\star}$ but a similar distribution to our observations. The blue line shows the best fit to our observational measurements.}
    \label{fig:zf_vs_mass}
\end{figure*}

Following this early phase, the median SFH of star-forming galaxies (blue line) tracks well the simple relation of the dark matter accretion rate, and it is also consistent with lower-$z$ SFR and $M_{\star}$ measurements by \citet{leja21}. We find that the quiescent galaxies at $z_{\rm obs}\approx0.8$ decouple from the SFH of star-forming galaxies at $z\sim2-3$ and the average quenching timescale is roughly $1.5-2.0$ Gyr. Interestingly, higher-mass galaxies ($\log(M_{\star}/M_{\odot})=11.0-11.5$) transition on average slightly earlier than the lower-mass galaxies ($\log(M_{\star}/M_{\odot})=10.5-11.0$), though the average quenching timescale seems not to depend strongly on stellar mass. We discuss this further in Section~\ref{subsec:t_quench}.

\subsection{Stellar age and formation redshift}
\label{subsec:age}

The SFHs of the galaxies in our sample show a large diversity. Nevertheless, there are some common features. SFHs can be described in three parts: a star-forming phase, a transition phase, and the quiescent phase. We now use a few simple parameters to describe the overall shape, including the mass-weighted age, the formation time, the star-formation timescale, and the quenching timescale (Table~\ref{tab:timescales}). 

First, we focus on the mass-weighted age $t_{\rm age}$, which can be directly computed from our SFHs. Importantly, $t_{\rm age}$ is defined as look-back time from the epoch of observation. Therefore, $t_{\rm age}$ depends on the epoch of observation ($z_{\rm obs}$); Fig.~\ref{fig:redshift_versus_age} shows $t_{\rm age}$ as a function of $z_{\rm obs}$ for massive galaxies ($\log M_{\star}/M_{\odot}>10.5$). We confirm the large diversity of SFHs: at a given $z_{\rm obs}$, we find a large variety of mass-weighted ages. At $z_{\rm obs}\approx0.8$, some galaxies have relatively young ages ($1-2$ Gyr), while other galaxies are rather old with ages of $>6$ Gyr. The colors of the points correspond to the mass doubling $\mathscr{D}$ (Eq.~\ref{eq:mass_double}), indicating that the quiescent galaxies (red points) are typically older than star-forming galaxies (blue points). However, there is a large overlap between these two kinds of galaxies, indicating that the star-formation activity at the epoch of observation cannot tell the full story about the past SFH.

The thin gray to thick black lines show the passive evolutionary paths of simple stellar populations (SSPs) with different formation redshifts, ranging from $z_{\rm form}=1$ to $z_{\rm form}=10$. Since our SFHs are typically not well described with SSPs, this is only meant to guide the eye, highlighting that our sample spans a wide range of formation redshifts (see the next figure). We find that several of our galaxies are close to being maximally old ($z_{\rm form}\ga10$), i.e., a few galaxies are close to the upper boundary of allowed ages given by the age of the universe. 

Fig.~\ref{fig:redshift_versus_age} also compares our measurements of the mass-weighted age with other estimates in the recent literature. At redshift 0, we indicate with the red shaded region the current age constraints of massive quiescent galaxies from a range of literature \citep{gallazzi05, gallazzi21, spolaor10, trussler20}. \citet{carnall19}, \citet{belli19}, \citet{estrada-carpenter20} and \citet{stockmann20} use a similar approach to that presented here, where extended (parametric and non-parametric) SFHs are fit to individual spectra. \citet{choi14} use full-spectrum stellar population synthesis modeling on a stack of quiescent galaxies quoting SSP-equivalent ages, while \citet{onodera15} use the Lick absorption-line indices to infer the light-weighted age on a stack of 24 massive quiescent galaxies. Similarly, \citet{gallazzi14} inferred light-weighted ages of individual galaxies at $z\sim0.7$ via Lick indices (we only plot the median of the $10^{11}~M_{\odot}$ bin). Finally, \citet{kriek16} perform full-spectrum fitting on an individual quiescent galaxy at $z=2.1$. Although this literature list is extensive, it is not complete, i.e. there are several other studies that constrain the ages of massive galaxies at intermediate redshifts that we have not plotted here \citep[e.g.,][]{jorgensen17, ferreras17}. In summary, our measurements are overall consistent with those measurements, which also indicate a large variety of formation redshifts. However, we acknowledge that the definition of ``age'' adopted in the different studies also leads to some spread and that -- in particular, SSP-equivalent ages -- are not straightforwardly translated in formation redshifts.

The difficulty with interpreting Fig.~\ref{fig:redshift_versus_age} is that $t_{\rm age}$ at two different epochs cannot directly be compared with each other since $t_{\rm age}$ evolves with epoch for a quiescently evolving system. It is therefore more informative to compute the epoch that corresponds to that age: the formation redshift $z_{\rm f}$. We plot $z_{\rm f}$ as a function of $M_{\star}$ in Fig.~\ref{fig:zf_vs_mass}. The left panel shows all of the galaxies in our sample, while the middle and right panels only plot the UVJ-quiescent objects. The points are colored according to their star-formation timescale $\tau_{\rm SF}$, which is defined by the time between $t_{20}$ and $t_{80}$ (time when 20\% and 80\% of the stellar mass was formed, respectively). 

Fig.~\ref{fig:zf_vs_mass} shows that galaxies span a wide range of $z_{\rm f}$ at fixed $M_{\star}$. As expected, star-forming galaxies have typically more recent formation redshifts. The most massive, quiescent galaxies ($\log(M_{\star}/M_{\odot})\ga11.5$) formed early with $z_{\rm f}\ga4.0$, with some having a formation of $z_{\rm f}\approx10$. We find only a weak trend with stellar mass: more massive galaxies have formed slightly earlier. Fitting galaxies with ($\log(M_{\star}/M_{\odot})>10.5$) leads to $t_{\rm form}/\mathrm{Gyr} = (2.1\pm0.1)-(1.3\pm0.4)\cdot \log(M_{\star}/(10^{11}~M_{\odot}))$, where $t_{\rm form}$ is the cosmic time that corresponds to $z_{\rm f}$. Beyond this trend, our low-mass sample of quiescent galaxies shows that several of those objects interestingly formed as early as the high-mass galaxies. Finally, Fig.~\ref{fig:zf_vs_mass} also shows a trend with the star-formation timescale $\tau_{\rm SF}$: galaxies that formed early formed their stars in a shorter amount of time than galaxies that formed later. We discuss this further in the next section.

We compare the formation redshift of our sample with literature values in the middle panel of Fig.~\ref{fig:zf_vs_mass}, focusing only on UVJ-quiescent galaxies. The measurements of \citet{estrada-carpenter20} span a similar range in $z_{\rm f}$ as us, though they are probing a narrower mass range (i.e., miss the highest and lowest-mass galaxies). \citet{belli19} spans a smaller range in $z_{\rm f}$, reporting no objects with $z_{\rm f}>6$ but finding a stronger trend with stellar mass. \citet{carnall19} describe a similar mass trend to that found in our sample (see below), but their formation redshifts are systematically lower (see also \citealt{siudek17}). The \citet{stockmann20} measurements, probing the most massive galaxies, lie between our measurements and the measurements of \citet{carnall19}.

The blue line in the right panel of Fig.~\ref{fig:zf_vs_mass} shows the best-fit relation for our sample: $t_{\rm form}/\mathrm{Gyr} = 2.6-1.5\cdot \log(M_{\star}/(10^{11}~M_{\odot}))$. As also reported in \citet{carnall19}, $z=0.7$ galaxies in IllustrisTNG \citep[TNG100;][]{nelson18_color, pillepich18} seem to be inconsistent with this mass trend: in TNG100, the formation redshifts of quiescent galaxies seem not to depend on stellar mass. As described in Section~\ref{subsec:tng}, we go one step further than \citet{carnall19} by projecting the TNG100 quantities into the observational space and perform the same measurements on the predicted spectra and photometry as in our observations (Section~\ref{subsec:tng}). The TNG100 results are shown in the right panel of Fig.~\ref{fig:zf_vs_mass} as gray crosses. We confirm the negligible dependence of $z_{\rm f}$ on stellar mass: quiescent TNG galaxies at $z=0.7$ seem to have a formation redshift of $z_{\rm f}\approx2-4$, independent of $M_{\star}$. Furthermore, TNG100 produces a similar wide distribution of $z_{\rm f}$ to that of our observations, but it seems to underproduce the earliest-forming galaxies with $z_{\rm f}\gtrsim6$. This also holds when looking at the true SFHs and ages, i.e. without the processing the simulated galaxies through the observational pipeline. This could point to missing physics in the IllustrisTNG model. However, the lack of early-forming galaxies could also be explained by resolution effects, i.e., the TNG100 box might not resolve the generation of the first galaxies. The higher-resolution TNG50 box could (at least partially) alleviate this problem, but the volume probed is limited (TNG50 is roughly an order of magnitude smaller than TNG100), making cosmic variance effects more severe.

\subsection{Star-formation timescale}
\label{subsec:tau_sf}

\begin{figure}
    \centering
    \includegraphics[width=\linewidth]{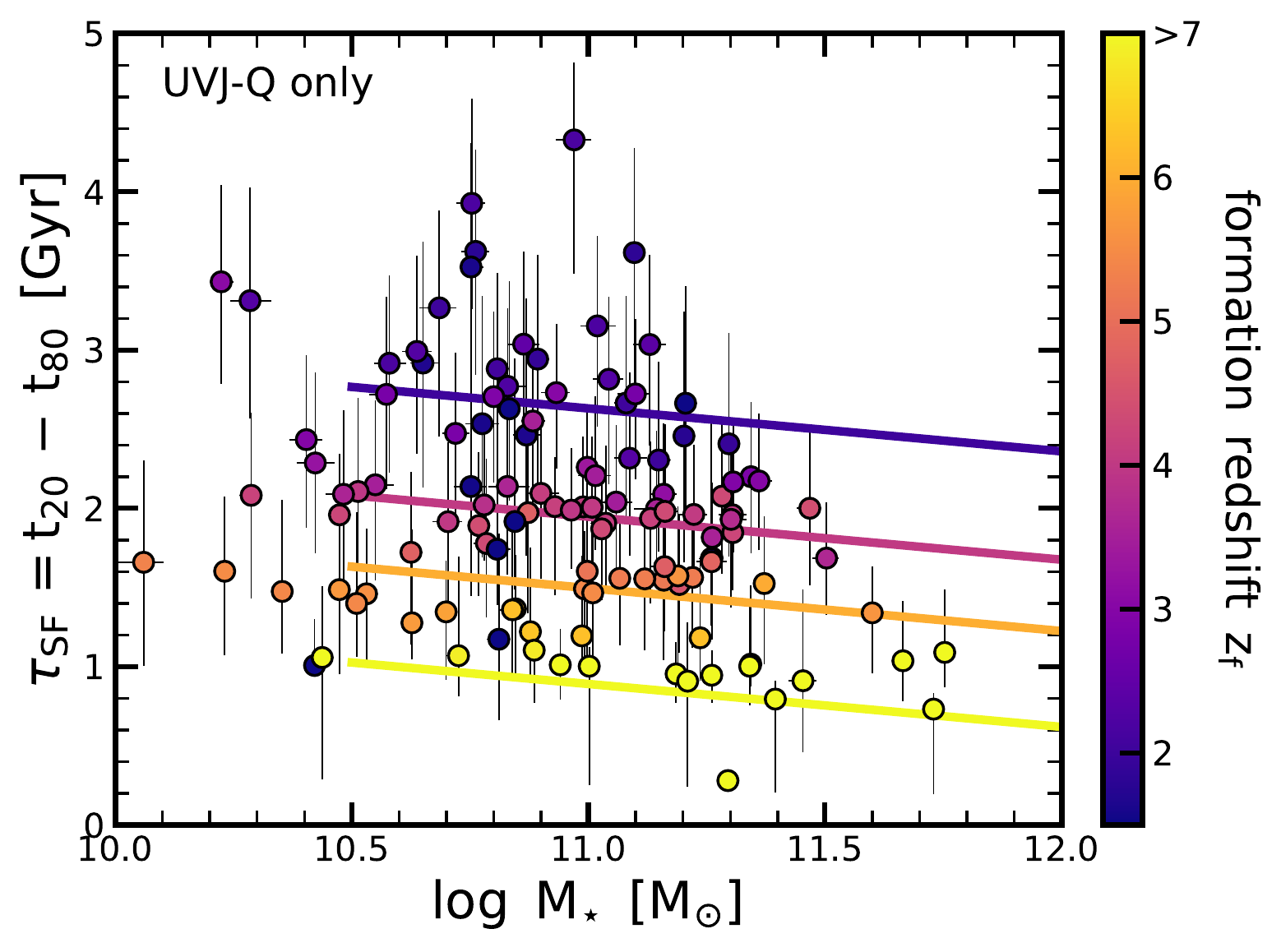}
    \caption{Star-formation timescale ($\tau_{\rm SF}$) as a function of stellar mass, color-coded by the formation redshift $z_{\rm f}$. We plot only UVJ-quiescent galaxies. The solid lines indicate the best fit of the 3D plane $M_{\star}-\tau_{\rm SF}-z_{\rm f}$ (Eq.~\ref{eq:tausf_zf}). We find that galaxies that formed early formed more rapidly.}
    \label{fig:tau_sf_vs_mass}
\end{figure}

\begin{figure*}
    \centering
    \includegraphics[width=\textwidth]{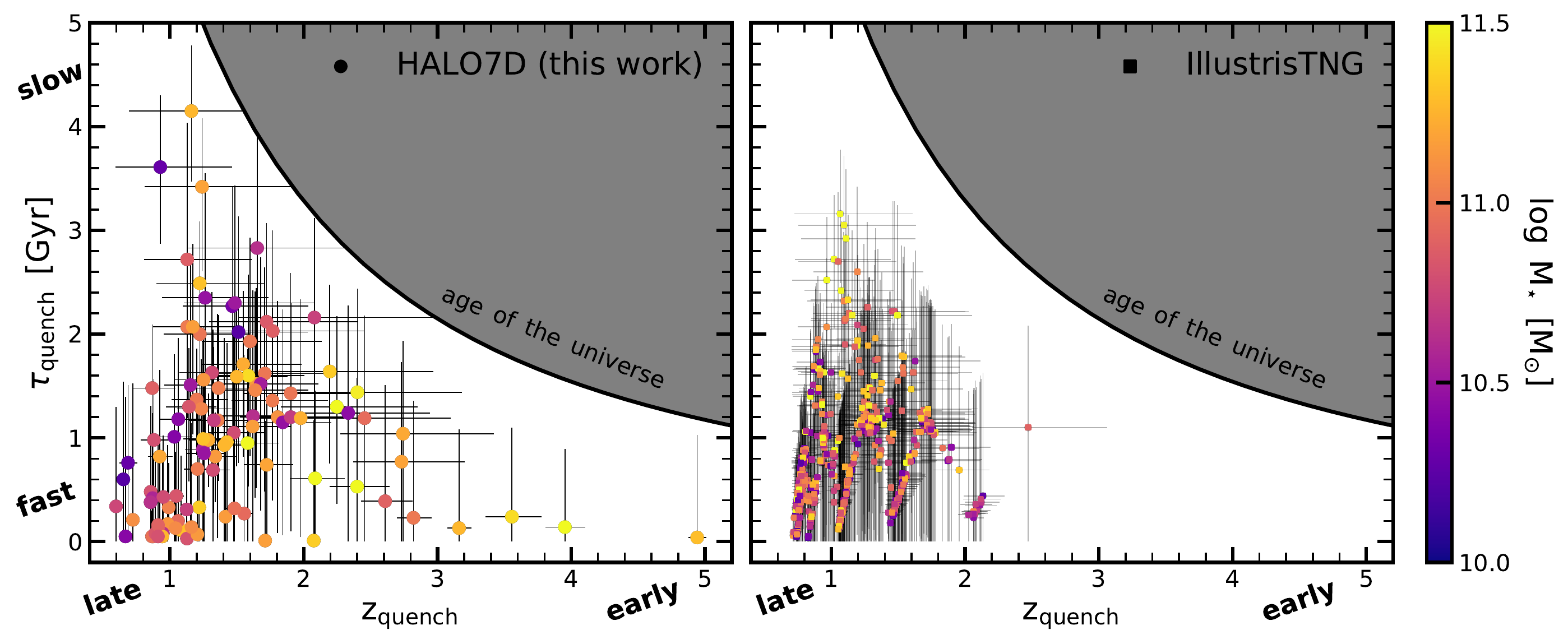}
    \caption{The quenching timescale $\tau_{\rm quench}$ as a function of the epoch of quenching $z_{\rm quench}$. The left panel shows our measurements, while the right panel shows the IllustrisTNG (TNG100) data measured with the same approach as our observations (Section~\ref{subsec:tng}). Both panels include all galaxies that quenched at some point in their past. The color-coding corresponds to stellar mass. The black solid line indicates the age of the universe, and the gray shaded region is the forbidden region (i.e., when the quenching timescale is longer than the age of the universe). The key message is that galaxies show a wide range of quenching epochs and quenching timescales: they basically fill the allowed parameter space with $z_{\rm quench}>3$ and $\tau_{\rm quench}>3$ Gyr. On the other hand, the 475 TNG100 galaxies are more confined to $\tau_{\rm quench}\approx0.1-3$ Gyr and $z_{\rm quench}\approx0.8-2$.}
    \label{fig:quenching_timescale}
\end{figure*}

We focus further on the relation between the star-formation timescale $\tau_{\rm SF}$, the formation redshift $z_{\rm f}$ and the stellar mass $M_{\star}$. Fig.~\ref{fig:tau_sf_vs_mass} plots $\tau_{\rm SF}$ as a function of $M_{\star}$, colored by $z_{\rm f}$, focusing on UVJ-quiescent galaxies. At $M_{\star}\approx10^{11}~M_{\odot}$, $\tau_{\rm SF}$ ranges from a few hundred Myr to 4 Gyr, highlighting again the diversity of pathways that achieve the same final stellar mass. As already indicated in Fig.~\ref{fig:zf_vs_mass}, there is a clear trend that early-forming galaxies have shorter star-formation timescales. While this is consistent with observational studies at $z=0$ that are based on the average ages and element abundance ratios of elliptical galaxies \citep[e.g.,][]{thomas05, graves10b}, our model assumptions (i.e., our SFH prior) allow for a different outcome (see also Appendix~\ref{app_sec:prior_impact}); this means that the data unequivocally prefer a correlation between early formation and shorter SFH timescales. As shown in Fig.~\ref{fig:corner_wo} and verified on the whole dataset, the degeneracy in the fitting is along ``older age (higher $z_{\rm f}$) -- larger $\tau_{\rm SF}$''. This can be understand by the following. The SFH is typically more constrained at recent times than at early times, implying that if one reduces the age, one needs to reduce the SFH at early times. This then leads to a shortening of $\tau_{\rm SF}$. This is the opposite trend shown in Fig.~\ref{fig:quenching_timescale}, where older galaxies (with larger $z_{\rm f}$) have typically a shorter $\tau_{\rm SF}$.

Furthermore, our measurements show that galaxies with recent formation redshifts ($z_{\rm f}<3$) all have rather long star-formation timescales ($\tau_{\rm SF}>2~\mathrm{Gyr}$), highlighting that galaxies that formed most of their mass in just 2 Gyr are uncommon at lower redshifts (``late bloomers''; Section~\ref{subsec:latebloomers}). Additionally, there is a weak trend that more massive galaxies have a shorter $\tau_{\rm SF}$. In summary, this three-dimensional plane can be described by:

\begin{equation}
\begin{split}
    \tau_{\rm SF}/\mathrm{Gyr} = & (4.1\pm0.2) - (0.3\pm0.2)\cdot\log\left(\frac{M_{\star}}{10^{11}~M_{\odot}}\right) \\ & - (3.1\pm0.2) \cdot\log(1+z_{\rm_f}),
\end{split}
\label{eq:tausf_zf}
\end{equation}

\noindent
confirming that the mass dependence is rather weak ($\propto M_{\star}^{-0.3}$), while the scaling with epoch of formation is strong ($\propto(1+z_{\rm f})^{-3.1}$). This fit is shown as solid lines in Fig.~\ref{fig:tau_sf_vs_mass} for $z_{\rm f}=2$, 4, 6, and 10, suggesting that this fit reproduces our measurements well.

\subsection{Quenching timescale}
\label{subsec:t_quench}

\begin{figure*}
    \includegraphics[width=\textwidth]{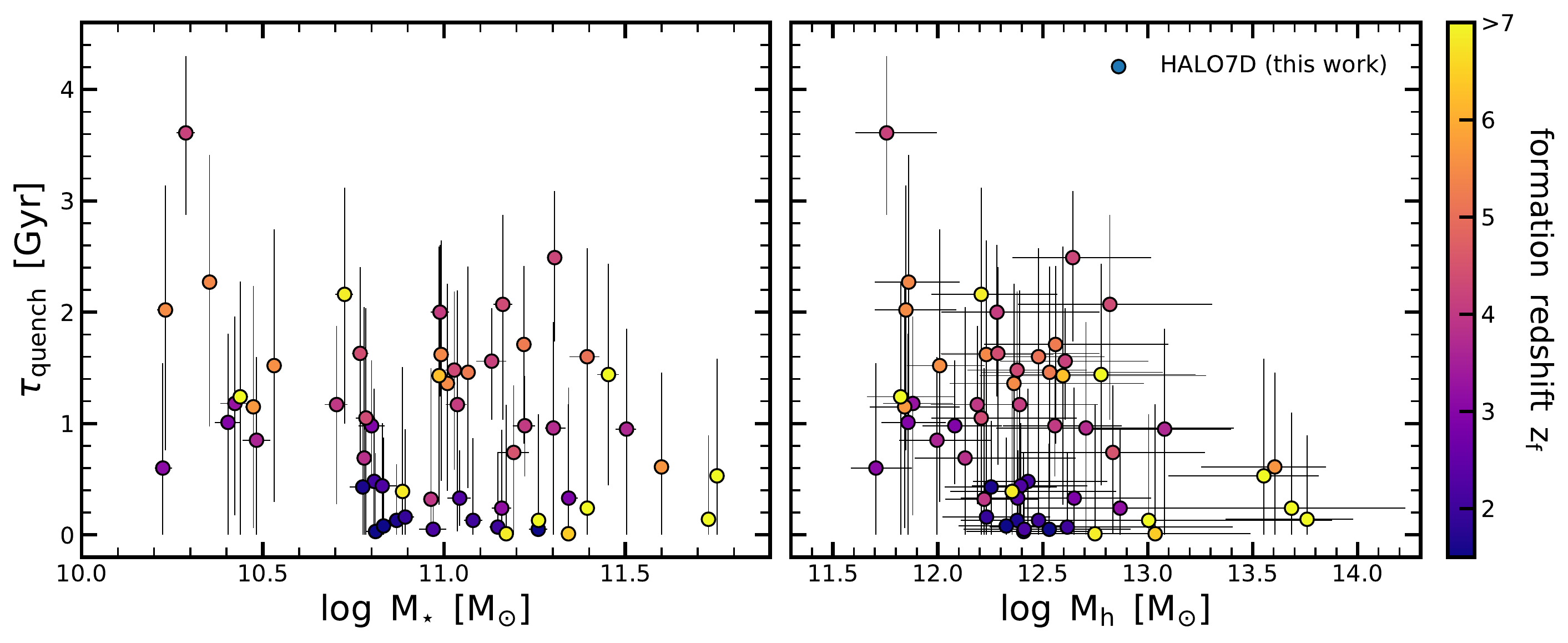}
    \caption{Quenching timescale ($\tau_{\rm quench}$) versus stellar mass ($M_{\star}$; left panel) and halo mass ($M_{\rm h}$; right panel). Both $\tau_{\rm quench}$ and $M_{\star}$ are obtained with our analysis, while $M_{\rm h}$ is estimated from the environmental density \citep{fossati17}. The color-coding of the points corresponds to formation redshift $z_{\rm f}$. Similar to the $M_{\star}-z_{\rm f}$ plane (Fig.~\ref{fig:zf_vs_mass}), the left panel shows no significant trend of $\tau_{\rm quench}$ with $M_{\star}$. In the right panel, although the uncertainties in both quantities are large, galaxies in massive halos ($\log(M_{\rm h}/M_{\odot})>13.0$) have short quenching timescales and high formation redshifts. }
    \label{fig:t_quench_vs_Mh}
\end{figure*}

Having described when and over which timescales star formation occurred in galaxies, we now move on to characterizing over which timescales galaxies cease their star formation. The analysis and figures in this section include all galaxies that quenched at some point in their past, i.e., mostly UVJ-quiescent galaxies, but also galaxies that went through a rejuvenation event recently. We quantify this by the quenching timescale $\tau_{\rm quench}$, which measures the time spent in the transition region between $1/[20t_{\rm H}(z)]<\mathrm{sSFR}<1/[3t_{\rm H}(z)]$ (between the dashed and solid black lines in Fig.~\ref{fig:sfh_all}). Furthermore, we measure the quenching epoch $z_{\rm quench}$ (Tab.~\ref{tab:timescales}), which represents when the galaxy is halfway through the transition region. Typically, the epoch of quenching is more tightly constrained than the quenching timescale (Appendix~\ref{app_sec:prior_impact}). Changing the definition of $z_{\rm quench}$ to the epoch when quenching starts has a negligible impact on our results in this section and the conclusion of this work.

Fig.~\ref{fig:quenching_timescale} shows $\tau_{\rm quench}$ as a function of $z_{\rm quench}$, color-coded by $M_{\star}$. The forbidden region, where galaxies quench on longer timescales than the age of the universe, is marked in gray. The left panel of Fig.~\ref{fig:quenching_timescale} shows our measurements, while the right panel shows the data from IllustrisTNG (TNG100). We find that the observed galaxies span a wide range of $\tau_{\rm quench}$ and $z_{\rm quench}$: consistent with Fig.~\ref{fig:sfh_massive_QG}, some galaxies transition early ($z_{\rm quench}>3$), while some galaxies quench recently ($z_{\rm quench} \approx 0.8-1.0$). Furthermore, $\tau_{\rm quench}$ spans basically the full allowed range at all $z_{\rm quench}$. We find a median quenching timescale of $\tau_{\rm quench}=1.1^{+1.2}_{-0.7}$ Gyr and $\tau_{\rm quench}/\tau_{\rm H}(z_{\rm quench})=0.23^{+0.24}_{-0.16}$ when normalized by the age of the universe. 

Our observational measurements are compared to our measurements from IllustrisTNG, processed in the same way (Section~\ref{subsec:tng}). The quiescent galaxies from TNG100 have a median $\tau_{\rm quench}=0.7^{+0.7}_{-0.4}~\mathrm{Gyr}$ and typically quench recently with $z_{\rm quench}\approx1-2$ -- consistent with the typical measurements of our observations. However, TNG100 galaxies span a narrower range in both $\tau_{\rm quench}$ and $z_{\rm quench}$. There are no galaxies that quenched early (i.e., $z_{\rm quench}>3$), consistent with the absence of early-forming galaxies ($z_{\rm f}>6$) shown in Fig.~\ref{fig:zf_vs_mass}. This may indicate that the quenching pathways in TNG100 are too monotonic and more diversity is needed. On the other hand, although the volumes probed by our observations and simulations are comparable, cosmic variance could still play a role concerning the abundance of rare objects. For example, \citet{merlin19} and \citet{valentino20} find quiescent objects at $z>3$ in the larger TNG300 box. However, those studies directly probe these high-redshift snapshots, while we perform (consistent with our observations) an archaeological stellar population approach (we expect that mergers lower $z_{\rm quench}$).

A median quenching timescale in TNG100 of $\tau_{\rm quench}=0.7^{+0.7}_{-0.4}~\mathrm{Gyr}$ for galaxies at $z=0.7$ is consistent with the timescales quoted in \citet{nelson18_color}, who find that the transition timescale through the green valley for massive $z\sim0$ galaxies is roughly 1 Gyr. They find that lower-mass $z\sim0$ galaxies transition on longer timescales ($\sim2$ Gyr). In our analysis at $z=0.7$, this trend is weaker, i.e. galaxies at all masses have similar quenching timescales: in our observations, we find $\tau_{\rm quench}=1.2^{+1.1}_{-0.8}~\mathrm{Gyr}$ and $\tau_{\rm quench}=1.0^{+0.6}_{-0.9}~\mathrm{Gyr}$ for galaxies with $\log(M_{\star}/M_{\odot})=10.5$ and $\log(M_{\star}/M_{\odot})=11.5$, respectively. This could point toward a difference due to epoch of observations (environmental quenching setting in at lower redshifts; e.g., \citealt{webb20}) or because of method (following the SFH evolution of the progenitor versus an archaeological stellar population approach) -- something to look into in more detail in the future. 

We investigate in Fig.~\ref{fig:t_quench_vs_Mh} how $\tau_{\rm quench}$ depends on stellar mass ($M_{\star}$; left panel) and halo mass ($M_{\rm h}$; right panel). The halo masses are obtained for a subsample (only those that are plotted in Fig.~\ref{fig:t_quench_vs_Mh}) of our galaxies from \citet{fossati17}, who leverage the spectroscopic and grism redshifts from the 3D-HST survey to derive densities in fixed apertures to characterize the environment of galaxies brighter than $JH_{140}<24$ mag in the redshift range $0.5<z<3.0$. The uncertainties on these halo masses are large. We compared these halo masses with the central stellar mass density ($\Sigma_1$) -- a proxy for the central velocity dispersion, which has been suggested to be correlated with the halo mass \citep{schechter15, zahid16, utsumi20}. We find indeed a correlation between $\Sigma_1$ and $M_{\rm h}$ and that galaxies follow the relation by \citet{zahid16}, although with a large scatter in $M_{\rm h}$ of about 0.5 dex at fixed $\Sigma_1$. The $\Sigma_1-M_{\rm h}$ connection has been extensively discussed in \citet{chen20}: smaller galaxies have smaller halos and lower-density centers, but there is a large scatter for star-forming galaxies. The scatter of the $\Sigma_1-M_{\rm h}$ relation gets even larger after quenching, when additional, post-quenching physical processes set in. This is something we will explore more in the future when studying the connection between the SFHs and the morphologies of the galaxies.

Although the uncertainties are large, Fig.~\ref{fig:t_quench_vs_Mh} shows that galaxies in massive halos with $\log(M_{\rm h}/M_{\odot})>13.0$ all have short quenching timescales ($\tau_{\rm quench}<1~\mathrm{Gyr}$), i.e., there is an absence of galaxies with long quenching timescales. The points in Fig.~\ref{fig:t_quench_vs_Mh} are color-coded by $z_{\rm f}$, highlighting that these massive halos also host galaxies that formed early ($z_{\rm f}>5$). At lower halo masses ($\log(M_{\rm h}/M_{\odot})\approx12.0-13.0$), $\tau_{\rm quench}$ is also correlated with $z_{\rm f}>5$, which is not surprising, since quiescent galaxies that formed recently need also to quench quickly, as described above. A similar, but weaker, trend can be found for $\tau_{\rm quench}-M_{\star}$.

\section{Discussion}
\label{sec:discussion}

We discuss here the implications of our key results. Our detailed fitting of both the spectroscopy and the photometry of 161 massive galaxies at $z\sim0.8$ indicates a large diversity of SFHs. Nevertheless, as we discuss below, our analysis supports the picture of ``grow \& quench'', where galaxies' early phase is dominated by star formation along the SFMS, followed by a transition phase to quiescence. We will end the discussion by highlighting some limitations of our work presented here and a short outlook.

\subsection{Diversity of star-formation histories}

One of the key results of our analysis is the large diversity of SFHs. Specifically, galaxies at a given epoch and stellar mass show a broad range of different paths in the sSFR versus look-back time plane (Fig.~\ref{fig:sfh_all}). In all bins, even in the most massive bin, we find galaxies of all types: star-forming, transitioning, quiescent, and rejuvenating. Furthermore, galaxies transition from the star-forming to the quiescent region at different epochs and over a range of timescales (Figs.~\ref{fig:sfh_massive_QG} and \ref{fig:quenching_timescale}). Consequently, there is a wide range of formation redshifts (Fig.~\ref{fig:zf_vs_mass}) and star-formation timescales (Fig.~\ref{fig:tau_sf_vs_mass}) at fixed stellar mass. 

Despite this diversity, there are general trends. Unsurprisingly, there is a relation between the formation redshift $z_{\rm f}$ and the star-formation timescale $\tau_{\rm SF}$: early-forming galaxies form their stars on shorter timescales (Fig.~\ref{fig:tau_sf_vs_mass} and Eq.~\ref{eq:tausf_zf}). This relation only depends weakly on stellar mass. Looking at the median evolution of the sSFR with cosmic time (Fig.~\ref{fig:sfh_avg}), we find that the sSFRs of star-forming galaxies follow $\propto(1+z)^{2.5}$, consistent with the specific accretion rate of dark matter halos and the SFMS measured independently at different epochs. This is already a first indication that the galaxies' SFMS describes an evolutionary track along which individual galaxies evolve on average. 

The similarity between the sSFRs of star-forming galaxies and the specific accretion rate of dark matter halos is expected from theoretical studies. Based on numerical simulations, \citet{dekel13} show that specific gas inflow rate and sSFR of galaxies scale with the specific dark matter accretion rate of their halos, as expected from analytical considerations using the Extended Press-Schechter approximation \citep{bond91}. Furthermore, simple empirical models that link the growth of galaxies to their dark matter halos are able to successfully describe galaxies at low and high $z$ \citep{behroozi13b, moster13, rodriguez-puebla17, tacchella18}. Importantly, this rule for star-forming galaxies ($\mathrm{sSFR}\propto(1+z)^{2.5}$) does not directly translate into a simple stellar-to-halo mass relation at earlier cosmic time because the star-formation efficiency (i.e. the slope of the stellar-to-halo mass relation) possibly depends itself on halo mass and redshifts \citep[see, e.g., Fig. 3 in][]{rodriguez-puebla16}.

\begin{figure*}
    \centering
    \includegraphics[width=\textwidth]{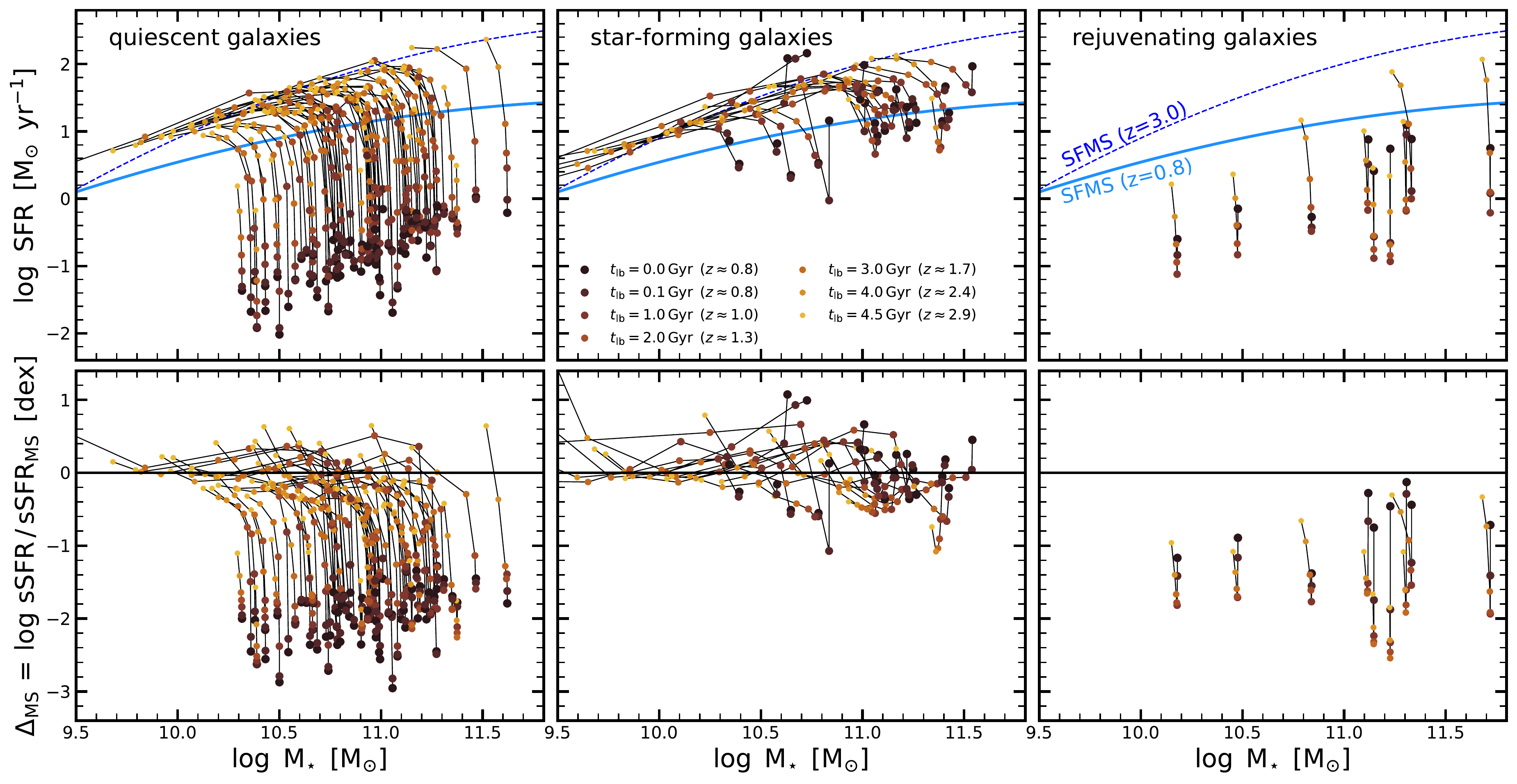}
    \caption{Evolutionary paths of galaxies in the $M_{\star}-\mathrm{SFR}$ plane (top panels) and relative to the SFMS ($\Delta_{\rm MS}=\log(\mathrm{sSFR}/\mathrm{sSFR}_{\rm MS})$, where $\mathrm{sSFR}_{\rm MS}$ is the sSFR of the SFMS at a given look-back time/redshift and stellar mass of the galaxy considered; bottom panels). We plot the $M_{\star}-\mathrm{SFR}$ and $M_{\star}-\Delta_{\rm MS}$ tracks as inferred from our SFHs for quiescent galaxies (left panel), star-forming galaxies (middle panel), and rejuvenating galaxies (right panel), classified according to Eq.~\ref{eq:mass_double}. We only plot galaxies at $z_{\rm obs}=0.6-1.0$. The points show a look-back time from the epoch of observations of $t_{\rm lb}=0.0$, 0.1, 1.0, 2.0, and 4.0 Gyr. To guide the eye, the solid light-blue (dashed dark-blue) line shows the SFMS at $z=0.8$ ($z=3.0$) \citep{leja21}. Most quiescent galaxies move back onto the SFMS by a look-back time of $\sim2-4$ Gyr, i.e., $z\sim3$. Star-forming galaxies show significant variations in the past 100 Myr of their lives, indicating variable star formation and evolution about the SFMS. Rejuvenation is significant in only 9 galaxies out of 161 galaxies of our sample, and it can take take place at all stellar masses, but it seems to prefer massive dark matter halos. }
    \label{fig:ms_evolution}
\end{figure*}

Focusing on quiescent galaxies, we find that they decouple on average from the SFMS early on ($z>3$) and quench around $z_{\rm quench}\approx1-2$ (Fig.~\ref{fig:sfh_avg}). This is consistent with recent findings of a quiescent galaxy population at $z>3$ \citep{gobat12, hill17, kubo18, schreiber18, girelli19, tanaka19, carnall20, deugenio20, forrest20a, forrest20b, saracco20, shahidi20,  valentino20, santini21}. In particular, \citet{saracco20} report a star-formation timescale of a few hundred Myr and a short quenching timescale ($\lesssim150~\mathrm{Myr}$) for a quiescent galaxy at $z=3.352$, which is consistent with our estimates. We find that the majority of those objects evolve passively to lower redshifts, but there are a few cases where rejuvenation plays a role (Section~\ref{subsec:rejuv}). Consistently, we find that some galaxies  are maximally old, meaning that they have formation redshifts of $z_{\rm f}\sim10$ (Fig.~\ref{fig:zf_vs_mass}). These super-old systems are the most massive galaxies in our sample ($\log(M_{\star}/M_{\odot})>11.2$) and also live in the highest mass halos ($\log(M_{\rm h}/M_{\odot})>13.0$) -- progenitors of today's slow rotators in the centers of clusters \citep[e.g.,][]{cappellari16}. Clearly, these early formation redshifts are, of course, exciting news for the \textit{James Webb Space Telescope} (\textit{JWST}), which will be able to shed new light onto the formation of those objects. These objects should be easily detected and characterized in detail, because we estimate them to have SFRs between $10-1000~\mathrm{M_{\odot}}~\mathrm{yr}^{-1}$ (Fig.~\ref{fig:sfh_all_sfr}). However, it is also possible that these early-forming galaxies were still in pieces at these high redshifts, making it harder for \textit{JWST} and other high-$z$ surveys to discover them.

We find that the formation redshift depends on stellar mass, but only weakly (Fig.~\ref{fig:zf_vs_mass}), which is consistent with previous results for galaxies at this and earlier epochs \citep{wu18_sfh, carnall19, morishita19, ferreras19, estrada-carpenter19}. As mentioned in the Introduction, local galaxies show a tighter relation between the stellar age and the stellar velocity dispersion than between the stellar age and stellar mass \citep{graves10b}. As previously noted by \citet{carnall19}, TNG100 is not consistent with the stellar mass trend and, overall, predicts lower formation redshifts (younger ages) than seen. It seems that TNG100 is missing early-forming quiescent objects. This is of interest since it could point toward diverse pathways to quiescence -- more diverse than what currently is implemented in TNG100, where the black hole mass is the main determinant of whether a galaxy quenches \citep[e.g.,][]{terrazas20}. On the other hand, as outlined in Section~\ref{subsec:tng}, effects of cosmic variance, sample selection, and resolution could also contribute to this absence of such galaxies.  

Although all the high-mass systems have formed early, the diversity increases toward lower masses. Interestingly, several of our low-mass, quiescent galaxies ($\log(M_{\star}/M_{\odot})=10.0-10.5$), which have not been probed by the literature previously, have formation redshifts of $z_{\rm f}\approx3-8$, similar to what we find in TNG100.

\subsection{Evolution about the star-forming main sequence}
\label{subsec:mainsequence}

As highlighted in the Introduction, a fundamental question concerning the SFMS is how galaxies evolve about this scaling relation. Are galaxies moving about this relation on short or on long timescales relative to the age of the universe? In the most extreme case, where galaxies actually do not evolve about the SFMS but follow independent (log-normal) SFH trajectories, the SFMS itself has no deeper implication and equal-mass galaxies on the SFMS need not to form an evolutionary cohort \citep{gladders13, abramson16}. In such a scenario, we would expect that the scatter of the SFMS shows a strong age gradient (see Figs. 9 and 10 in \citealt{abramson16}), because galaxies at the upper envelope of the SFMS can only have formed recently. On the other hand, if galaxies fluctuate about the SFMS ridgeline on short timescales, similarly massed galaxies have similar SFHs and the evolution of the SFMS can be used to track the SFH of galaxies. 

We first focus on tracking how galaxies evolve in the $\mathrm{SFR}-M_{\star}$ plane. Fig.~\ref{fig:ms_evolution} plots SFH tracks in the space of $\mathrm{SFR}-M_{\star}$ (top panels) and $\Delta_{\rm MS}-M_{\star}$. $\Delta_{\rm MS}$ is the log-distance from the SFMS, i.e., $\Delta_{\rm MS}=\log(\mathrm{sSFR}/\mathrm{sSFR}_{\rm MS})$, where $\mathrm{sSFR}_{\rm MS}$ is the sSFR of the SFMS at a given look-back time/redshift and stellar mass of the galaxy' history. We use the SFMS prescription of \citet{leja21}, which is based on the 3D-HST photometry and stellar population analysis presented in \citet{leja19}, meaning that the SFMS is estimated from direct measurements of SFR and $M_{\star}$ of galaxies at $z=0.5-3.0$. The left, middle, and right panels show the tracks of quiescent, star-forming, and rejuvenating galaxies, respectively. The classification is based on Eq.~\ref{eq:mass_double}. Rejuvenating galaxies are galaxies that were quiescent in the past and have $\mathscr{D}>1/20$ at the epoch of observation. Each track consists of a range of look-back times $t_{\rm lb}$, ranging back to $t_{\rm lb}=4.5$ Gyr, which corresponds to $z\approx3$. It is important to keep in mind when interpreting these tracks that they do not necessarily reflect the SFH of a single galaxy, since galaxies can merge and the resulting stellar population mixes in situ and ex situ formed stars. We discuss this further in Section~\ref{subsec:limitations}.

\begin{figure}
    \includegraphics[width=\linewidth]{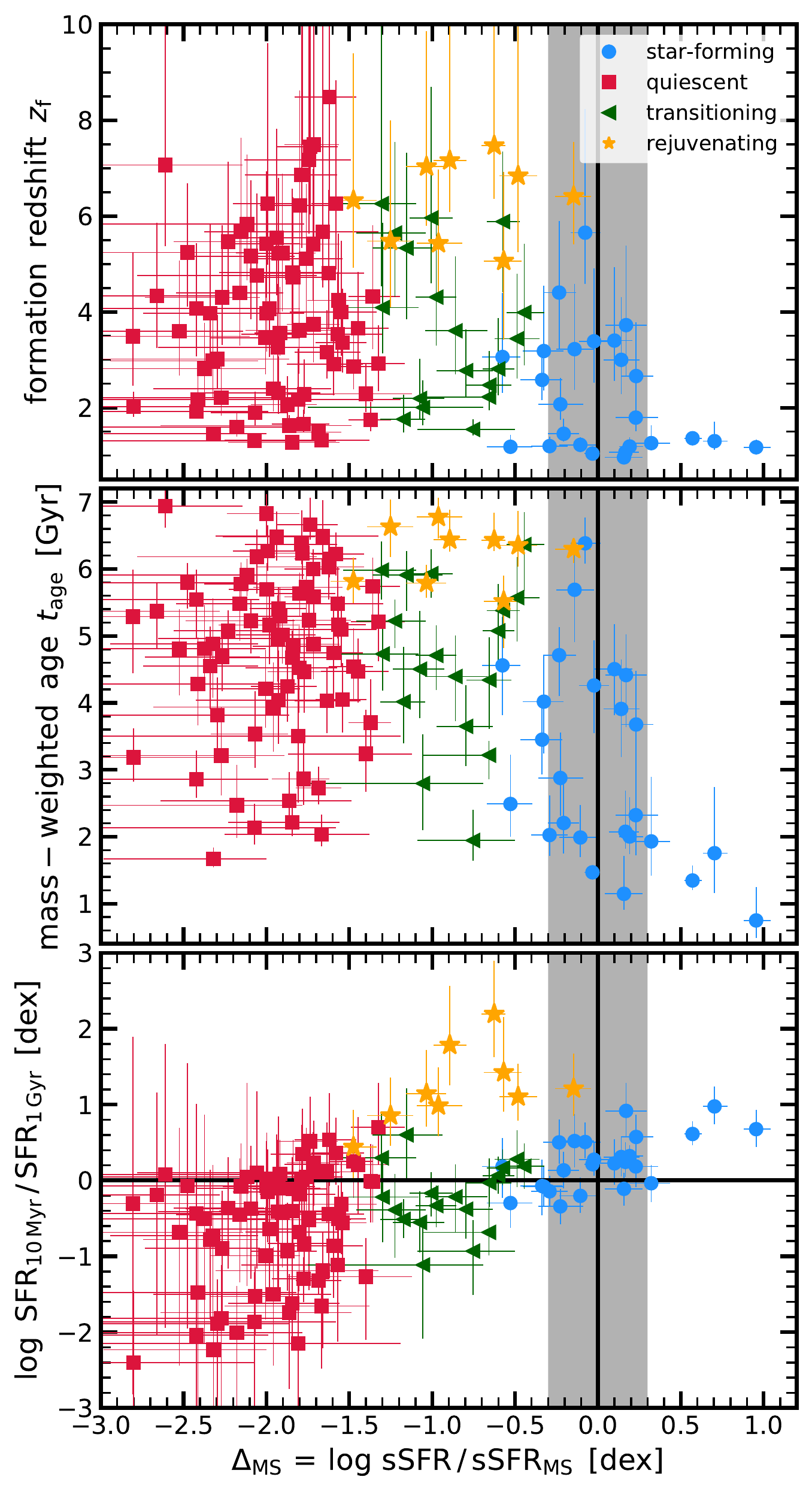}
    \caption{Formation redshift $z_{\rm f}$ (top panel), mass-weighted age $t_{\rm age}$ (middle panel), and ratio of the SFR measured over 10 Myr and 1 Gyr ($\mathrm{SFR}_{\rm 10Myr}/\mathrm{SFR}_{\rm 1Gyr}$; bottom panel) as a function of the distance from the SFMS $\Delta_{\rm MS}$, which is based on the SFR measured over 100 Myr. The blue circles, red squares, green triangles and orange stars indicate star-forming, quiescent, transitioning, and rejuvenating galaxies, respectively, classified according to Eq.~\ref{eq:mass_double}. We only plot galaxies at $z_{\rm obs}=0.6-1.0$. The vertical black line marks the SFMS ($\Delta_{\rm MS}=0.0$), while the gray band shows the SFMS's scatter (roughly 0.3 dex). There is no significant gradient for $z_{\rm f}$ and $t_{\rm age}$ across the SFMS. Only galaxies above the SFMS have all young ages, indicating that they have formed most of their mass recently. This is overall consistent with a picture where galaxies evolve about the SFMS on timescales significantly shorter than the age of the universe. We find a correlation between $\mathrm{SFR}_{\rm 10Myr}/\mathrm{SFR}_{\rm 1Gyr}$ and $\Delta_{\rm MS}=0.0$, implying that transitioning and quiescent galaxies have a declining SFH.}
    \label{fig:age_vs_MS}
\end{figure}

Fig.~\ref{fig:ms_evolution} shows that quiescent galaxies have similar tracks. This is not too surprising since when galaxies significantly reduce their SFR they will not move further to the right in this figure. Obviously, mergers can increase $M_{\star}$ of quiescent galaxies, but our SFH measurements are insensitive to this: we are only able to infer the combined SFH of the main progenitor and all of its previously merged subcomponents. We find that by $z\sim3$, most of the quiescent galaxies are on the SFMS, consistent with what we found in Fig.~\ref{fig:sfh_avg}. We will discuss these tracks and quenching more generally in Section~\ref{subsec:quenching}. Similarly, we will follow up on the rejuvenating galaxies in Section~\ref{subsec:rejuv}. 

We now focus on star-forming galaxies (middle panel of Fig.~\ref{fig:ms_evolution}), which unfortunately only includes a limited number of galaxies in our sample (31 galaxies, out of which 25 are at $z_{\rm obs}=0.6-1.0$ and are displayed in the figure). The key point to notice is that the star-forming galaxies show significant movement relative to the SFMS in recent times. Some galaxies moved more than 1 dex in the past 100 Myr. Galaxies on average cross the SFMS twice, while only one galaxy never crosses the SFMS and three galaxies cross the SFMS four times. This also holds when considering the uncertainties on the inferred SFHs. Galaxies move from above to below the SFMS on a timescale of $t_{\rm cross}=0.33_{-0.28}^{+1.74}~\mathrm{Gyr}$, while they move from below to above on a timescale $t_{\rm cross}=4.02_{-3.58}^{+1.08}~\mathrm{Gyr}$. This only considers the first crossing after the observations. Later oscillations are more difficult to estimate because of our time bins, which leads to oscillation timescales of $1-2$ Gyr. We also notice that galaxies slightly lie above the SFMS at large look-back times (see also quiescent galaxy panels on the left), which could be related to a systematic difference between the SFMS prescription and this work, or the accretion of smaller galaxies with time, leading to a bias high in SFR relative to the SFMS\footnote{The inferred SFHs might place galaxies systematically above the SFMS in the past if mergers are frequent. Assume that the SFMS can be parameterized as $\mathrm{SFR}=C \cdot M_{\star}^{\alpha}$ \citep[e.g.,][]{speagle16}. Consider two galaxies (galaxy 1 with $M_{\star,1}$ and SFR$_1$ and galaxy 2 with $M_{\star,2}$ and SFR$_2$) merging into one galaxy at time $t^{\prime}$. We then infer SFH of the galaxy. At time $t^{\prime}$, we infer an SFR from SED modeling of $\mathrm{SFR}_{\rm SFH}=\mathrm{SFR}_1+\mathrm{SFR}_2=C \cdot (M_{\star,1}^{\alpha}+M_{\star,2}^{\alpha})$. On the other hand, the SFR on the SFMS at time $t^{\prime}$ for $M_{\star,1}+M_{\star,2}$ is $\mathrm{SFR}_{\rm MS}=C \cdot (M_{\star,1}+M_{\star,2})^{\alpha}$. This leads to the ratio of
\begin{equation}
    \mathcal{R}=\frac{\mathrm{SFR}_{\rm SFH}}{\mathrm{SFR}_{\rm MS}} = \frac{(M_{\star,1}^{\alpha}+M_{\star,2}^{\alpha})}{(M_{\star,1}+M_{\star,2})^{\alpha}}.
\end{equation}
An SFMS slope in the range of $0.5<\alpha<1.0$ leads to a ratio $1.0<\mathcal{R}<1.4$, i.e., the SFR inferred from the SFH is larger than the one from the SFMS. Additionally, this could be further complicated by the different light weight of the two galaxies on the combined SED.}. 

To further study this, we look explicitly for an age gradient across the SFMS to understand whether there is some long-term correlation of whether galaxies are above or below the SFMS ridgeline. Fig.~\ref{fig:age_vs_MS} shows the formation redshift $z_{\rm f}$ (top panel), the mass-weighted age $t_{\rm age}$ (middle panel), and the ratio of the SFR measured over the previous 10 Myr and 1 Gyr (bottom panel) as a function of the distance from the SFMS $\Delta_{\rm MS}$. The vertical black line marks the SFMS, while the gray band indicates its scatter of $\sim0.3$ dex. We plot star-forming (blue circles), transitioning (green triangles), quiescent (red squares), and rejuvenating (orange stars) galaxies, classified by our mass-doubling criterion specified in Eq.~\ref{eq:mass_double}.

The top panel of Fig.~\ref{fig:age_vs_MS} shows that there is no significant gradient in $z_{\rm f}$ with $\Delta_{\rm MS}$. We notice that the galaxies above the SFMS ($\Delta_{\rm MS}>0.5~\mathrm{dex}$) have all formed recently ($z_{\rm f}<1.5$), which can be naturally explained by a burst of star formation adding significantly stellar mass, leading to a decrease of $z_{\rm f}$. However, recent formation redshifts $z_{\rm f}$ are also not uncommon on the SFMS, which shows a large diversity of $z_{\rm f}$. This is also true when we study this trend at fixed stellar mass.

The middle panel of Fig.~\ref{fig:age_vs_MS} indicates a weak, not significant trend of $t_{\rm age}$ with $\Delta_{\rm MS}$. This holds also when studying this trend at fixed stellar mass. Again, galaxies above the SFMS are the youngest galaxies in our sample, indicating that this recent burst of star formation significantly adds stellar mass. These results point toward galaxy oscillations about the SFMS on timescales significantly shorter than the age of the universe. 

The bottom panel of Fig.~\ref{fig:age_vs_MS} shows the logarithm of the ratio of the SFR measured over short (10 Myr) and over long (1 Gyr) timescales, i.e. a positive number indicates that the SFR has increased, while a negative number indicates that the SFR has decreased over the past 1 Gyr. There is no significant trend across the SFMS for star-forming galaxies; only galaxies above the SFMS have recently increased their SFRs significantly. As expected, transition and rejuvenating galaxies -- roughly at the same distance from the SFMS -- have significantly different ratios, consistent with the idea that transition galaxies are reducing the SFRs, while rejuvenating galaxies are increasing their SFRs. Quiescent galaxies typically have a negative ratio, indicating decreasing SFHs.

In summary, Fig.~\ref{fig:ms_evolution} (together with Fig.~\ref{fig:sfh_avg}) indicates that galaxies naturally move along the SFMS. Importantly, we find clear indications of short-term variations in the SFH of star-forming galaxies, indicating that at least some of these galaxies move on rather short timescales (few hundred Myr) about the SFMS. Fig.~\ref{fig:age_vs_MS} shows no strong trends of the formation redshift or galaxy age across the SFMS; only ``star-bursting'' galaxies show younger ages. It is important to stress that we focus here on $z\approx0.8$. We expect that these gradients in $z_{\rm f}$ and $t_{\rm age}$ across the SFMS are expected to increase in strength toward lower redshifts, because environmental processes will play a larger role in regulating star formation \citep[e.g.,][]{poggianti06, sanchez-blazquez09, jablonka13, cantale16, webb20, van-der-burg20}. An indication for this is, for example, the dependence of the clustering strength on the distance from the SFMS \citep{berti21}.

These results are consistent with numerical simulations, which show that galaxies oscillate about the SFMS \citep{tacchella16_MS, matthee19}. On which timescales galaxies oscillate is still uncertain observationally, but a fruitful avenue to characterize this is the use of the temporal power spectrum density \citep{caplar19}. This framework allows us to assess the importance of SFR fluctuations on a wide range of timescales and it can tell us about different physical processes regulating star formation that act on different timescales, ranging from molecular clouds to gas accretion \citep{tacchella20}. Interestingly, the predictions for the power spectrum density from galaxy formation models (numerical as well as semianalytical models) vary widely, highlighting that this is an interesting space to constrain those models \citep{iyer20}. Future, more detailed measurements of the power spectrum density via different SFR and stellar age indicators will be important to assess how galaxies evolve about the SFMS \citep[e.g.,][]{guo16, broussard19, caplar19, emami19, faisst19, wang20_p1}.

\subsection{Late Bloomers}
\label{subsec:latebloomers}

\begin{figure}
    \includegraphics[width=\linewidth]{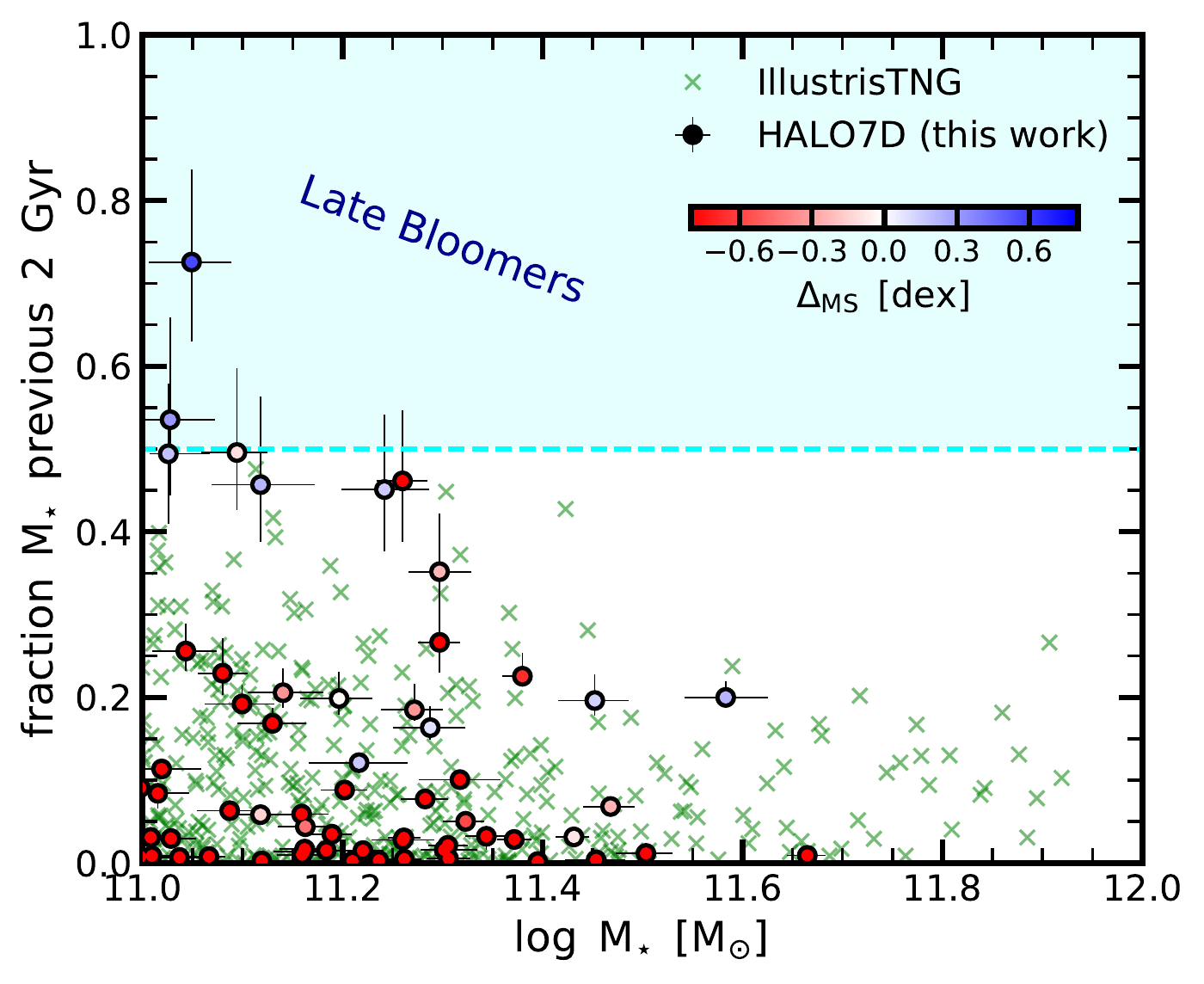}
    \caption{Fraction of mass formed in the previous 2 Gyr (look-back time) as a function of stellar mass $M_{\star}$. We focus on galaxies with $\log(M_{\star}/M_{\odot})>11.0$ since this sub-sample is unbiased with respect to SFR. Our sample is color coded by the distance from the SFMS ($\Delta_{\rm MS}$). The green crosses mark the IllustrisTNG (TNG100) galaxies at $z=0.7$. The horizontal dashed line indicates $50\%$, i.e., galaxies above this line formed the majority of their stars in the previous 2 Gyr and are classified as ``late bloomers''. Only a few of our galaxies are late bloomers, consistent with our ages estimates of $t_{\rm age}>2~\mathrm{Gyr}$.}
    \label{fig:late_bloomer}
\end{figure}

It is challenging for the ``grow \& quench'' framework to explain a large population of galaxies that are extremely young. To this end, \citet{dressler16, dressler18} searched for ``late bloomers'', which are massive ($\log(M_{\star}/M_{\odot})>10.0$) galaxies at $z<1$ that formed the majority of their stars within $\sim2$ Gyr of the epoch of observation. \citet{dressler18} found that late bloomers account for $\sim20\%$ of $z\sim0.6$ galaxies with masses of the modern ($z\approx0$) Milky Way, with a moderate dependence on mass. They conclude that this strongly contradicts paradigms in which galaxies are thought to simply grow along the SFMS and quench. \citet{mamon20}, using SDSS spectra, have also looked for such late-bloomer galaxies in the local universe (but called them very young galaxies [VYGs]), finding that they are common ($\sim50\%$) at $\log(M_{\star}/M_{\odot})\approx8.0$, but rare at high stellar masses ($\sim0.1\%$ at $\log(M_{\star}/M_{\odot})\approx11.5$).

Fig.~\ref{fig:late_bloomer} shows the fraction of stellar mass formed in the previous 2 Gyr as a function of $M_{\star}$. We focus on $\log(M_{\star}/M_{\odot})>11$ since this is the mass range where our sample is unbiased (see Section~\ref{subsec:selection}). Our measurements are indicated with circles (color-coded by $\Delta_{\rm MS}$), while the green crosses show the TNG100 data processed in the same way as our data. We find the expected correlation: galaxies above the SFMS have a higher fraction of stellar mass formed in the previous 2 Gyr.

Only a few of our galaxies are late bloomers (are above the horizontal dashed black line), which is consistent with our mass-weighted age estimates (Fig.~\ref{fig:redshift_versus_age}). We find an observed late-bloomer fraction of $f_{\rm LB}=3.6_{-1.8}^{+8.9}\%$ in the mass range of $\log(M_{\star}/M_{\odot})=11.0-11.5$. In order to understand the nature of these late bloomers, we correlate $t_{20}$ (look-back time by when 20\% of the stellar mass was formed) and $\tau_{\rm SF}$ with the fraction of mass formed in the previous 2 Gyr ($y$-axis of Fig.~\ref{fig:late_bloomer}). We find that $t_{20}$ smoothly decreases with this mass fraction, which indicates that late bloomers seem to have started their star formation more recently than non-late bloomers. Interestingly, late bloomers seem to show some diversity in $\tau_{\rm SF}$, i.e., there are objects with $\tau_{\rm SF}\sim1.5$ Gyr as well as objects with $\tau_{\rm SF}>3$ Gyr. This indicates that late bloomers can form through both a recent burst of star formation and a more continuously increasing SFH. This is also confirmed when looking at the SFHs themselves (i.e. SFR versus time, Fig.~\ref{fig:sfh_all_sfr}).

In comparison with TNG100, we do not find a significantly different distribution in Fig.~\ref{fig:late_bloomer}, but there are no late bloomers at all in TNG. The fraction is consistent with 0, i.e., $f_{\rm LB}=0.0_{-0.0}^{+1.3}\%$. \citet{dressler18} quote at $\langle z \rangle =0.6$ $f_{\rm LB}\approx5-10\%$ for galaxies with $\log(M_{\star}/M_{\odot})=11.0$, which is slightly higher than what we find in our observations and TNG100. The weak disagreement between \citet{dressler18} and our measurement could point toward an overestimate of the late-bloomer fraction in \citet{dressler18} -- in particular toward lower $M_{\star}$ -- caused by the assumed prior in their work. They fit for the mass formed in $N$ fixed time bins, which is not agnostic about the shape of $\mathrm{SFR}(t)$ but instead prefers rising SFHs and high instantaneous sSFRs \citep{leja19_nonparm}. This can be understood intuitively by imagining a series of random draws for each of the $N$ time bins that add up to a fixed total mass by construction (typically, this total mass is loosely constrained by the normalization of the SED). Most of the SFHs thus constructed will have significant mass in only one or two time bins, which naturally results in ``bursty'' SFHs and in SFHs for star-forming galaxies that are rising. These effects could lead to an overestimation of the late-bloomer fraction. A separate issue is the uncertainty in the assumed dust attenuation law. In order to estimate the stellar mass formed in the previous 2 Gyr (in particular, the age range of $0.5-2.0$ Gyr) accurate estimates of the far-UV and near-UV luminosities are needed (see discussion in Appendix~\ref{app_sec:gain}). This means performing an accurate dust correction. We know that the dust attenuation law varies from galaxy to galaxy \citep{kriek13, reddy15, salim20, shivaei20}, as expected from theoretical models \citep{smith15, narayanan18}, and there is a need for a flexible attenuation law when reproducing the physical properties of a large variety of objects \citep{lo-faro17}. We marginalize over different slopes of the dust attenuation law, while \citet{dressler18} assume a fixed \citet{calzetti00} law but allow for different dust amounts in each SFH time bin. These different approaches could also lead to a difference in $f_{\rm LB}$.

\subsection{How are galaxies quenching? Fast or slow?}
\label{subsec:quenching}

Quiescent galaxies decouple from the median SFH of star-forming galaxies and the typical accretion rate of dark matter (Fig.~\ref{fig:sfh_avg}), thereby falling significantly below the SFMS ($>2~\mathrm{dex}$). We show that galaxies transition over a wide range of timescales, from a few tens of Myr to a few Gyr (Fig.~\ref{fig:quenching_timescale}). The typical quenching timescale in our sample is $1.1_{-0.7}^{+1.2}~\mathrm{Gyr}$. About 25\% of our galaxies quench on short timescale with $\tau_{\rm quench}<500~\mathrm{Myr}$. This is consistent with complementary studies of post-starburst galaxies, which infer that such fast-quenching galaxies are responsible for about $20-50\%$ at $0.5<z<1.5$ \citep{belli19, wild20}.

Focusing on the left panel of Fig.~\ref{fig:ms_evolution}, we find that galaxies quench over a wide range of stellar masses. The lowest-mass systems quench at $\log(M_{\star}/M_{\odot})\approx10.3$, while the highest-mass systems have $\log(M_{\star}/M_{\odot})\approx11.6$. As mentioned before, there is a high probability, in particular for the most massive systems, that these galaxies have merged and consist of multiple progenitors. This means that the quenching mass range in reality might be narrower. For example, in TNG100, galaxies typically quench in the range $\log(M_{\star}/M_{\odot})\approx10.5-11.0$. Above this mass range, the ex situ mass fraction increases \citep{tacchella19}. Furthermore, these tracks are nearly vertical, indicating little mass growth during quenching. Specifically, we measure a relative stellar mass growth (in comparison with final stellar mass) in the ``green valley'' of $f_{\rm \Delta M_{\star}}=7.9_{-7.6}^{+25.0}\%$, which, by definition, is correlated with the quenching timescale. 

In the literature, several studies highlight the need for two quenching paths -- slow and fast -- in both observations \citep[e.g.,][]{yesuf14, barro16_inact, wu18, belli19, carnall19} and simulations \citep[e.g.,][]{rodriguez-montero19}. We do not find a bimodal distribution of quenching timescales in Fig.~\ref{fig:quenching_timescale}, which could be related to the rather large uncertainties of $\tau_{\rm quench}$. At face value, our quenching timescale distribution and the wide quenching mass range can be explained by a combination of a quenching mechanism with a ``gate keeper'' \citep{tacchella16_MS}: while the galaxy grows in stellar mass on the SFMS, the galaxy tries to quench and depart the SFMS via stellar and black hole feedback, which itself can be activated through mergers and violent gas instabilities \citep{hernquist89, zolotov15}. However, as long as the replenishment time of fresh gas is short enough, i.e. there is enough fuel available for sustaining star formation, the galaxy is forced to keep forming stars and stay on the SFMS. Only when the halo is hot enough (the halo mass plays the role of the aforementioned ``gate keeper'') can fresh gas not be supplied after another quenching attempt, and the galaxy is able to quench. A result of this combination of a quenching mechanism with a ``gate keeper'' is a wide range of quenching timescales and quenching masses since the exact timing of when the quenching mechanism is activated (e.g., a merger takes place) and the timing of when the critical halo mass is reached differ from galaxy to galaxy. This is also consistent with Fig.~\ref{fig:t_quench_vs_Mh}, where the most massive halos (presumably the halos with the most shock-heated gas) host galaxies that were able to quench rapidly.

Consistently, \citet{chen20} argue that the wide quenching mass range can be explained by varying black halo masses that beat against halo masses that also vary across this mass range. This picture accounts nicely for the width and stable mean mass of the quenching mass range, and it naturally accommodates a ``fast'' quenching channel precipitated by gas-rich mergers. However, this work has only little to say about how long it takes to quench, since it focuses mainly on preventive feedback, which has a natural timescale of the order of the halo dynamical timescale. Quenching timescales should be shorter at early times since halos are denser, but we do not find a such trend here. Ejective feedback might lead to more variation in the quenching timescale. 

An additional observational avenue to constrain the quenching mechanism(s) is to study the spatial progression of quenching within galaxies. The first steps in this direction have been done in high-redshift galaxies, both for star-forming galaxies by studying the distribution of star formation \citep{wuyts13, tacchella15_sci, nelson16_insideout, abdurrouf18, morselli19} and for quiescent galaxies by studying the stellar populations on spatially resolved scales \citep{akhshik20, jafariyazani20}.

\subsection{Rejuvenation is uncommon}
\label{subsec:rejuv}

As mentioned above, rejuvenation in our sample is uncommon. We use a rather stringent criterion, where galaxies are rejuvenating if they were quiescent in the past and are either in the star-forming or the transition region at the time of observation (based on the criterion in Eq.~\ref{eq:mass_double}). Specifically, only two galaxies in our sample of 161 galaxies rejuvenate back to be star-forming, corresponding to $1.2\%$. There are seven other galaxies ($4.3\%$) that rejuvenate, but only to the transition region. Interestingly, as highlighted in Fig.~\ref{fig:age_vs_MS}, these rejuvenating galaxies all have high formation redshifts ($z_{\rm f}\approx5-8$) and are rather old (half-mass age of $t_{\rm age}>5~\mathrm{Gyr}$), which indicates that only little mass has formed in these recent rejuvenation events ($<10\%$). Also, we do not see any indication of AGNs in these systems ($f_{\rm AGN}<1\%$).

Consistent with high formation redshift and old age, we find that rejuvenated galaxies prefer to lie in massive halos. In particular, we find that the fraction of rejuvenating galaxies at $\log(M_{\rm h}/M_{\odot})<13.0$ is only $f\approx4\%$, while at $\log(M_{\rm h}/M_{\odot})>13.0$, it is $f\approx30\%$. This could signal a cooling flow failure as seen in local massive ellipticals such as Perseus A (NGC 1275; e.g., \citealt{canning14}) and Abell 1795 \citep[e.g.,][]{ehlert15}, which are massive central cluster galaxies but are hosting star formation. Galaxy-galaxy merger, which can bring in new fuel for star formation, might also play a role, but because of the high relative velocity of galaxies in these high-density environments (i.e., (proto)clusters), mergers should be less common.

The rarity of rejuvenating galaxies is consistent with theoretical expectations \citep{trayford16, pandya17} and other observational measurements  \citep{belli17, chauke19, akhshik21}. TNG100 predicts a rejuvenation fraction above $\log(M_{\star}/M_{\odot})>10.5$ of $1.6\%$. In \citet{chauke19}, the authors infer that 16\% of the $z\sim0.8$ quiescent population has experienced rejuvenation events in the redshift range $0.7 < z < 1.5$ after reaching quiescence at some earlier time. This is a different criterion than the one we are using. With their criterion, a double-peaked SFH (separated by an $\mathrm{SFR}\approx0$) is considered rejuvenated (see Fig. 4 of their work). This raises, of course, the interesting question of the exact definition of rejuvenation. Since it is challenging to recover a short burst of star formation several Gyr back in time and our prior weights against such a behavior \citep{leja19_nonparm}, our definition of rejuvenation is a conservative one and focuses on the current epoch only. 

Furthermore, we believe that the quoted 16\% in \citet{chauke19} is an overestimation of rejuvenation events. Firstly, a double-peaked SFH (with quiescence in between) does not imply rejuvenation. It could also be that two galaxies, each with a different formation redshift, have merged. The resulting merger remnant will have a double-peaked SFH, but no actual rejuvenation took place. Secondly, \citet{chauke19} fit for the mass formed in $N$ fixed time bins, which leads to a prior in SFHs that tends to form the majority of the mass in one or two time bins (see Section~\ref{subsec:latebloomers}). Nevertheless, we agree on their main conclusion that these rejuvenation events only contribute little to the total stellar mass and that they are not an important evolutionary channel when considering the growth of the quiescent galaxy population. 

Clearly, it would be be of great interest to study the morphology of these rejuvenating galaxies, which we will do in the future. These galaxies might be consistent with objects such as presented in \citet{mancini19}, where the SFHs have been measured for bulge and disk components separately for 10 galaxies at $0.45 < z < 1$. These authors have argued that the bending of the SFMS is due significantly to quiescent galaxies acquiring star-forming disks, which rejuvenate the systems.

\subsection{Limitations and outlook}
\label{subsec:limitations}

A limitation in our study is the sample size of 161 galaxies, which makes it difficult to split the sample by physical properties, such as mass, star formation, or formation redshift. Large samples of both star-forming and quiescent galaxies will allow us, for example, to shed more light on the correlation between quenching timescale and other galaxy properties. Furthermore, our sample only includes a few star-forming galaxies; increasing the sample size of star-forming galaxies will help us to understand how galaxies evolve about the SFMS in more detail. 

In the future we are planning to expand the sample size by including galaxies from other spectroscopic surveys, such as LEGA-C \citep{van-der-wel16} and VANDELS \citep{mclure18_vandels, pentericci18}. In addition, for increasing the number of objects, the inclusion of these surveys will help to span a wider range of epochs of observation. This will be particularly useful for understanding how galaxies evolve post-quenching, including both major and minor mergers. As highlighted above, a major limitation when studying SFHs is the uncertainty related to mixing several galaxies into a single galaxy and misinterpreting the measured SFH as the evolutionary path of one single object. Comparing the SFH of in situ and ex situ stars in TNG100 does not actually show a significant difference, which might alleviate some of these concerns. It seems that galaxies, which will later be accreted by the main progenitor, have similar SFHs to the main progenitor itself, i.e., there is some overarching coherence scale for the star-formation in galaxies in a certain environment. Nevertheless, a more detailed investigation is needed.

Understanding how representative a measured SFH is in comparison to the evolutionary path of the main progenitor is an open question. Probing galaxies at different epochs could help with this (but see \citealt{abramson16}). Furthermore, connecting the measured SFHs to the morphology of the galaxies is also of great interest \citep{belli15, fagioli16, almaini17, faisst17_size, williams17, maltby18, wu18, estrada-carpenter20, suess20} -- something we will focus on in an upcoming publication. This is of interest not only for understanding the evolution of galaxy morphology with epoch but also because the morphology of galaxies can give us clues about the occurrence of mergers.

\section{Conclusions}
\label{sec:conclusion}

We present a detailed stellar population analysis of 161 massive galaxies at a redshift $z_{\rm obs}\approx0.8$ from the Keck/DEIMOS survey HALO7D. We use \texttt{Prospector} \citep{johnson21} to fit a 27-parameter model to both UV-to-IR photometry and rest-frame optical spectroscopy from Keck/DEIMOS. The model parameters (Table~\ref{tab:parameters}) describe a wide range of different physical components, including a 10-bin non-parametric SFH, a flexible dust attenuation law, and emission by an AGN component, emission lines, and dust. Since the \texttt{Prospector} framework is fully Bayesian and forward-models many aspects of spectroscopic data analysis and calibration, we are able to capture parameter degeneracies and to marginalize over them, providing us with realistic uncertainties. We show that fitting both photometry and spectroscopy with a flexible dust attenuation law is necessary to fully break the dust-age-metallicity and constrain the SFH of galaxies on short (few tens of Myr), intermediate (few hundreds of Myr), and long (few Gyr) timescales. 

The key result of our analysis is the diversity of SFHs: at a given epoch and stellar mass, galaxies of all types are present (star-forming, transitioning, and quiescent), and we find a large range of star-formation timescales, quenching timescales, and quenching epochs (Figs.~\ref{fig:sfh_all} and \ref{fig:zf_vs_mass}). Quiescent galaxies, even at fixed stellar mass and redshift, quench fast and slow, early and late. Nevertheless, there are some intriguing, uniform trends in our data: galaxies that form early formed their stellar mass on shorter timescales, while galaxies that form late formed their stars on longer timescales (Fig.~\ref{fig:tau_sf_vs_mass}). Star-forming galaxies' SFHs follow on average the expected trend from dark matter accretion histories (Fig.~\ref{fig:sfh_avg}): the sSFRs of star-forming galaxies tracks the specific accretion rate of their dark matter halos ($\mathrm{sSFR}\propto(1+z)^{2.5}$), which is expected from numerical and analytical models. Individual SFHs of star-forming galaxies show variability, thereby crossing the SFMS ridgeline several times (Fig.~\ref{fig:ms_evolution}). We do not find a correlation between the distance from the SFMS and formation redshift (Fig.~\ref{fig:age_vs_MS}), implying that the SFMS oscillations take place on shorter timescales than the age of the universe. 

Quiescent galaxies decouple from the SFMS and quench on short and long timescales: about 25\% of our sample quenches on timescales $\tau_{\rm quench}<500~\mathrm{Myr}$ that are short relative to the dynamical timescale of their dark matter halos (Fig.~\ref{fig:quenching_timescale}). The large range of $\tau_{\rm quench}$ in our sample points toward a combination of galaxy-internal (such as stellar and black hole feedback) and galaxy-external (such as a hot dark matter halo) mechanisms to quench galaxies. Furthermore, green valley (transitioning) galaxies grow only little in stellar mass: the relative mass growth (in comparison with final stellar mass) amounts to $f_{\rm \Delta M_{\star}}=7.9_{-7.6}^{+25.0}\%$ (Fig.~\ref{fig:ms_evolution}). 

We presented several comparisons with the galaxy formation model IllustrisTNG (specifically, medium-sized box TNG100). TNG100 overall produces similar formation redshifts as our observations but seems to miss the earliest-forming galaxies ($z_{\rm f}>6$), which is expected because of the finite resolution of the simulation, but effects related to cosmic variance and sample selection can also contribute. The typical timescale for quenching is in agreement with our observations, though the overall distribution is narrower than what we measure (Fig.~\ref{fig:quenching_timescale}), which could point to less diversity in quenching pathways compared to the real universe.

Overall, our analysis supports a ``grow \& quench'' framework where galaxies grow along the SFMS and then quench on a wide range of timescales. We find a wide and continuously populated diversity of quenching timescales. Our observations show that rejuvenation of star formation in quiescent galaxies is rare. However, we find that galaxies do not simply grow exactly on the SFMS ridgeline but rather oscillate about it, and understanding the exact timescales of these oscillations is of great interest to constrain the regulation of star formation in galaxies \citep{tacchella20}. In the future, we will connect the derived SFHs in this work with galaxies' morphology and metal abundances to further constrain their formation and evolution. Another avenue to explore these HALO7D spectra is to investigate spatial gradients in colors and stellar populations within galaxies. Finally, as shown in this work, we expect quiescent galaxies to be present at early cosmic time, which opens the door for JWST to unravel the physics of quenching of these first quiescent galaxies.

\section*{Acknowledgements}

We thank the referee for a thorough report on a lengthy paper that greatly improved and strengthened this work. We are grateful to Margaret Geller, Sirio Belli, Reinhard Genzel, Kartheik Iyer, Dylan Nelson, Erica Nelson, and Annalisa Pillepich for productive discussions and comments. ST is supported by the Smithsonian Astrophysical Observatory through the CfA Fellowship and by the 2021 Research Fund 1.210134.01 of UNIST (Ulsan National Institute of Science \& Technology). SMF, GB, YG, DCK, and HMY received  partial support from NSF grants AST-0808133 and  AST-1615730. ECC is supported by a Flatiron Research Fellowship at the Flatiron Institute, which is supported by the Simons Foundation.

The computations in this paper were run on the FASRC Cannon cluster supported by the FAS Division of Science Research Computing Group at Harvard University. This research made use of NASA's Astrophysics Data System (ADS), the arXiv.org preprint server, the Python plotting library \texttt{matplotlib} \citep{hunter07}, \texttt{astropy}, a community-developed core Python package for Astronomy \citep{astropycollaboration13, astropycollaboration18}, and the Python binding of \texttt{FSPS} \citep{foreman_mackey14}.

This work has made use of the Rainbow Cosmological Surveys Database, which is operated by the Centro de Astrobiología (CAB/INTA), partnered with the University of California Observatories at Santa Cruz (UCO/Lick, UCSC). Some of the data presented herein were obtained at the W. M. Keck Observatory, which is operated as a scientific partnership among the California Institute of Technology, the University of California and the National Aeronautics and Space Administration. The Observatory was made possible by the generous financial support of the W. M. Keck Foundation. The authors wish to recognize and acknowledge the very significant cultural role and reverence that the summit of Maunakea has always had within the indigenous Hawaiian community.  We are most fortunate to have the opportunity to conduct observations from this mountain.

\appendix

\section{Instrumental Line Spread Function (LSF)}
\label{app_sec:lsf}

\subsection{Measurement of the LSF}

\begin{figure}
    \centering
    \includegraphics[width=0.5\textwidth]{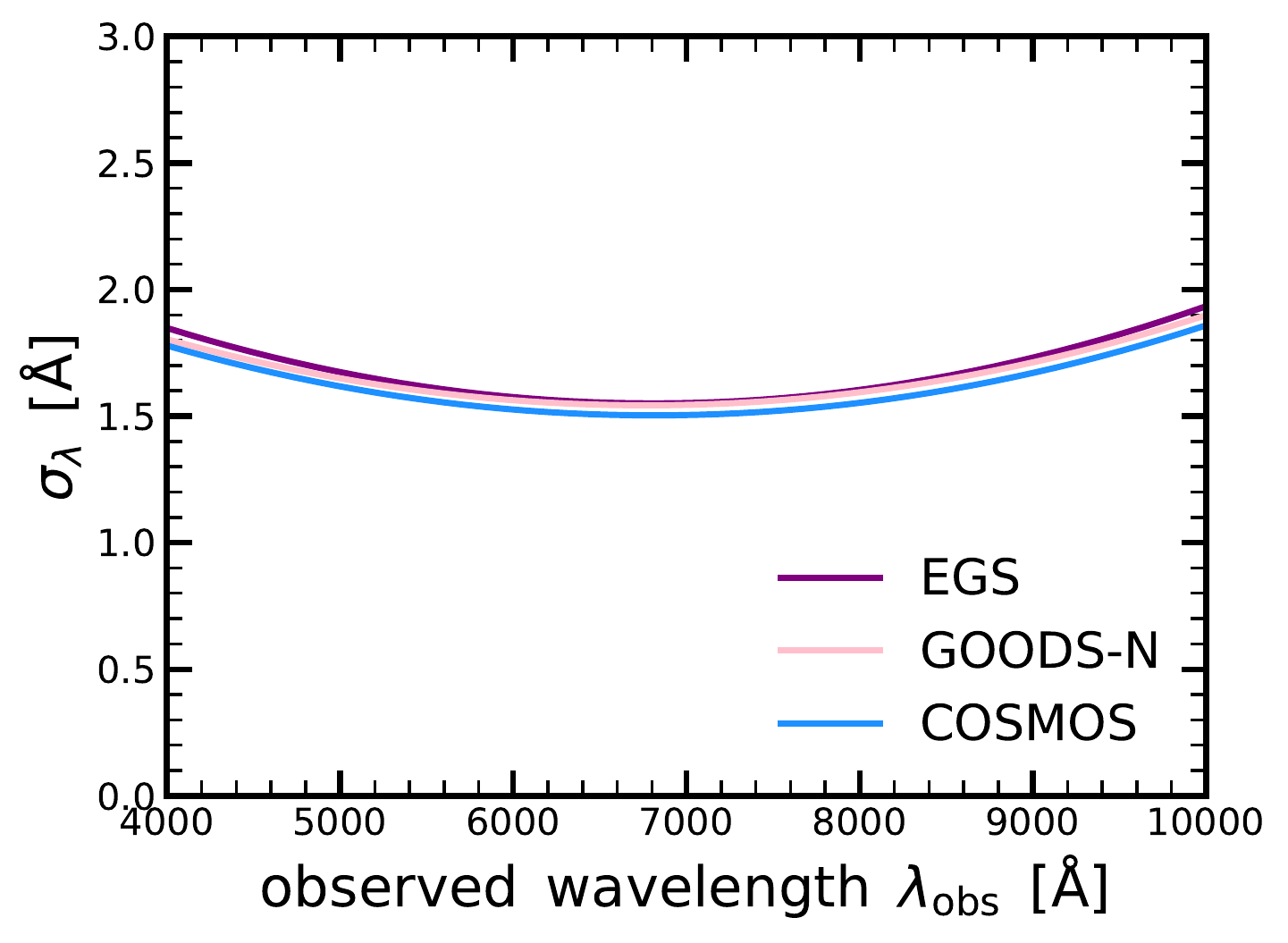}
    \caption{Instrumental line spread function (LSF) adopted in this work. We assume that the LSF can be described as a Gaussian with $\sigma_{\rm lambda}$. The solid lines show the quadratic LSF for each individual field. We find only little field-to-field variations. }
    \label{fig:lsf}
\end{figure}

As described in Section~\ref{sec:physical_model}, the high-resolution empirical stellar library MILES has a comparable resolution to our observed galaxy spectra. In particular, the observed spectra of our low-$z$ galaxies have a lower resolution than the MILES templates. Therefore, the MILES templates need to be matched to the LSF of our observations by convolving them with a kernel. This instrumental LSF of the observation describes the observed shape of an intrinsically very narrow spectral line at the wavelength $\lambda$, due to purely instrumental broadening, including the effect of integrating over the detector pixels. 

We construct the LSF for each HALO7D/CANDELS field separately. Specifically, we measure the Gaussian width $\sigma_{\lambda}$ as a function of wavelength from fitting Gaussians to arc lamps and night skylines for all the slits that fell in each field. We then fit a quadratic function to all the width measurements as a function of wavelength. These LSFs are shown in Figure~\ref{fig:lsf}. The solid lines show the quadratic LSFs for each individual field. We measure a typical width of $\sigma_{\lambda}\approx1.5-2.0~\mathrm{\AA}$ and find little difference from field to field. 

Since we assume that the LSF is Gaussian, the kernel with which the templates need to be convolved is also Gaussian with a width $\sigma_{\rm diff}$. Specifically, we determine the broadening $\sigma_{\rm diff}$ of the rest-frame MILES library spectra required to match the observed LSF by:

\begin{equation}
    \sigma_{\rm diff}^2 = \sigma_{\rm v}^2-\sigma_{\rm v, MILES}^2
\end{equation}

\noindent
where $\sigma_{\rm v} = c \times \sigma_{\lambda}/\lambda_{\rm obs}$ and $\sigma_{\rm v, MILES}=c \times \mathrm{FWHM}_{\rm MILES} / 2.355 / \lambda_{\rm rest}$ with $\mathrm{FWHM}_{\rm MILES}=2.54~\mathrm{\AA}$. As noted in the main text, $\sigma_{\rm diff}$ becomes negligibly small for high-redshift galaxies in our sample. This means that the observed LSF is dominated by the physical stellar velocity dispersion -- which is fitted -- and the ``uncorrectable'' difference between the MILES and instrumental LSF is not important to the final total LSF. Finally, the effects of the LSF on the inferred stellar population parameters are negligible.

\section{Gain in fitting UV-IR photometry and optical spectroscopy together}
\label{app_sec:gain}

\begin{figure*}
    \centering
    \includegraphics[width=1.0\linewidth]{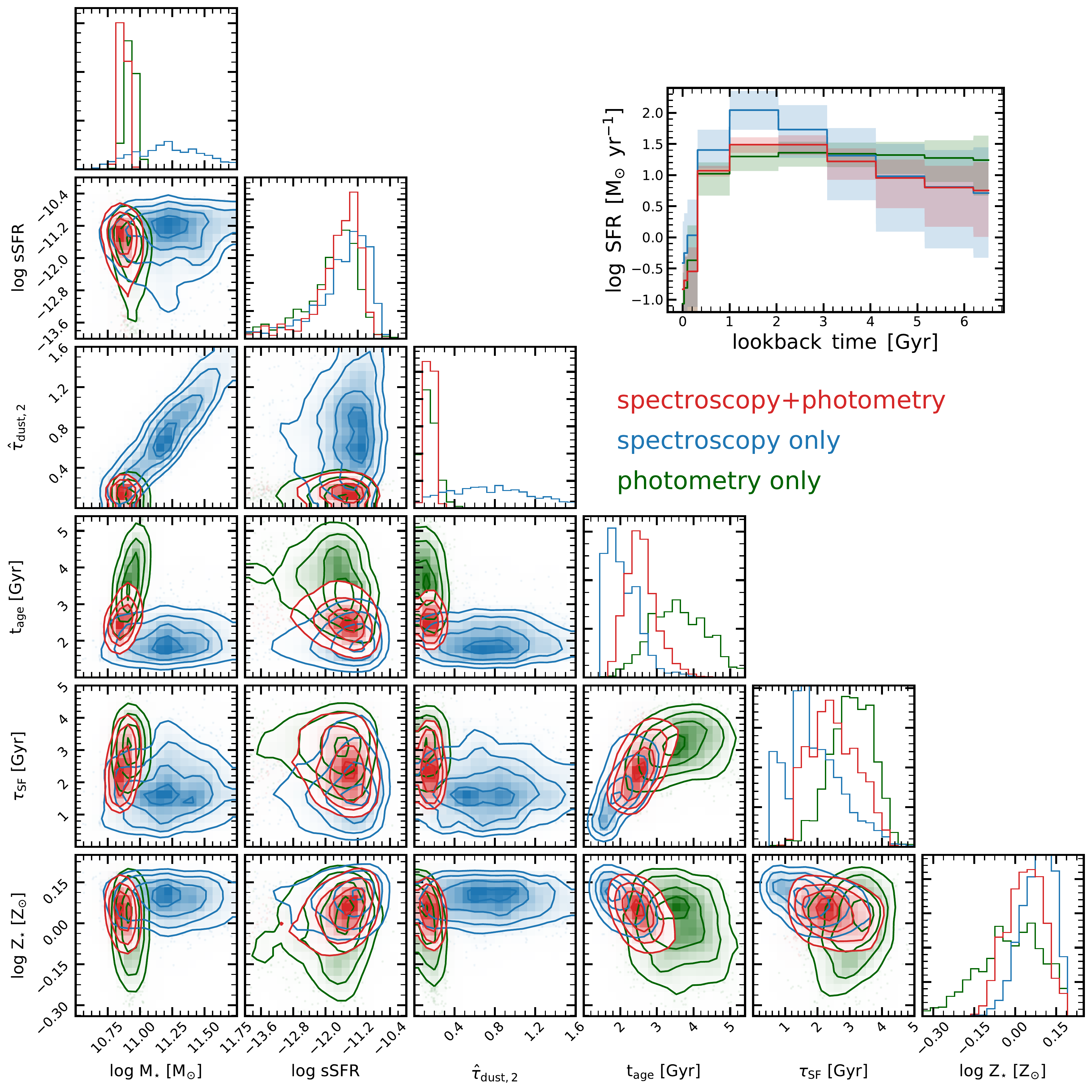}
    \caption{The need for spectroscopy and photometry. This joint posterior figure for the galaxy ID 16490 (same as shown in Fig.~\ref{fig:example_corner}) shows the posteriors of $M_{\star}$, sSFR (measured over the past 100 Myr), $\hat{\tau}_{\rm dust,2}$, $t_{\rm age}$, $\tau_{\rm SF}$, and $Z_{\star}$ for fitting our standard data (spectroscopy + photometry; in red), only spectroscopy (blue), and only photometry (red). The spectroscopy constrains the stellar age and metallicity, while the photometry constrains the dust attenuation. The photometry, together with the spectroscopy, is able to break the dust-age-metallicity degeneracy.}
    \label{fig:corner_wo}
\end{figure*}

\begin{figure*}
    \centering
    \includegraphics[width=\textwidth]{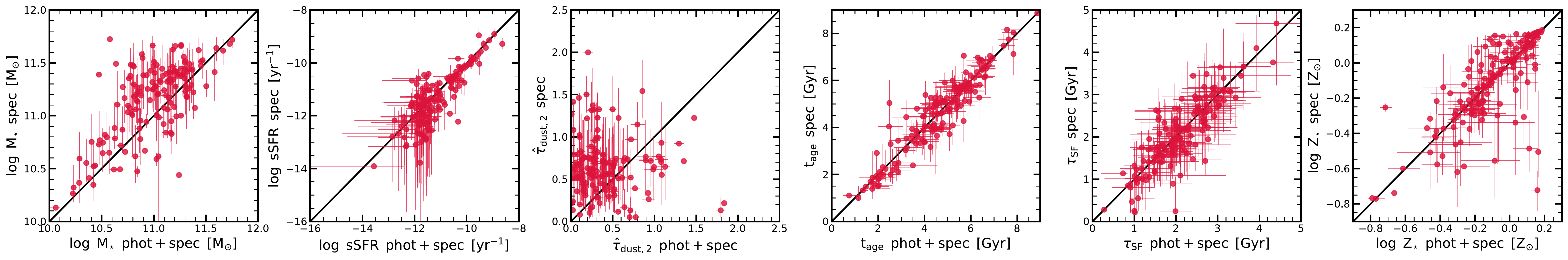}
    \includegraphics[width=\textwidth]{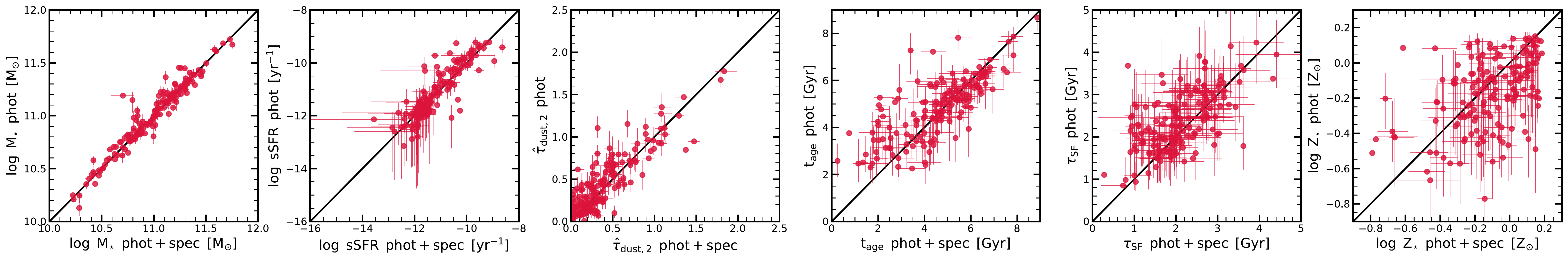}
    \caption{The combination of photometry and spectroscopy is able to break the dust-age-metallicity degeneracy. The top panels compare results for fits to our standard data (spectrum + photometry) with fits to spectroscopy only, while the bottom panels compare our standard results with the ones to photometry only. The top panels show that spectroscopy alone (with observed $I$-band flux) is able to accurately determine the sSFR, age, star-formation timescale $\tau_{\rm SF}$ and metallicity, while there are large uncertainties concerning stellar mass and dust opacity (see also Fig.~\ref{fig:key_quant_wo_unc}). The bottom panels show that the photometry alone (assuming a fixed spectroscopy redshift) is able to put stringent constraints on the stellar mass, sSFR, and dust opacity, while the age and particularly the star-formation timescale remain largely unconstrained.}
    \label{fig:key_quant_wo}
\end{figure*}

\begin{figure*}
    \centering
    \includegraphics[width=\textwidth]{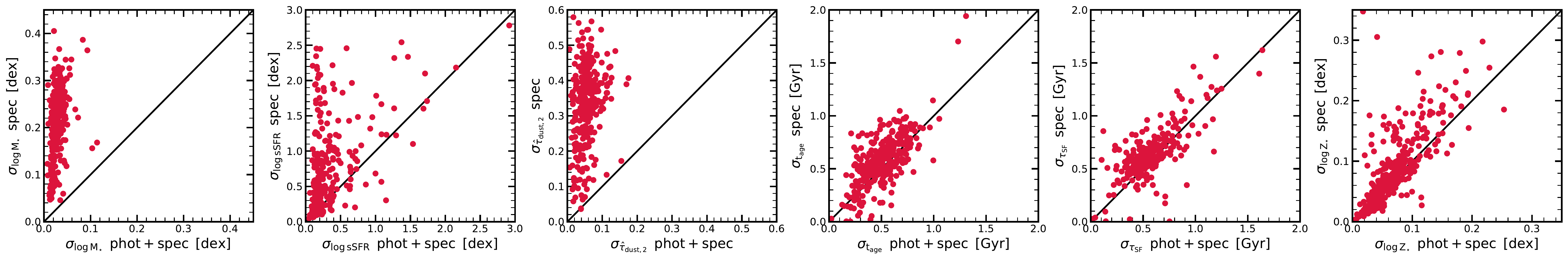}
    \includegraphics[width=\textwidth]{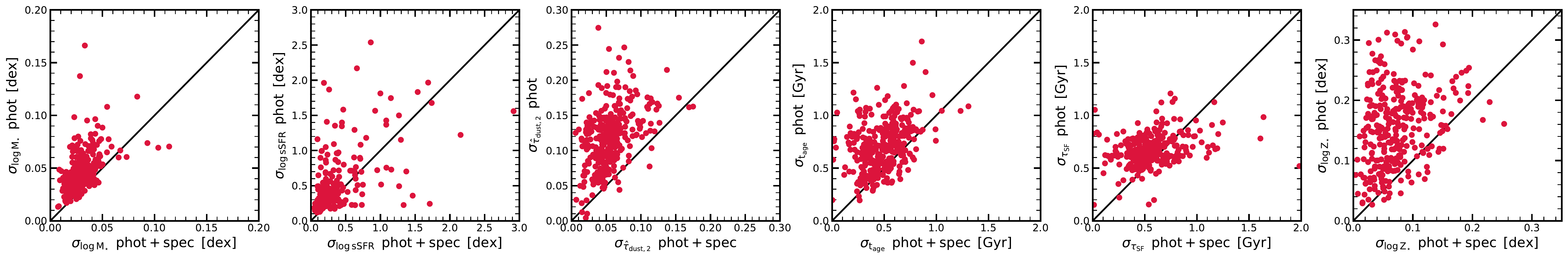}
    \caption{Comparison of the inferred uncertainties using only spectroscopy, only photometry, and the combination of spectroscopy and photometry. This figure follows the same layout as Fig.~\ref{fig:key_quant_wo}. The points typically lie above the 1-to-1 line, indicating that the uncertainties inferred using a combination of spectroscopy and photometry are smaller than those using photometry or spectroscopy alone. There are large uncertainties concerning stellar mass and dust opacity when using spectroscopy alone, while age, star-formation timescale $\tau_{\rm SF}$ and metallicity are well constrained and have comparable errors to the spectroscopy+photometry fits. Using photometry alone, we find that the stellar mass and dust opacity have small uncertainties, while the age and particularly the metallicity have larger uncertainties than the spectroscopy+photometry fits.}
    \label{fig:key_quant_wo_unc}
\end{figure*}

Assessing in detail the gain in fitting both photometry and spectroscopy together is beyond the scope of this paper. Nevertheless, we would like to stress that detailed constraints on the SFHs are only possible when fitting both photometry and spectroscopy. Focusing on photometry alone, \citet[][see also \citealt{wuyts09sim}]{pforr12} study in detail the recovery of galaxy properties from SED fitting. They show that coverage from the rest-frame ultraviolet to the rest-frame near-infrared appears to be optimal in recovering age, stellar mass, and dust reddening. Similarly, \citet{leja17} discuss how different SED parameters shape the overall SED, specifically showing that the UV wavelength range is sensitive to the SFH, but also that they are degenerate with dust attenuation and metallicity. \citet{johnson21} provide a more rigorous analysis on combining spectra with photometry by performing a full mock simulation. The present analysis shows that combining photometry and spectroscopy significantly improves derivation of parameters.

This analysis focuses on the spectroscopic data available in this work. Although the redshift range probed by our galaxies is rather large ($z_{\rm obs}=0.4-1.2$), thanks to the broad wavelength coverage ($\lambda_{\rm obs}=4600-9500~\mathrm{\AA}$), key absorption features are covered by all galaxies (see Section~\ref{subsec:observations}): the hydrogen absorption lines from H10 (found at $3799~\mathrm{\AA}$) to H$\beta$ (at $4863~\mathrm{\AA}$), the Calcium H and K lines (at $3934~\mathrm{\AA}$ and $3969~\mathrm{\AA}$), the CN line (at $4160~\mathrm{\AA}$), the MgIb triplet (at $5176~\mathrm{\AA}$), and several other Mg (at $5530~\mathrm{\AA}$), Ca (including at $4227~\mathrm{\AA}$ and $4455~\mathrm{\AA}$) and Fe lines (including at $4383~\mathrm{\AA}$, $4531~\mathrm{\AA}$, $4668~\mathrm{\AA}$, and $5270~\mathrm{\AA}$). Only galaxies probing the highest redshifts ($z_{\rm obs}>1$) do not have coverage of the MgIb triplet and the H$\beta$ line.

We now show how our results change when fitting only photometry, only spectroscopy, and both photometry and spectroscopy together. For all those fits, we assume the same model. Fig.~\ref{fig:corner_wo} compares the posterior distributions from fitting photometry only (green), from fitting spectroscopy only (blue), and from combining photometry and spectroscopy (red). We again consider the galaxy with ID 16490 (same as shown in Figs.~\ref{fig:example_spec} and \ref{fig:example_corner}). In the ``spectroscopy only'' case, we actually include the observed $I$-band to normalize for the otherwise unconstrained stellar mass. We confirm the expected trends \citep[e.g.,][]{gallazzi05, conroy13_rev}: the dust attenuation and the stellar mass are mainly constrained by the photometry, while the age and metallicity are mainly driven by the spectroscopy. Fig.~\ref{fig:key_quant_wo} confirms that the trends for galaxy ID 16490 also hold throughout our sample. Specifically, Fig.~\ref{fig:key_quant_wo} compares the spectroscopy-only with the photometry+spectroscopy run (top panels) and the photometry-only with the photometry+spectroscopy run (bottom panels). The specific sSFR (second panels from the left) is averaged over 100 Myr and both photometry and spectroscopy contribute about equally to its determination. The second panels from the left highlight that the star-formation timescale $\tau_{\rm SF}$ is mainly constrained by the spectroscopy. Fig.~\ref{fig:key_quant_wo_unc}, following the same layout as Fig.~\ref{fig:key_quant_wo}, compares the inferred uncertainties using only spectroscopy, only photometry, and the combination of spectroscopy and photometry. For each galaxy, we estimate (and plot) the uncertainty by both computing the difference between the $16^{\rm th}$ and $50^{\rm th}$ percentiles and the difference between the $50^{\rm th}$ and $84^{\rm th}$ percentiles. There are large uncertainties concerning stellar mass and dust opacity when using spectroscopy alone, while age, star-formation timescale $\tau_{\rm SF}$ and metallicity are well constrained and have comparable errors to the spectroscopy+photometry fits. Using photometry alone, we find that the stellar mass and dust opacity have small uncertainties, while the age and particularly the metallicity have larger uncertainties than the spectroscopy+photometry fits.

A closer look, however, shows an interesting detail when focusing on the age posterior in Fig.~\ref{fig:corner_wo}: we find that the age posterior of the spectroscopy+photometry lies in between the photometry-only and spectroscopy-only cases, while the individual photometry-only and spectroscopy-only constraints are not formally very consistent with each other. There is important age information in the photometry as well. This is confirmed when looking at the SFH (top right inset in Fig.~\ref{fig:corner_wo}), which shows that the SFH lies actually closer to the photometry-only posterior than the spectroscopy-only posterior in the intermediate age range of $\sim0.5-3~\mathrm{Gyr}$. This is not too surprising since we know that the rest-frame UV is sensitive to a wide range of ages \citep[e.g.,][]{conroy13_rev}, but it highlights the importance of being able to model the UV accurately. This means that we need to fully marginalize over possible variations in the dust attenuation law (see Section~\ref{subsec:dust_attenuation_model}).

\section{Variations of the model}
\label{app_sec:variations}

We briefly present here variations to our physical model (Section~\ref{sec:physical_model}), in particular focusing on the number of time bins describing the SFH and the effect of assuming a parametric and a non-parametric SFH. 

\subsection{Increasing the number of time bins of the SFH}

\begin{figure*}
    \centering
    \includegraphics[width=\linewidth]{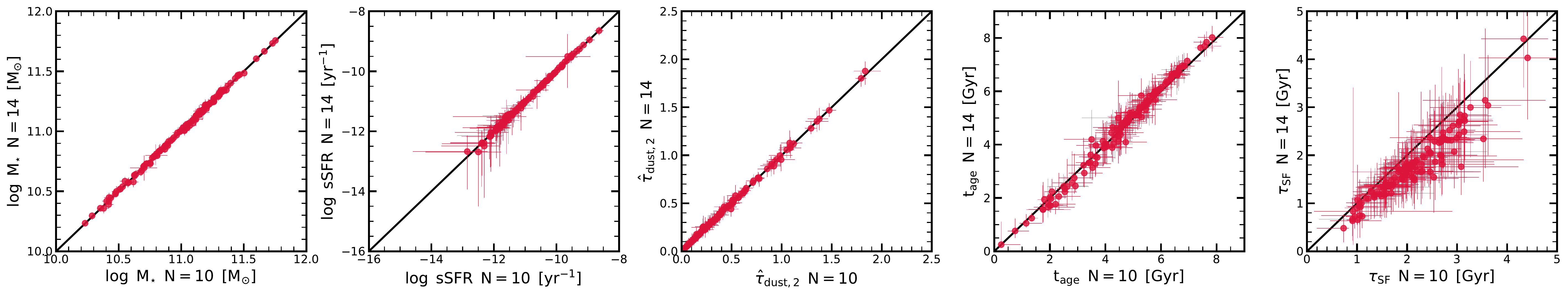}
    \caption{Comparison of $M_{\star}$, sSFR, $\tau_{\rm V}$, $t_{\rm age}$, and $\tau_{\rm SF}$ (from left to right) for fits assuming a non-parametric SFH with 10 time bins ($x$-axis) and with 14 time bins ($y$-axis). Each point is a galaxy from our sample. Changing the number of time bins has only a small effect on our inferred quantities. The largest effect concerns $\tau_{\rm SF}$, which is typically shorter when more time bins are used.}
    \label{fig:key_quant_np_np14}
\end{figure*}

As mentioned in the main text, we assume a non-parametric SFH for fitting the photometric and spectroscopic data. We call it non-parametric, because we do not assume any particular parametric form for the shape of the SFH. Nevertheless, the number of time bins ($N_{\rm SFH}$) describing the SFH is a parameter in our prescription. Our fiducial assumption is $N_{\rm SFH}=10$. We have explored fewer and more bins. Obviously, more bins increase the time resolution, which is important for determining, among other quantities, the star-formation and quenching timescales. 

Fig.~\ref{fig:key_quant_np_np14} compares the result of some of our key quantities (stellar mass, sSFR, dust attenuation, age, and star-formation timescale $\tau_{\rm SF}$) for our fiducial $N_{\rm SFH}=10$ run with our $N_{\rm SFH}=14$ run. The 1-to-1 line is indicated as a solid black line. This figure shows that these two runs deliver consistent results, which is comforting. Furthermore, this is consistent with the analysis by \citet{leja19_nonparm}, who explored varying the number of time bins between $N_{\rm SFH}=4-14$ and showed that the results of the mock analysis are largely insensitive to the number of bins as long as $N_{\rm SFH}\ga4$. We choose our fiducial number of time bins to be 10 because the $N_{\rm SFH}=14$ run leads to non-convergence of 30 galaxies within a reasonable amount of time (i.e., 14 CPU days).

\subsection{Parametric versus non-parametric SFH}

\begin{figure}
    \centering
    \includegraphics[width=0.75\textwidth]{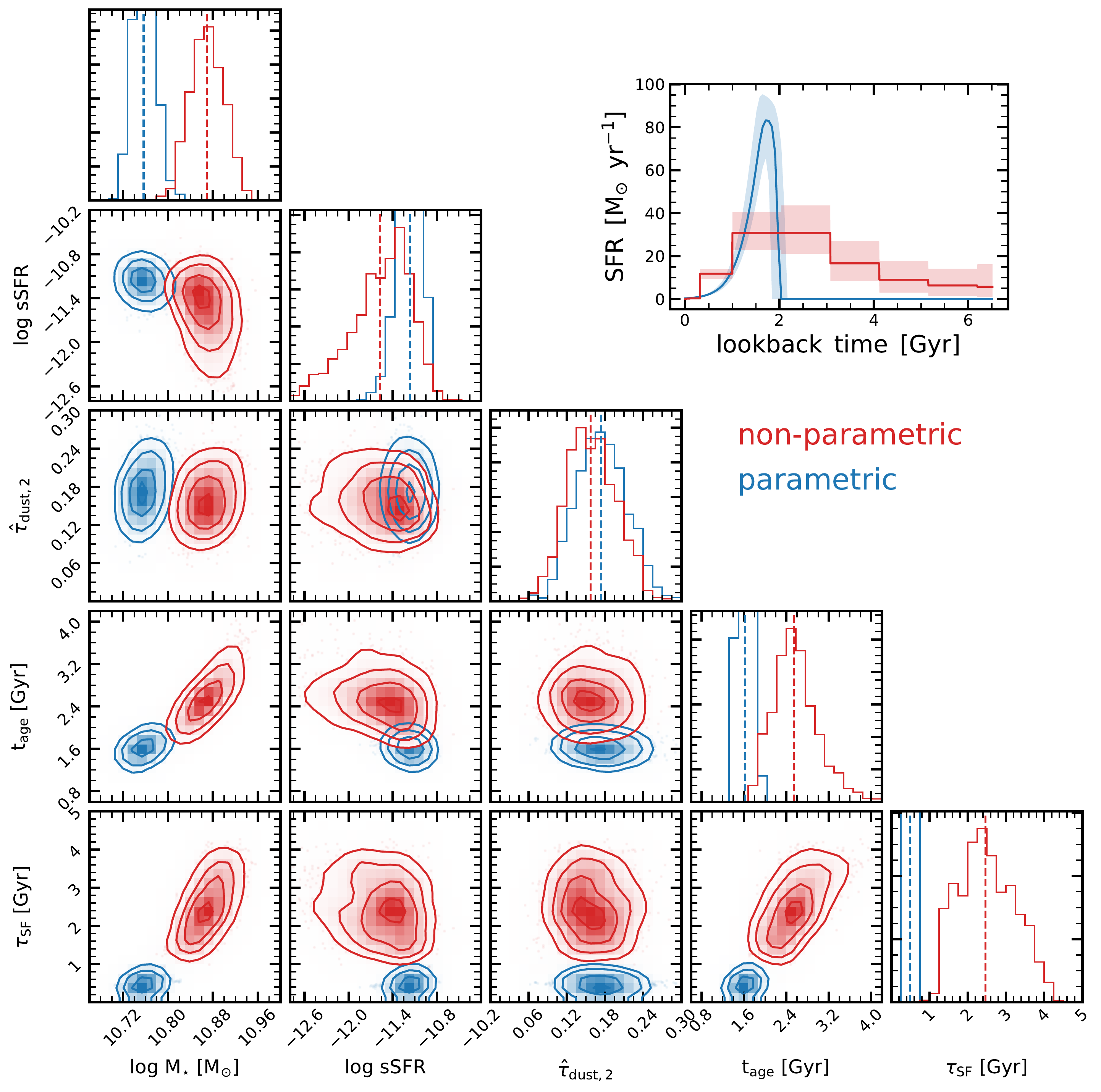}
    \caption{Joint posterior plot of key quantities ($M_{\star}$, sSFR, $\hat{\tau}_{\rm dust,2}$, $t_{\rm age}$, and $\tau_{\rm SF}$) assuming a non-parametric (red) and a parametric (blue) SFH. This is the same galaxy (ID 16490) as shown in Fig.~\ref{fig:example_corner}. The parametric SFH is a delayed-tau model (Eq.~\ref{eq:delayed_tau}). The parametric SFH is only consistent with the non-parametric SFH in the last $\sim1.5$ Gyr; at earlier times, the parametric SFH is 0 as given by the functional form, while the non-parametric SFH is significantly above 0. This results in an older age for the non-parametric SFH ($t_{\rm age}=2.54_{-0.40}^{+0.40}~\mathrm{Gyr}$) than for the parametric SFH ($t_{\rm age}=1.62_{-0.22}^{+0.14}~\mathrm{Gyr}$), which leads to a larger stellar mass and lower sSFR for the non-parametric SFH than for the parametric SFH. Similarly, the star-formation timescale $\tau_{\rm SF}$ is smaller for the parametric SFH than for the non-parametric SFH. Importantly, due to its parametric shape, the uncertainty of $\tau_{\rm SF}$ is probably underestimated in the parametric SFH case. }
    \label{fig:corner_np_p}
\end{figure}

\begin{figure*}
    \centering
    \includegraphics[width=\linewidth]{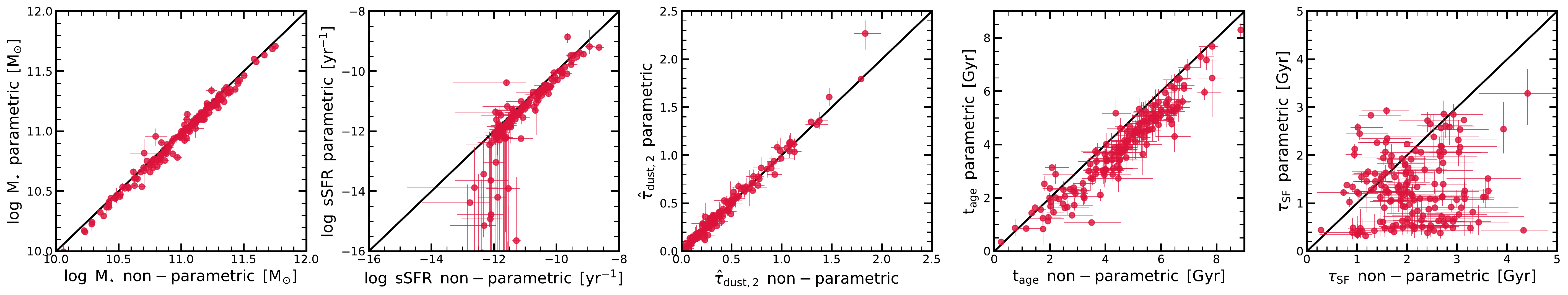}
    \caption{Comparison of $M_{\star}$, sSFR, $\hat{\tau}_{\rm dust,2}$, $t_{\rm age}$, and $\tau_{\rm SF}$ (from left to right) for fits assuming a non-parametric SFH ($x$-axis) and a parametric SFH (i.e., delayed-tau model; $y$-axis). Each point is a galaxy from our sample. Non-parametric SFHs result in older ages ($t_{\rm age}$), larger $M_{\star}$, and higher sSFR than parametric SFHs. The non-parametric sSFRs plateau around $\log~\mathrm{sSFR}\sim-12$, while the parametric sSFRs can fall to much smaller values. There is a large scatter in $\tau_{\rm SF}$ between the parametric and non-parametric approach, which indicates that the parametric prior is too narrow (see Fig.~\ref{fig:prior_tausf}). The dust opacity $\hat{\tau}_{\rm dust,2}$ is consistent for the two different SFH assumptions.}
    \label{fig:key_quant_np_p}
\end{figure*}

We highlight here some of the differences that arise when assuming a parametric SFH instead of our fiducial non-parametric SFH. We assume a delayed-tau model for the parametric SFH as defined in Eq.~\ref{eq:delayed_tau}. As shown in Appendix~\ref{app_sec:prior_impact}, assuming a parametric SFH imposes strong priors on several key quantities, including age (formation redshift), star-formation timescale, and quenching timescale (Figs.~\ref{fig:prior_tausf} and \ref{fig:prior_tauq}). 

Fig.~\ref{fig:corner_np_p} shows the posteriors for key quantities ($M_{\star}$, sSFR, $\hat{\tau}_{\rm dust,2}$, $t_{\rm age}$, and $\tau_{\rm SF}$) assuming a non-parametric (red) and a parametric (blue) SFH. These are the posteriors of the same galaxy (ID 16490) as shown in Fig.~\ref{fig:example_corner}. The posterior SFHs are significantly different in two cases: while the SFHs of the non-parametric and parametric are similar in the last $\sim1.5$ Gyr (where the SNR is the highest), they diverge at earlier times. In particular, the parametric SFH results in a much higher peak SFR of $>80~M_{\star}~\mathrm{yr}^{-1}$ than the non-parametric one (peak SFR of $\sim30~M_{\star}~\mathrm{yr}^{-1}$). Given by its parametric form, the parametric SFH declines rapidly at earlier times and is consistent with 0 at a look-back time beyond 2 Gyr. This leads to a short star-formation timescale, consistent with Fig.~\ref{fig:prior_tausf}, which shows that long star-formation timescales are omitted by this prior. On the other hand, the non-parametric SFH peaks and declines more slowly toward earlier cosmic times. This results in older ages, larger $M_{\star}$, and lower sSFR in the non-parametric SFH case than in the parametric SFH case.

Fig.~\ref{fig:key_quant_np_p} compares the key quantities of the non-parametric SFH run ($x$-axis) with the ones of the parametric SFH run ($y$-axis). Consistent with Fig.~\ref{fig:corner_np_p}, we find a significant difference for $M_{\star}$, sSFR and $t_{\rm age}$. There are no significant differences for the dust opacity $\hat{\tau}_{\rm dust,2}$. This is overall consistent with \citet{carnall19_sfh}, who showed that parametric SFH models (including delayed-tau models) bias measurements such as SFR and stellar masses by imposing a strong prior preference for young stellar populations. They conclude -- and we support it with our analysis here -- that this makes it challenging for parametric SFH models to understand mass assembly in galaxies. Non-parametric SFH models are better suited because they are less biased and return more accurate errors than parametric SFHs \citep{leja19_nonparm}.

\section{Prior imprint on results}
\label{app_sec:prior_impact}

\begin{figure*}
    \centering
    \includegraphics[width=\textwidth]{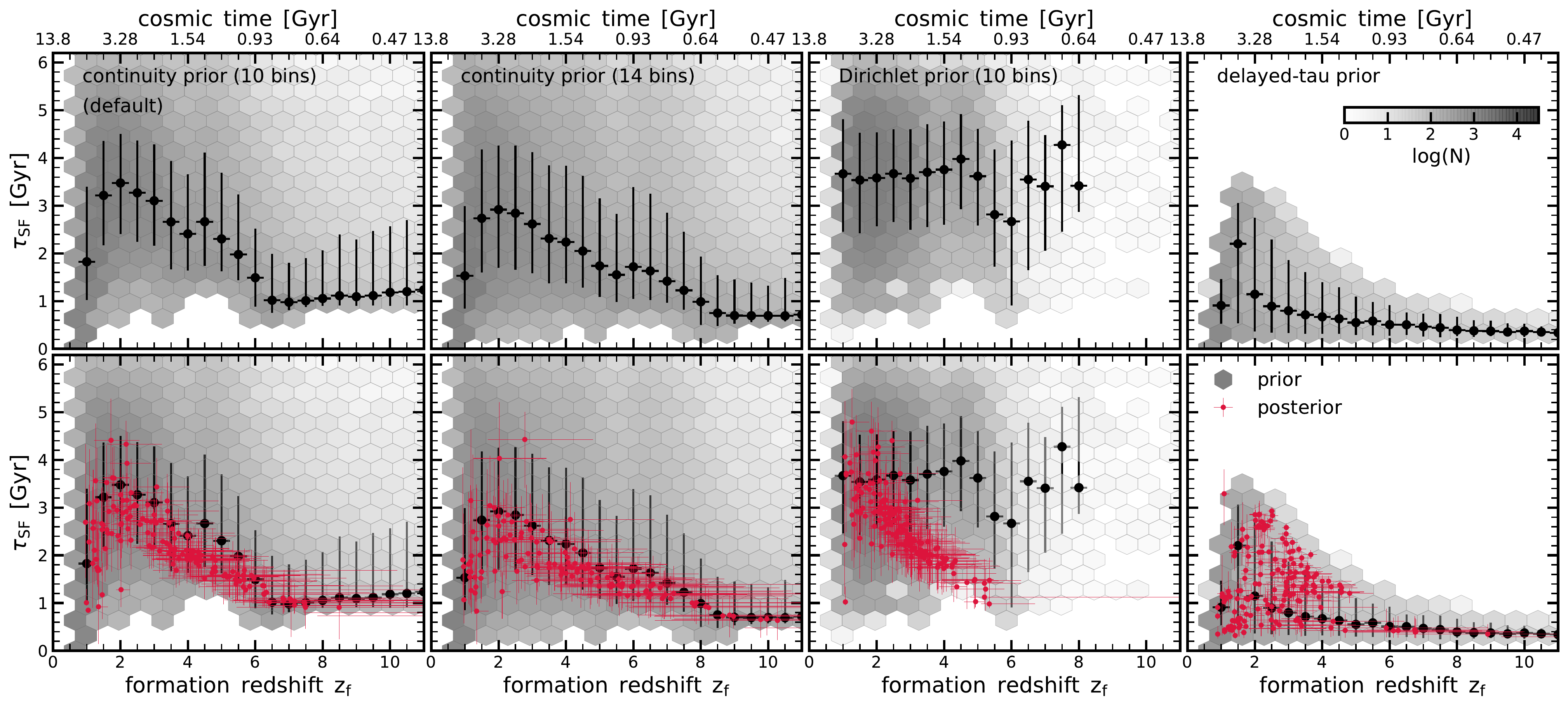}
    \caption{Comparison of priors and posteriors in the $\tau_{\rm SF}-z_{\rm f}$ plane. We investigate how adopting different priors on the SFH impacts our results. The upper panels show the prior distributions for both star-forming and quiescent galaxies at the epoch of observations, while the lower panels add the resulting posteriors (each galaxy is represented by a red point with the associated uncertainty). The black points with errorbars show the median and $16^{\rm th}-84^{\rm th}$ percentile of $\tau_{\rm SF}$ in bins of $z_{\rm f}$ of the prior. The adopted priors include, from left to right, the continuity prior with 10 time bins (our standard prior), the continuity prior with 14 time bins, the Dirichlet prior with 10 time bins, and the parametric delayed-tau prior. The continuity prior covers the $\tau_{\rm SF}-z_{\rm f}$ plane more homogeneously than do the Dirichlet and delayed-tau priors, and is also preferred according to the Bayes factor. Importantly, although the inferred trend in the posterior of shorter $\tau_{\rm SF}$ toward larger $z_{\rm f}$ is similar in the continuity prior distribution, this trend is also present in the inference where Dirichlet prior is assumed, which itself does not invoke any trend of $\tau_{\rm SF}$ with $z_{\rm f}$.}
    \label{fig:prior_tausf}
\end{figure*}

\begin{figure*}
    \centering
    \includegraphics[width=\textwidth]{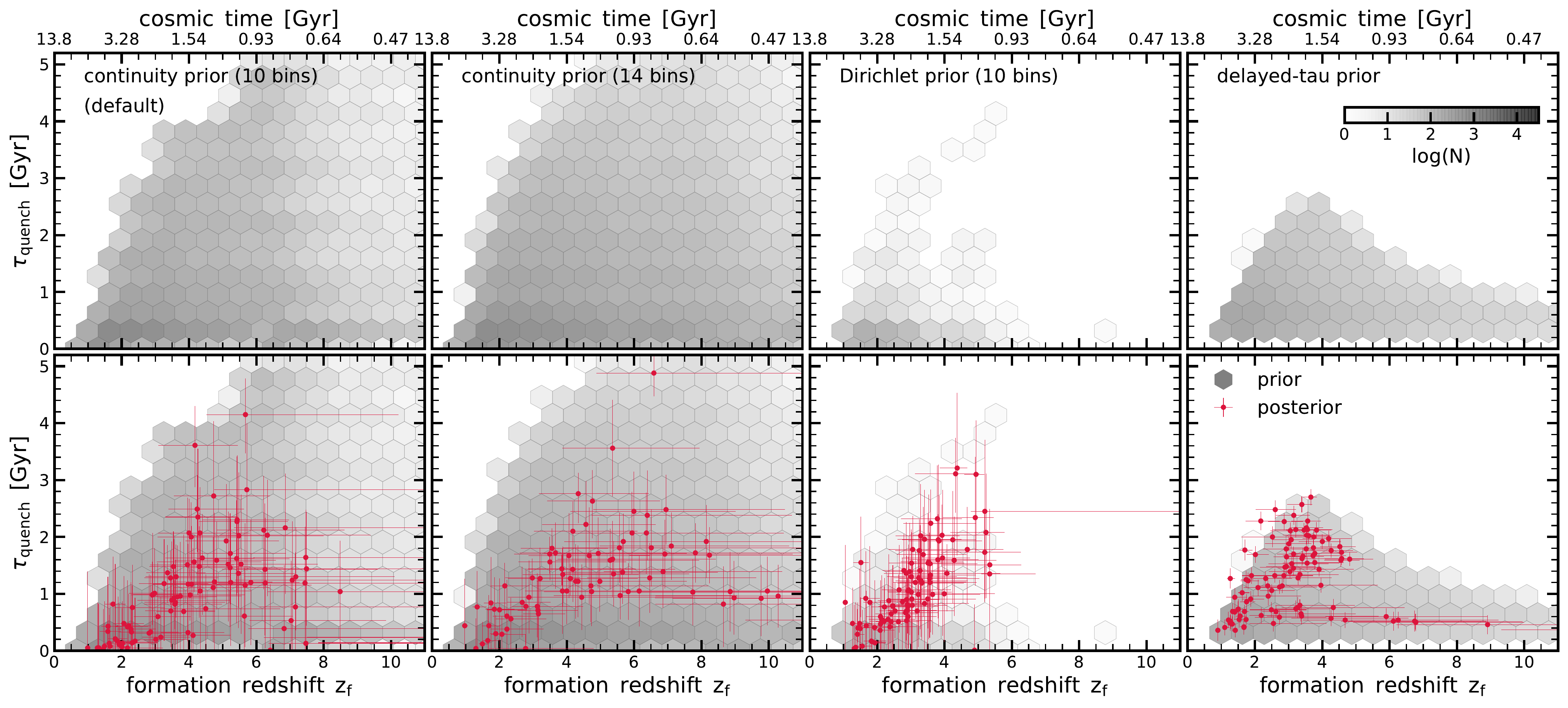}
    \caption{Similar to Fig.~\ref{fig:prior_tausf}, but for the $\tau_{\rm quench}-z_{\rm f}$ plane. The priors here only include quiescent galaxies (at the epoch of observations). The continuity prior also samples this space more homogeneously than do the Dirichlet and parametric delayed-tau priors. The prior spaces exclude slowly quenching galaxies that formed recently since such objects are nonphysical. }
    \label{fig:prior_tauq}
\end{figure*}

In addition to the exquisite photometric and spectroscopic data, our analysis relies on utilizing the sophisticated Bayesian inference framework \texttt{Prospector} \citep{johnson21}. An important step in the analysis is defining the priors of the different parameters and also understanding what their effect is on our results. Crucially, we want to give the parameters enough flexibility so that a wide range of different, physically plausible outcomes are allowed and probed. 

We focus this on the priors for the SFH, since our key results on the star-formation timescale $\tau_{\rm SF}$ and quenching timescale ($\tau_{\rm quench}$) are closely related to it. In Fig.~\ref{fig:prior_sfh} we show, in the space of these two quantities, or fiducial SFH prior, which is the continuity prior that fits directly for $\Delta\log(\mathrm{SFR})$ between adjacent time bins. This prior explicitly weights against sharp transitions in $\mathrm{SFR}(t)$ (see Table~\ref{tab:parameters}). 

We now investigate two additional SFH priors: the non-parametric Dirichlet prior and the parametric delayed-tau prior. The Dirichlet non-parametric prior specifies that the fractional sSFR for each time bin follows a Dirichlet distribution \citep{leja17}. The Dirichlet distribution additionally requires a concentration parameter ($\alpha_{\rm D}$), which controls the preference to put all of the weight in one bin ($\alpha_{\rm D}<1$, i.e., bursty SFH) versus distributing the weight evenly between all bins ($\alpha_{\rm D}\geq1$, i.e., smooth SFH). We set $\alpha_{\rm D}=1.0$. The parametric delayed-tau prior is the well-known delayed-tau model for the SFH (see Eq.~\ref{eq:delayed_tau}), with a log-uniform prior between 0.1 and 10 Gyr for $\tau$ and a uniform prior between 1 Myr and the age of the universe at the epoch of observations for $t_a$. 

Each panel in Fig.~\ref{fig:prior_tausf} shows the star-formation timescale $\tau_{\rm SF}$ versus formation redshift $z_{\rm f}$. Different priors are shown from left to right: continuity prior with 10 time bins, continuity prior with 14 time bins, Dirichlet prior with 10 time bins, and the parametric delayed-tau prior. The gray hexagons show the prior distribution obtained from randomly sampling the prior (assuming a fixed redshift $z_{\rm obs}=0.8$). The upper panels show solely the prior, while the bottom panels overlay our measurements in red (each point represent a galaxy). The black points with errorbars show the median and $16^{\rm th}-84^{\rm th}$ percentiles of $\tau_{\rm SF}$ in bins of $z_{\rm f}$ of the prior. Fig.~\ref{fig:prior_tauq} follows the same layout but plots the quenching timescale $\tau_{\rm quench}$ versus formation redshift $z_{\rm f}$, focusing only on quiescent galaxies (i.e., after randomly sampling from the prior, we only keep the samples that represent a quiescent galaxy at the time of observation). 

Ideally, we want to have a prior that spans homogeneously the widest possible range in the planes of $\tau_{\rm SF}-z_{\rm f}$ and $\tau_{\rm quench}-z_{\rm f}$. There are physical boundaries, namely, the age of the universe (i.e., $z_{\rm f}<\infty$) and the quenching timescale cannot be too large in comparison with the stellar age (no prior volume in the upper left regions of the panels in Fig.~\ref{fig:prior_tauq}). Figs.~\ref{fig:prior_tausf} and \ref{fig:prior_tauq} show that the continuity prior spans quite homogeneously the largest volume in the considered spaces. The Dirichlet prior overall prefers lower formation redshifts (younger ages) and slightly longer star-formation timescales than found by the continuity prior. Our measurements (red points) reflect this: we find smaller formation redshifts and longer $\tau_{\rm SF}$. It is, however, reassuring that the overall distribution of the measurements in both the $\tau_{\rm SF}-z_{\rm f}$ and $\tau_{\rm quench}-z_{\rm f}$ planes is similar.

The inferred trend in the posterior of shorter $\tau_{\rm SF}$ toward larger $z_{\rm f}$ (see Fig.~\ref{fig:tau_sf_vs_mass}) is also present in the continuity prior distribution (black points), though the inferred trend lies at slightly smaller $\tau_{\rm SF}$ at fixed $z_{\rm f}$ than what is given by the prior. More importantly, the Dirichlet prior does not show such a trend, i.e. the median $\tau_{\rm SF}$ as a function of $z_{\rm f}$ is constant for this prior. Nevertheless, the observational data clearly prefer a shorter $\tau_{\rm SF}$ at larger $z_{\rm f}$, giving rise to a similar $\tau_{\rm SF}-z_{\rm f}$ trend despite the different priors. Therefore, we conclude that the $\tau_{\rm SF}-z_{\rm f}$ trend is robustly detected in our data.

The parametric delayed-tau prior probes a much smaller region in both the $\tau_{\rm SF}-z_{\rm f}$ and $\tau_{\rm quench}-z_{\rm f}$ planes: this prior, by construction, does not allow for long star-formation and quenching timescales. Furthermore, galaxies that formed early need to have short star-formation timescales and quenching timescales. Our measurements basically ``fill'' the full prior volume, which raises the worry this prior might be too restrictive. 

Focusing on the continuity prior, we find that the prior also does not allow short $\tau_{\rm SF}$, but this prior is still broader than the Dirichlet prior. This is a fundamental problem when assuming time bins of finite width. We are basically limited by the width of the time bins, which can be as wide as 1 Gyr when assuming 10 SFH bins. This problem is significantly alleviated when increasing this to 14 bins, but a more flexible approach might be needed to fully solve this problem \citep{iyer17, iyer19}. Therefore, assuming 10 time bins, one has to worry that we only infer upper limits for $\tau_{\rm SF}$, in particular at $z_{\rm f}>4$. However, it is reassuring to find that, when increasing to 14 time bins, the measurements do not change, in particular, in the intermediate range of $z_{\rm f}\approx4-6$. Therefore, we conclude that the $\tau_{\rm SF}$ measurements out to $z_{\rm f}\approx6$ are robust. Beyond this redshift, $\tau_{\rm SF}$ should be interpreted with care and shorter $\tau_{\rm SF}$ cannot be ruled out by our measurements. 

Do our observations prefer any prior? For this, we look at the ratio of the evidences, i.e., Bayes factor. We find $\mathcal{Z}_{\rm np,14}/\mathcal{Z}_{\rm np,10}=1.3$, $\mathcal{Z}_{\rm np,dir}/\mathcal{Z}_{\rm np,10}=6\times10^{-4}$ and $\mathcal{Z}_{\rm delayed-tau}/\mathcal{Z}_{\rm np,10}=0.14$. This shows that the non-parametric SFH model with a continuity prior is significantly preferred over the one with a Dirichlet prior and it is substantially preferred over the parametric delayed-tau model. We find that -- unsurprisingly -- more SFH bins ($N=14$) are slightly preferred over fewer SFH bins ($N=10$), but fitting more than 10 bins make the analysis more expensive to run. In summary, our fiducial SFH prior seems to be preferred from the data and provides enough flexibility to the quantities of key interest in this study.

\section{Trends in the UVJ color-color diagram}
\label{app_sec:uvj}

\begin{figure}
    \centering
    \includegraphics[width=\textwidth]{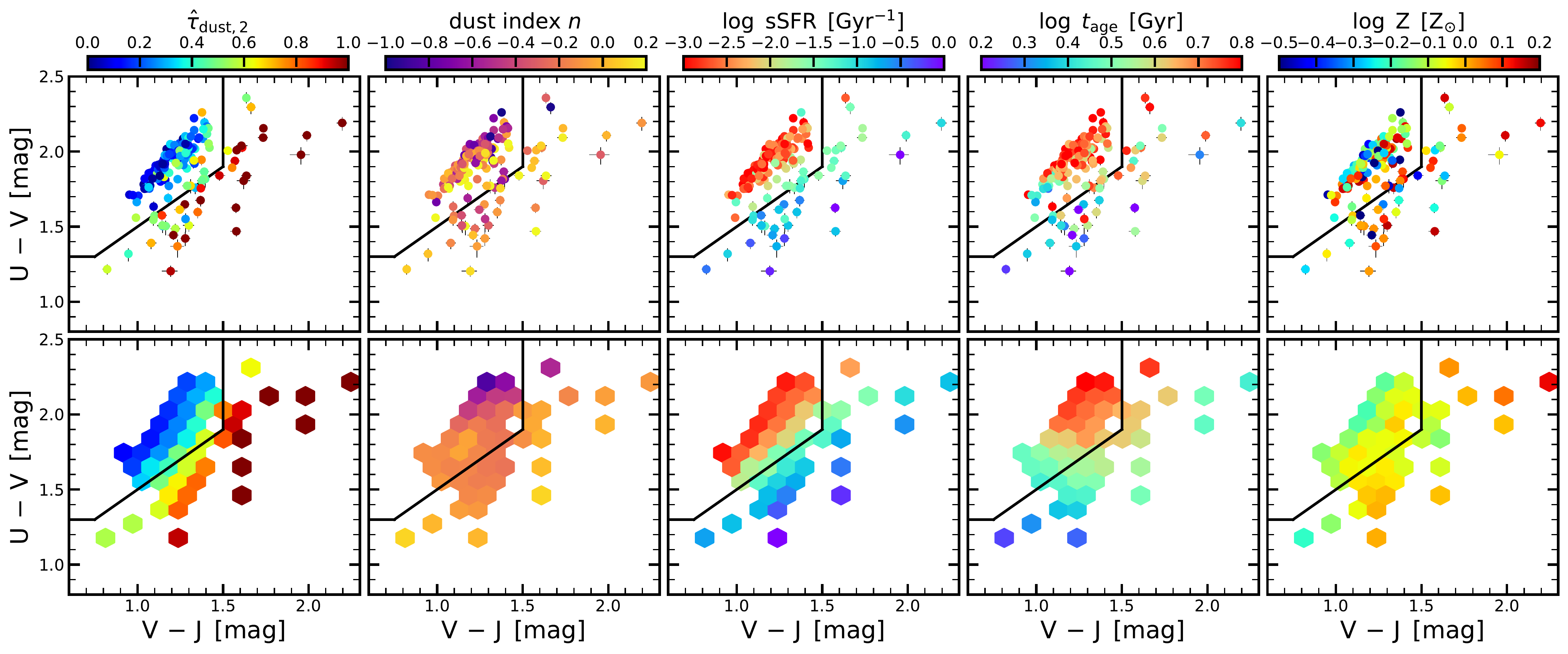}
    \caption{Median stellar population properties in the UVJ diagram of our sample. From left to right: dust optical depth, dust index (relative to \citealt{calzetti00}), sSFR averaged over the most recent 100 Myr, average stellar age, and stellar metallicity. The top panels show the individual data points, while the bottom panels show the average trends using the LOESS method. We recover the well-known trends that UVJ-quiescent galaxies are less dusty, have lower sSFR, and are older. We find a weak trend of increasing age toward the upper right corner in the UVJ-quiescent box. However, we miss galaxies in the bottom left of the UVJ quiescent box (i.e., post-starburst galaxies). }
    \label{fig:uvj_plane}
\end{figure}

\begin{figure}
    \centering
    \includegraphics[width=0.9\textwidth]{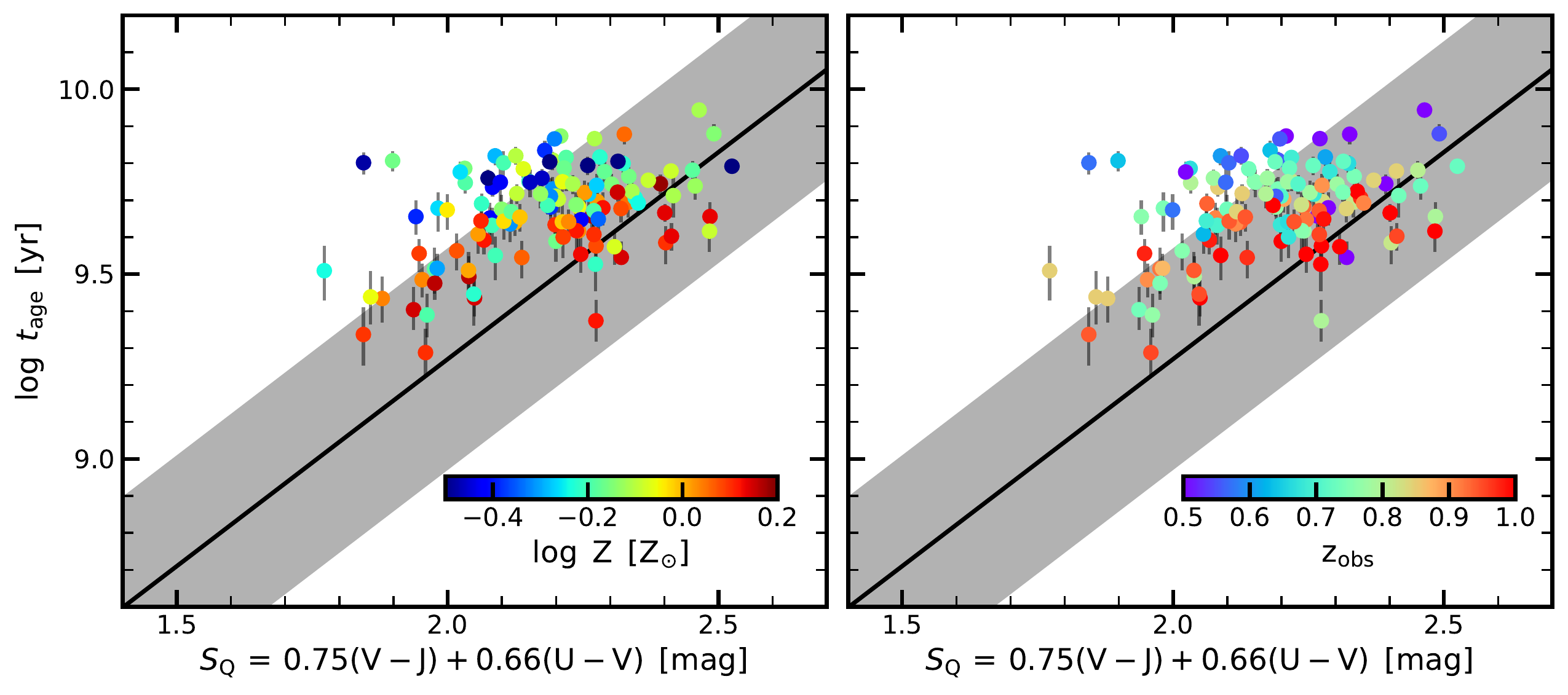}
    \caption{Projection within the UVJ-quiescent box. Focusing now solely on UVJ-quiescent galaxies, we plot the average age ($t_{\rm age}$) as a function of $S_{\rm Q}$, which is a projection along the UVJ-quiescent sequence. The black solid line is the correlation found by \citet{belli19}, who studied galaxies at $1.5<z_{\rm obs}<2.5$. We do not find a relation between age and $S_{\rm Q}$. In particular, the relation is flat for galaxies with low $z_{\rm obs}$.}
    \label{fig:uvj_projection}
\end{figure}

The UVJ color-color diagram has been popular for separating star-forming and quiescent galaxies \citep[e.g.,][]{williams09, muzzin13, whitaker13} and also, in recent times, for informing galaxy evolutionary pathways \citep[e.g.,][]{belli19, carnall19}. \citet{leja19_uvj} show that the actual $U-V$ and $V-J$ colors alone only marginally constrain the stellar populations (small difference between the prior and the posterior), but the underlying scaling relations between age, dust, and metallicity give rise to strong trends in the UVJ space. 

We confirm these general trends with our measurements. Fig.~\ref{fig:uvj_plane} shows our sample in the UVJ space color-coded by dust opacity, dust index, sSFR, age, and metallicity from left to right. The top panels show the individual galaxies, while the bottom panels show the average trends inferred with the LOESSS method. We find that the galaxies in the UVJ-quiescent box have typically less dust, have lower sSFRs, and are older. There is only a weak trend with metallicity and dust index; the reddest quiescent galaxies seem to have the steepest attenuation law. 

We find that 19.8\% of UVJ-quiescent galaxies show signs of star formation with $\mathscr{D}>1/20$, i.e., belong to the transition or star-forming population. This contamination fraction is consistent with recent studies using using spectroscopic information \citep[e.g.,][]{belli17, schreiber18} or SED fitting \citep[e.g.,][]{moresco13, fang18}, which find that UVJ-quiescent selection includes $\sim10\%-30\%$ contamination from star-forming galaxies. We find that the main reason for this contamination stems from the dust attenuation law: non-quiescent galaxies (high-sSFR objects) are in the UVJ-quiescent region because they have a steep attenuation law ($n\approx-1$ in Eq.~\ref{eq:diffuse_dust}) with non-negligible amounts of dust and star formation.

Fig.~\ref{fig:uvj_plane} also shows an age trend within the quiescent box: UVJ-quiescent galaxies toward the upper right tend to be older. This is consistent with the direction of slow aging after quenching and with earlier findings \citep{whitaker13, leja19_uvj, belli19}. In particular, \citet{belli19} present a calibration between the mass-weighted age and $S_{\rm Q}$ at $z=1.5-2.5$, where $S_{\rm Q}=0.75(\mathrm{V}-\mathrm{J}) + 0.66(\mathrm{U}-\mathrm{V})$ is the color along the UVJ quiescent box. 

Fig.~\ref{fig:uvj_projection} shows our galaxies in the plane of age versus $S_{\rm Q}$, colored by the stellar metallicity (left panel) and observed redshift (right panel). The black solid line is the relation found by \citet{belli19}. Fig.~\ref{fig:uvj_projection} shows that galaxies do not follow the trend found by \citet{belli19}. We suspect two reasons for this. First, related to the methodology, \citet{belli19} assume a rather tight prior for the stellar metallicity (Gaussian prior with mean of solar metallicity and $\sigma=0.005$), while we allow a much wider range (flat prior in the range $-1.0<\log(Z/Z_{\odot}<0.2$; Table~\ref{tab:parameters}). The left panel of Fig.~\ref{fig:uvj_projection} shows that most deviant galaxies from \citet{belli19} relation are the ones with rather low metallicities, which hints that this effect could play a role. Another reason for this disagreement might be evolutionary processes within the quiescent galaxy population itself: the sample of \citet{belli19} includes galaxies at $z>1.5$, while we focus on galaxies at $z\approx0.8$. The right panel of Fig.~\ref{fig:uvj_projection} indeed shows that galaxies with $z_{\rm obs}>0.8$ are more consistent with the relation by \citet{belli19} than galaxies with $z_{\rm obs}<0.8$. In summary, this shows that one needs to be careful to apply this relation -- which itself might evolve with cosmic epoch -- to infer evolutionary paths of galaxies.

\end{document}